\pdfoutput=1

\documentclass[twocolumn]{aastex62}

\usepackage{graphicx}
\usepackage{epstopdf}
\usepackage{natbib}
\usepackage{amssymb}
\usepackage{amsmath}
\usepackage{float}
\usepackage{chngcntr}

\def\beq{\begin{equation}}
\def\eeq{\end{equation}}

\received{10 July 2019}
\revised{30 July 2019}
\accepted{1 August 2019}
\submitjournal{ApJS}

\shorttitle{M4 K2 Variable Catalog} 
\shortauthors{Wallace, Hartman, Bakos et al.}

\begin{document}

\title{A Search for Variable Stars in the Globular Cluster M4 with K2}

\author[0000-0001-6135-3086]{Joshua J. Wallace}
\affiliation{Department of Astrophysical Sciences, Princeton
  University, 4 Ivy Ln, Princeton, NJ 08544, USA}

\author[0000-0001-8732-6166]{Joel D. Hartman}
\affiliation{Department of Astrophysical Sciences, Princeton
  University, 4 Ivy Ln, Princeton, NJ 08544, USA}

\author[0000-0001-7204-6727]{G\'asp\'ar \'A. Bakos}
\altaffiliation{MTA Distinguished
Guest Fellow, Konkoly Observatory}
\affiliation{Department of Astrophysical Sciences, Princeton
  University, 4 Ivy Ln, Princeton, NJ 08544, USA}

\author[0000-0002-0628-0088]{Waqas Bhatti}
\affiliation{Department of Astrophysical Sciences, Princeton
  University, 4 Ivy Ln, Princeton, NJ 08544, USA}


\correspondingauthor{Joshua Wallace}
\email{joshuawallace800@gmail.com}

\begin{abstract}
We extract light curves for 4554 objects with $9{<}G{<}19$ in the K2
superstamp observations of the globular cluster M4, including 3784
cluster members, and search for
variability. Among cluster member objects, we detect 66 variables, of
which 52 are new discoveries.  Among objects not belonging to the
cluster, we detect 24 variables, of which 20 are new discoveries.  We
additionally discover 57 cluster-member suspected variables, 10
cluster-non-member suspected variables, and four variables with
ambiguous cluster membership.  Our light curves reach
sub-millimagnitude precision for the cluster horizontal branch,
permitting us to detect asteroseismic activity in six horizontal
branch stars outside the instability strip and one inside the strip
but with only ${\sim}$1 mmag amplitude variability.  19 additional
stars along the red giant branch also have detected asteroseismic
variability. Several eclipsing binaries are found in the cluster,
including a 4.6-day detached eclipsing binary and an EW-class eclipsing
binary, as well as an EW with uncertain cluster membership and three
other candidate EWs.  A 22-day detached eclipsing binary is also found
outside the cluster. We identify a candidate X-ray binary that is a 
cluster member with
quiescent and periodic ${\sim}$20 mmag optical variability.  We also
obtain high-precision light curves for ten of the previously known RR
Lyrae variables in the cluster and identify one as a candidate Blazhko
variable with a Blazhko period in excess of 78 days.  We make our
light curves publicly available.

\end{abstract}



\section{Introduction}
\label{sec:intro}

The globular cluster (GC) M4 (NGC 6121), located in the constellation
Scorpius, is the closest GC to Earth at a distance of
${\sim}$1.8 kpc (\citealt{kaluzny2013b,braga2015,neeley2015}).  M4 is an
old GC, with recent age measurements falling between ${\sim}$11--12 
Gyr (\citealt{bedin2009,kaluzny2013b,vandenberg2013})
 and it has a metallicity of [Fe/H]${\approx}-1.2$
(\citealt{harriscatalog}, 2010 edition).  Given its relative
proximity to us
and also the relative sparseness of its core, M4 is a prime target for
the detailed study of individual GC member stars.

M4 is rich in variable objects---90 in the current count
of \cite{clement2001}, June 2016 edition---such as pulsating 
variables (including 
dozens of RR Lyrae variables), eclipsing binaries, and cataclysmic
variables
(\citealt{clement2001,bassa2004,kaluzny2013a,kaluzny2013b,stetson2014,
samus2017,watson2017} and references therein). Some recent examples of
the scientific utility of these variables include using RR Lyrae
variables for 
 an M4 distance determination 
(e.g., \citealt{braga2015}) and using M4 eclipsing binaries to provide constraints
on the mechanism of formation of close binaries in GCs
(\citealt{kaluzny2013a}). 
Given the large number of variable objects already known in M4 and the 
 scientific impact of both better understanding known variables and
discovering new ones, any data that permits such  is of great value.

\begin{figure*}
\begin{center}
\includegraphics[width=0.85\textwidth]{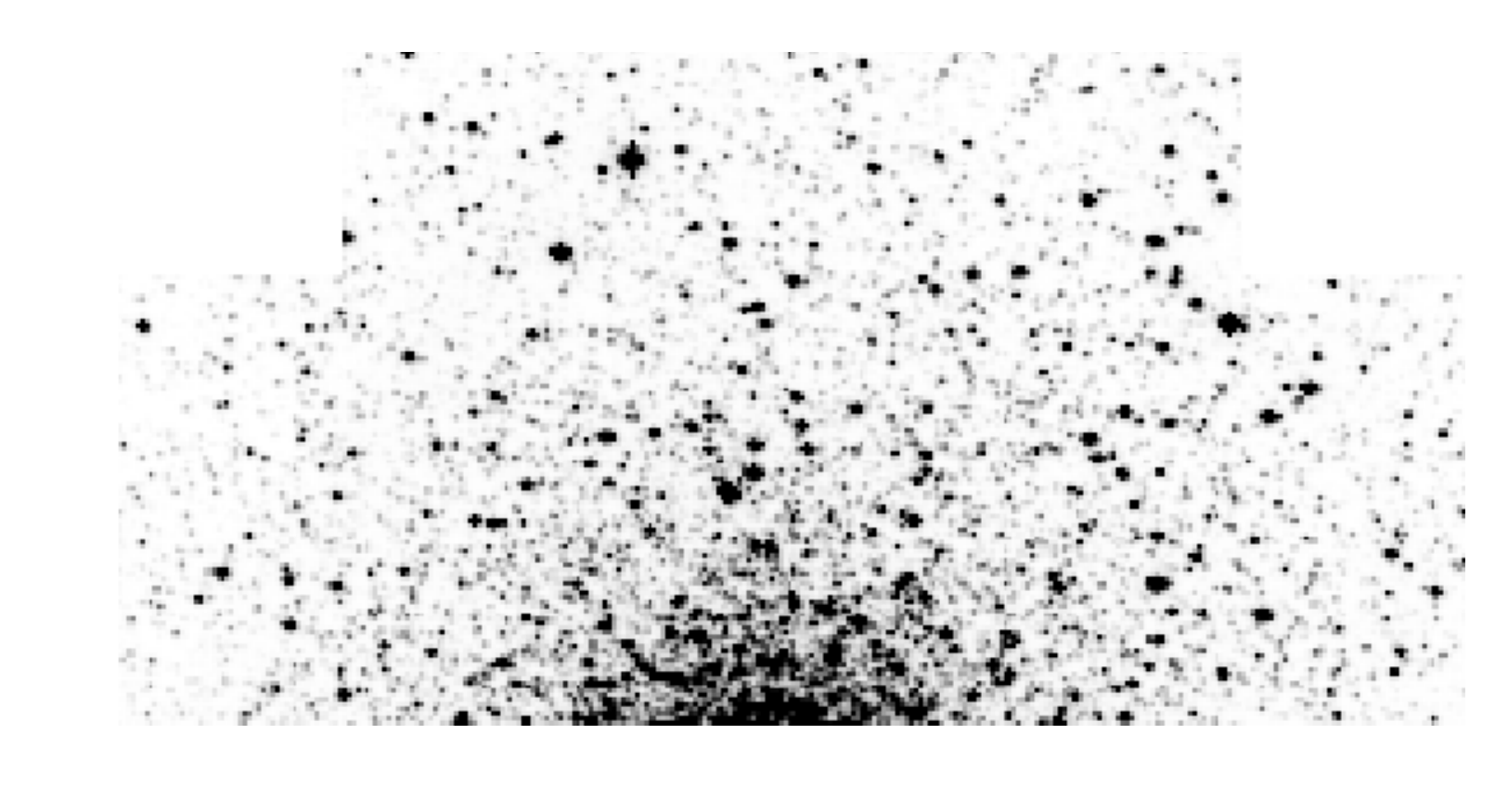}
\end{center}
\caption{\label{fig1} The astrometric reference image of the K2 superstamp of
  M4.  The image is 300 pixels by 150 pixels, or approximately
  20\arcmin\ by 10\arcmin, and is displayed 
with arbitrary z-scale and colors inverted.  The white regions in the
  upper left and right corners are  regions that were not included in the superstamp.  The core of the
  cluster is ${\sim}1$\arcmin\ off of the bottom edge of 
  the image.} 
\end{figure*}

M4 was in the field of
view of the {\it Kepler} telescope during Campaign 2 
(running from 2014 Aug 23 to 2014 Nov 10) of the K2
mission \citep{howell2014}, and continuous observations of a 
portion of this cluster in the form of a ``superstamp'' 
were included in the data downloaded from the observatory.  
These and other K2 observations of GCs represent, by far, the longest
continuous observations of GCs to date, and in the case of M4, the
longest continuous observation of what happens to be the closest
GC. Additionally, these observations were taken by a space-based
observatory designed and built with high-precision photometry as its goal.  This is
a prime data set for an object of great scientific interest and will
likely be the best time series data we have
for a GC for a while to come.

Unfortunately, {\it Kepler}'s design was not optimized for observing
GCs.  Its 3\farcs98/pixel pixel scale leads to significant blending in the 
images, particularly close to the core.  Fortunately, techniques exist
to partially mitigate the effects of the blending, and
given the 
expected richness and value of the derived light curves, the effort to
work 
through these issues is still worthwhile. The present work uses image 
subtraction \citep{imagesubtraction} among other techniques to deal
with the blending, and,  
building off of \cite{wallace2019}, it is, as far as we are aware, the
first general analysis of the K2 observations of a GC. Previous work on
these images were limited in scope: \cite{miglio2016}
looked at asteroseismic oscillations in K giants and 
\cite{kuehn2017} looked at the RR Lyrae variables, but 
that has been it so far.  The results from these limited
searches
demonstrate the incredible potential of the M4 K2 superstamp
data. This work is focused 
more on breadth (production of quality light curves and identification
of variables) rather than depth (full characterization of individual
variable objects) and is only a starting point for analysis of these data.
We describe our methods to extract and analyze
data from the images in Section~\ref{sec:method}, and in
Section~\ref{sec:results} we present the
results of our variability search.  A discussion is presented in
Section~\ref{sec:discussion} and we conclude in
Section~\ref{sec:conclusion}.


\section{Method}
\label{sec:method}
We present here a detailed description of our data reduction
and variable identification pipeline.

\subsection{Image Preparation}

The images we used are the 16 target pixel files (TPFs) that make up the M4
superstamp from the Mikulski Archive for Space Telescopes.  Each is 50
pixels by 50 pixels in dimension. These
files had the K2 EPIC ID numbers 200004370 -- 200004385.  
We stitched the TPFs together using
\texttt{k2mosaic} \citep{barentsen2016}, producing a series of images
with dimensions of 150 pixels
by 300 pixels, each missing two 50 pixel by 50 pixel notches.
These images were ${\sim}$10\arcmin\ by  ${\sim}$20\arcmin\  on the sky.
One of the images is shown in
Figure~\ref{fig1}. The superstamp is not centered on the cluster, but
rather avoids the cluster center, and is focused more on the cluster
outskirts on one side of the cluster.
A total of 3856 superstamp images are produced, one for each
cadence.  
By mission design, 39 of the images had no data recorded as they took
place during 
resaturation events (major thruster fires used to spin up the
reaction wheels) that occurred every 96 cadences and were thus not
usable in our analysis.

Our data extraction and reduction pipeline is very similar to that of
\cite{soaresfurtado}.  After assembling the superstamp images, we used
the \texttt{fistar} tool from
the open-source \texttt{FITSH} software package
(\citealt{pal})
for source detection in
preparation for image registration.  We used an asymmetric Gaussian
model for the point spread function (PSF), 
a detection threshold of 400 ADUs, the default uplink
candidate extraction algorithm, and two symmetric and one general
iterations. From this, we generated a list of source positions,
fluxes, and PSF shape and width parameters for each detected source.
The image with the smallest median PSF full width at half maximum
(FWHM) across all the detected 
sources was chosen as the astrometric reference image.  This smallest
median FWHM was 1.457 pixels, and the collection of median FWHM values
across the 
images  had 
a mean of 1.503 pixels and a standard deviation of 0.018
pixels.   The selected astrometric reference
frame image---the 1197th cadence in the campaign, which is shown in
Figure~\ref{fig1}---also had one of the most
symmetric FWHMs of all the images.

The \texttt{grmatch} tool from \texttt{FITSH} was then used
to match the detected sources in each image to the selected
astrometric reference image and calculate a transformation to register
each image to the astrometric reference image.  To determine the best
parameters for the 
match, a grid was employed consisting of two different transformation
orders (1 and 2) and many different values (170--500) for the maximum
number of sources to select from the reference and image source lists
(ordered by greatest flux to least)
to use for the triangle matching.  We ran the \texttt{grmatch} code
for each 
image for all the parameters on this grid.
For each image, we adopted the set of parameters which maximized the
number of matched objects normalized by the square of the weighted
residual, subject to the restriction that at least 100 objects were
matched, and that the match was accurate (i.e., the weighted residual
reported by \texttt{grmatch} was less than $0.001$ and the reported
unity was greater than $0.015$).
 The
\texttt{FITSH} tool \texttt{fitrans} was then used to register each
image to the frame of the astrometric reference image using the
selected transformation calculated by \texttt{grmatch}.

After registering the images, the next task was to create a
photometric reference image to use for image subtraction. For each
image, the
Euclidean distance (in pixels) of the transformation of a point at the
center of the image to the astrometric reference image and the closeness of the
PSF size and shape (as measured by the median S, D, and K parameters) of the
image to the astrometric reference image were calculated.  Cutoff
values for  the transformation distance and the SDK closeness 
(respectively 0.0998 pixels and 0.1) were selected such that there were 100
images chosen to be used in the creation of a photometric reference
image. The chosen images were taken mostly during the first half
of the campaign, which is unsurprising considering the much larger
drift in the second half of the campaign.  These 100 images were then
median combined using \texttt{ficombine} from \texttt{FITSH} to create
the master 
photometric reference image.  

\subsection{Image Subtraction and Photometry Extraction}
\label{sec:photometry}

\texttt{FITSH}'s \texttt{ficonv} tool 
was then used
to subtract the master photometric reference image from each of the K2
images.  A first-order polynomial was fit to the background and also
subtracted. A constant discrete convolution kernel with a half-size of 4 pixels
was used to match the PSF and flux scale of the reference image to
that of each individual K2 image.  This unfortunately
meant that objects that were within 4 pixels of the edge of the image
(a little less than 1\% of the image, referred to in this work as
``the edge region'') 
were not included in the image subtraction calculation and objects
near to the edge region with parts of their images cut off did not get their
photometry calculated.  Nine isolated,
relatively bright stars across the least crowded portions of the super
stamp (left, right, and upper portions) were selected by eye and used
to optimize the parameters of the background transformation and the
convolution kernel.

\begin{figure*}
\begin{center}
\includegraphics[width=0.85\textwidth]{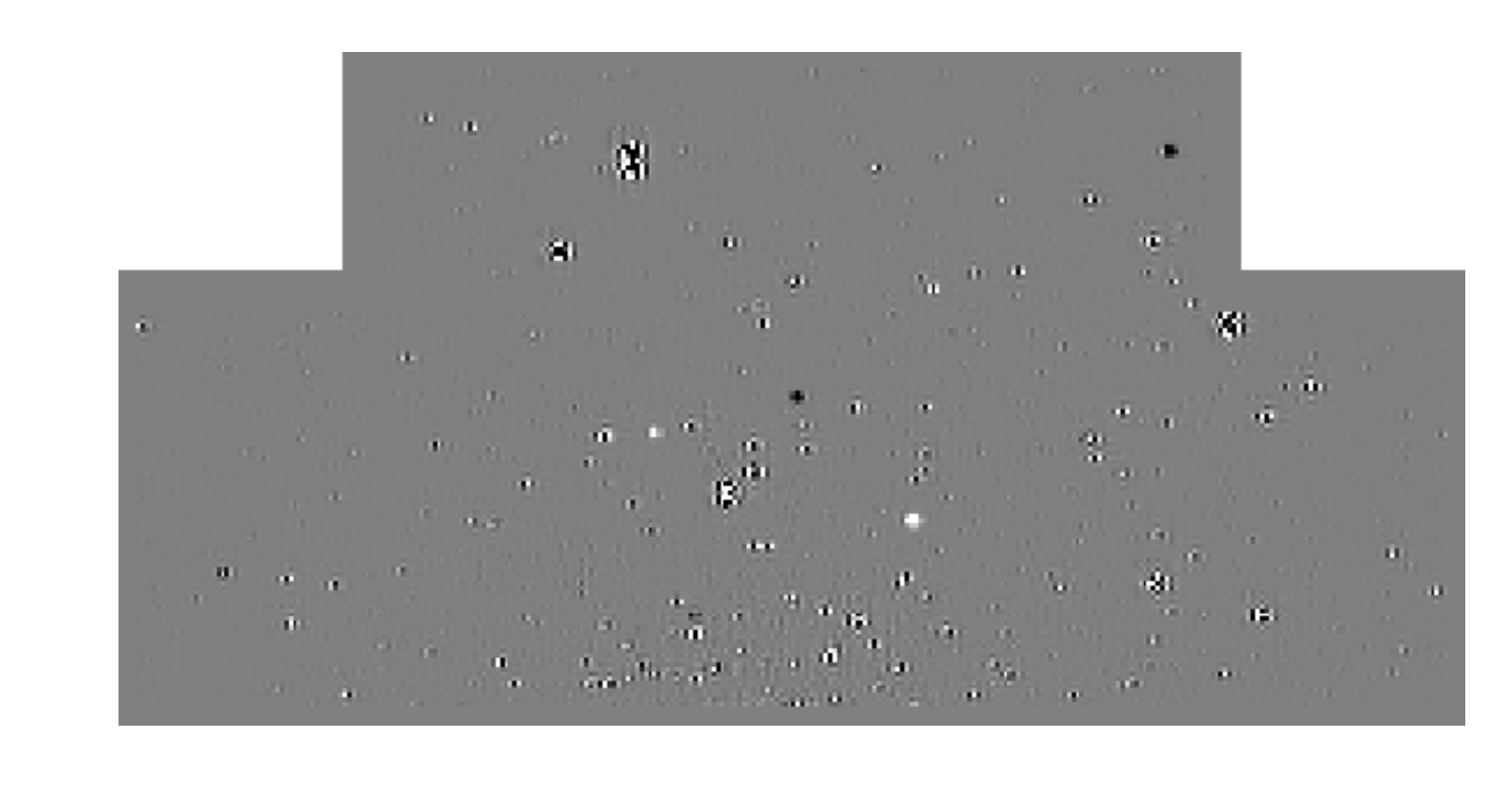}
\end{center}
\caption{\label{fig2} Subtracted image for the image in
  Figure~\ref{fig1}, with arbitrary
  z-scale of same dynamic range as Figure~\ref{fig1} and no color
  inversion.  The white regions in the upper left and right corners are 
  the same as the white regions in Figure~\ref{fig1}, 
  regions that were not included in the superstamp. 
  The RR Lyrae variables are of sufficiently large amplitude to be
  visible to the naked eye in the subtracted image: the black
  ``holes'' in the middle and in the upper right
  of the image are two RR Lyrae variables, as are the bright spots
  (i.e., no dark pixels in the star's image)
  left and slightly down as well as right and down from the middle hole.
  Residual noise and saturation artifacts are visible.
The 4-pixel border
  of zero-value pixels filling the edge region, as described in the
  text, is also present, and can be made out at the
  bottom of the image.
}
\end{figure*}

What remains after the image subtraction (barring any uncorrected
systematics and/or an incorrect background fit) is an image free of any
non-variable sources with random 
scatter about a statistical average of zero.  Stars leave behind larger
magnitudes of scatter than the source-less background, and saturated stars
leave behind visible artifacts.  Figure~\ref{fig2} shows the same
image as Figure~\ref{fig1} after subtracting the master photometric
reference image as described above.

Extracting photometry from the subtracted images requires a catalog of
source positions as well as reference fluxes/magnitudes for each
source to properly calibrate the amplitude of the variable signals
found in the subtracted images.  We used the {\it Gaia} first data
release (DR1) source catalog
(\citealt{gaiadr1,gaiamission}) as
both our astrometric (\citealt{gaiaastrometry}) and photometric
(\citealt{gaiaphotometry}) reference catalog.  Our analysis was
sufficiently progressed at the release of the {\it Gaia}
second data release (DR2; \citealt{gaiadr2}) that we chose to stick
with the {\it Gaia} DR1 data despite DR2's superior quality.
That being said, data from 
{\it Gaia} DR2 were used as part of our analysis (for example, its
identification of duplicate DR1
sources).

The {\it Gaia} DR1 source
catalog is virtually complete at the magnitude range of the main
sequence turnoff stars in M4 ($G{\approx}$16--17)
and its excellent astrometry allows for precise source position
determination and aids in identifying and disentangling close
neighbors that are impossible to differentiate in the K2
images. That being said, crowded regions limited {\it Gaia}'s
completeness in both DR1 and DR2 (e.g., \citealt{gaiadr1}).  Given
that these limitations in completeness correlate with crowdedness and
that in the most crowded regions of our images, any star missing from
our astrometric reference catalog is
likely to appear in some other star's photometric processing aperture,
and we proceeded anyway 
despite the potential completeness issues.
{\it Kepler}'s and {\it Gaia}'s bandpasses are also very comparable, 
which we found eliminated any need to derive more than 
an additive conversion from our instrumental magnitudes to {\it Gaia}
$G$ magnitudes.

From the {\it Gaia} DR1 archive, we extracted those sources
that fell inside or near to the region of the M4 superstamp and had
a  $G$ magnitude  brighter than 19.  
This cutoff does not go deep enough to cover
all the stars in the cluster, nor does it go deep enough to cover the
possible variable stars in the background, many of which may
be sufficiently unblended in the images to detect variability. The choice
of this magnitude cutoff was based on the photometric performance
of \cite{soaresfurtado} and our initial goal to primarily search for
transiting exoplanets rather than larger-amplitude variables.
The right ascension and
declination values obtained for the {\it Gaia} DR1 sources were projected
onto a pixel-based image coordinate system and then matched using
\texttt{grmatch} with the extracted sources of the selected
astrometric reference image.  The matching, similar to before, was
performed over a grid of spatial orders 
and number of
objects 
to include in the triangle matching.
The best
transformation was then chosen as the match with at least 100 matched
objects, weighted residual less than 0.001, and unitarity greater
than 0.015 that had the largest number of matched objects normalized
by the square of the weighted residual.  
We then transformed the coordinates of the {\it Gaia}
DR1 sources to the astrometric reference image's frame using
\texttt{grmatch} based on the transformation calculated above.  After removing
those sources with transformed coordinates that fell outside the
astrometric reference image, there were 5914 
sources. We refer to this as our source position catalog.

The next step was to calculate
the photometry for each of the 5914 sources from the subtracted
images. This required first deriving a conversion from the
the $G$ magnitudes of
the photometric reference catalog to the instrumental magnitudes of the K2
images. To accomplish this, we first used the
\texttt{FITSH} tool \texttt{fiphot} to obtain photometry from the
master photometric reference image for 
a set of circular apertures, with 15 apertures 
ranging from 1.15 to 2.55 pixels.
These
radii were selected to obtain a good measure of 
how changing the aperture size affected the amount of flux measured
for a given source over a range relevant to where the bulk of the flux
falls in the PSF (the median FWHM of the PSF across the images was
${\sim}$1.5 pixels, with the range 1.45--1.55 pixels covering nearly
all the median PSF widths).  
The apertures were centered at each of the positions of the 1024
objects that had been directly matched between the {\it Gaia} DR1 source catalog
and the astrometric reference image.  (Since \texttt{fiphot} had found
only 1073 sources directly from the images, probably due to inability
to disentangle highly 
blended sources, that is why there were far fewer matched sources
than the total available from just the {\it Gaia} DR1 source catalog.)
For this calculation, the sky was
subtracted based on the mode of pixel values in an annulus with inner
radius of 17 pixels and 
outer radius of 30 pixels.   A
radius of 3 pixels around any source in the set of 1024 matched sources
was excluded from this background calculation, and the pixel values
were sigma-clipped (3$\sigma$, two iterations) prior to the calculation.

After performing this reconnaissance photometry, we determined a
 transform from {\it Gaia} $G$
to {\it Kepler} instrumental magnitudes. As mentioned previously, we
found that an additive 
transform was all that was needed for this conversion, likely because
of the very similar 
bandpasses of the two instruments. Since there is significant blending
of the sources in the K2 images, we first selected out those K2
sources for which we thought there were negligible contributions from
neighbors.  Several unblended sources, as well as a few unsaturated
bright sources for which any blending from neighbors would be small,
were selected from the astrometric reference image by eye and were
verified to be negligibly blended by using the {\it Gaia} DR1 source
catalog. After this, sources with instrumental magnitudes in
a narrow range around the transformed $G$ magnitudes (and thus
presumably negligibly blended on the images)  were selected and then
fit to determine a more precise value
for the additive constant.  For all this, we used a  2.5-pixel radius
 aperture to calculate the instrumental
magnitudes.  
Next, we determined the effect that
changing the aperture size had on this conversion factor.  For the
brightest unblended and unsaturated stars, we normalized the fluxes
 calculated over a range of aperture sizes to
the flux in the 2.5-pixel aperture and then determined the median
normalized flux for each aperture size across the selected stars.  We
then fit the integral of a Gaussian function to the median normalized
fluxes to determine a conversion from the flux at a given aperture
size to that of the 2.5-pixel aperture.

We then ran aperture photometry on the master photometric reference
image for all the positions in the astrometric source catalog. As
before, the sky background was
calculated as the mode of pixels values in  an annulus with inner
radius of 17 pixels and 
outer radius of 30 pixels, with the same sigma clipping and source
exclusion as before. The background was then subtracted.
We
performed the photometry calculation with apertures 
 1.5, 1.75, 2.0, 2.25, 2.5, 2.75, and 3.0 pixels in
radius. Then, using the {\it G} magnitudes from the photometric
reference catalog, we substituted the reference
fluxes for each object with the  values determined from the converted
{\it G} magnitudes, additionally modified based
on the aperture size.  This provided reasonably 
accurate and unblended reference fluxes for each of the objects.  

We then calculated the image subtraction photometry using \texttt{fiphot}
and the derived reference fluxes.  The sky
background, having been previously subtracted when the subtracted
images were created, was not fit in this step.  We also used the same
convolution kernels  calculated for the creation
of the subtracted images.  
At this point, {\it Kepler} BJDs (KBJDs; BJD$-2454833.0$) were assigned
to each cadence for each object. 
Each of the original 16 TPFs was assigned only a single KBJD for each
cadence, calculated 
along the center of the TPF.
We assigned to each object the KBJDs from the TPF
image in which it was found.

After the photometry calculation from the image subtraction, we
obtained light curves for 4601 objects. The reason for the reduction
from the original 5914 we were calculating photometry for was that some objects were excessively blended with much
brighter neighbors and were unable to have photometry measured, and
that some of the objects  fell in or excessively overlapped with the
excluded edge region.  
The brightest  stars (for cluster members, this corresponds to many of
the
giant stars) were saturated.  We did not perform any special treatment
of saturated stars, though because they were so bright the
largest apertures employed in our processing (3 pixel radius) were used.
Additionally, there was one
previously know RR Lyrae variable, V27 of \cite{clement2001}, that was not a
{\it Gaia} DR1 source and thus did not get a light curve through the
above method.  We separately
extracted a light curve for this star following the procedure
described above and based on the transformed {\it Gaia} DR2 position for
this object. The light curve for V27 did not undergo any of the
following post-processing procedures since large-amplitude variables
were not served well by the roll decorrelation, described in Section~\ref{sec:postprocessing}.
 Including V27, we produced 4602 light curves in total.  The light
 curves at this stage are what we refer to as the ``raw light curves''
 throughout the rest of this work.  All of our raw and processed light curves
 are published and publicly available at
 \cite{lightcurves}\footnote{Published at Princeton University's
   DataSpace and licensed under a Creative Commons Attribution 4.0
   International License, accessible via the permanent URL \url{http://arks.princeton.edu/ark:/88435/dsp01h415pd368}}.

\subsection{Photometry Post-processing}
\label{sec:postprocessing}

The roll of the telescope during the K2 mission introduced
systematic variations to the brightness of objects as they moved across
the detector \citep{howell2014}. This is due to differences in pixel sensitivity
unaccounted for in the K2 data reduction.  These brightness variations
are correlated with the object position on the detector, and are not
fully corrected by the image subtraction photometry. The remaining
systematic variations can be decreased by performing a decorrelation
of flux variations against object position with a procedure 
based on 
\cite{vanderburg2014} and \cite{vanderburg2016}.
We divided the light curves, normalized to their median values, into
the same eight time chunks 
as \cite{vanderburg2016} did for Campaign 2 
(A. Vanderburg, private communication).  
To determine the drift position of each object, the positions in the
source position catalog were transformed for each cadence using the
inverse of the transformation originally used to register each
cadence's image to the astrometric reference frame.
Since the drift of objects across the detector was primarily
in one direction, for each object we used a principal components
analysis (PCA) to determine this primary direction of drift.  The object
positions for each cadence were transformed to the axes defined by the
PCA and then a fifth-order polynomial was fit to the positions. Each
object's drift's arc length along the polynomial at each cadence was
calculated and stored for later decorrelation. 

For
each time chunk, we iterated over fitting long-term trends with a B-spline
fit and decorrelating against the roll.  For the B-spline, we had
breakpoints set nominally every 1.5 days.
The 1.5-day breakpoint spacing was adjusted to allow for knots to be
distributed evenly across the time chunk.  Also, where possible, 0.75
days from adjacent time chunks were included to improve the smoothness
and accuracy of the spline fit across time-chunk boundaries.
We then
excluded 3-$\sigma$ outliers to the B-spline fit, refit the spline,
and repeated this until no outliers remained to be removed. The
median-normalized light
curve was then divided by the spline fit.  The spline fit is not ever
reintroduced into our light curves, so smoothly varying signals with
timescales longer than the 1.5-day knot placement are likely
to either be altered or removed. Objects with such signals are best
studied from our data using the raw light curves.

After this, the fluxes of each chunk of the light curve
were binned into 15 bins in arclength.  3-$\sigma$ in flux
outliers were excluded in each bin, and then a linear interpolation
was made using the mean flux values of each non-empty bin.  In cases where
bins had only a single point, an interpolation between adjacent bins
was made.  If there was a single point in the last bin (usually
corresponding to outliers in the pointing), no fit was made for that
point.
The light curve
was then divided by this interpolation.  This process of fitting a
spline to the longer trends and decorrelating against position was
repeated eight times or until convergence, whichever came first.  We
selected eight to be the maximum number of times because we found
those light curves that required more than eight iterations were
usually oscillating between two close fits to the data that were 
 not quite close enough to be counted as converging.
If less than 10 points were in a time chunk, the decorrelation against
drift position was not performed.

\begin{figure}
\begin{center}
\includegraphics[width=\columnwidth]{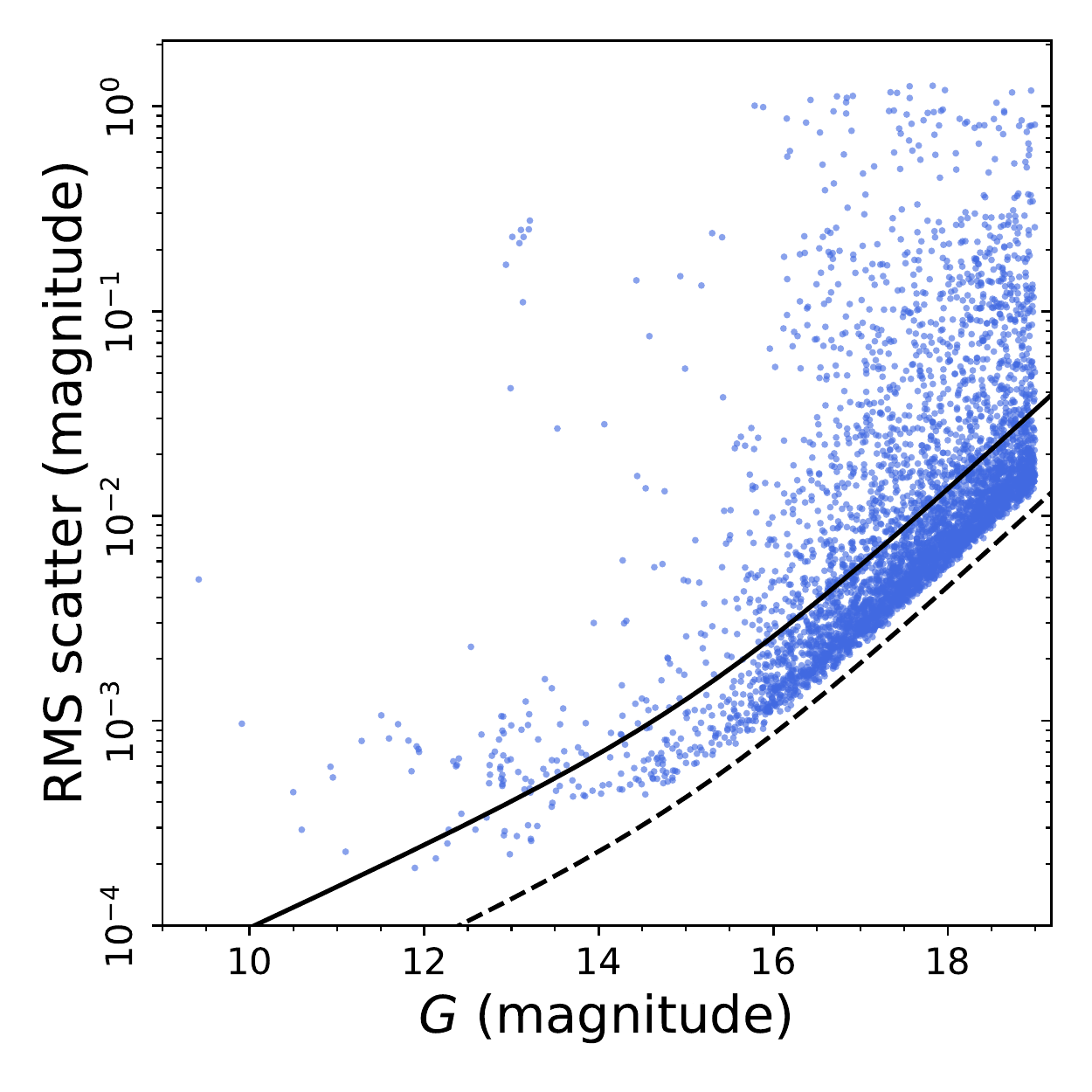}
\end{center}
\caption{\label{fig:rms} RMS scatter as a function of {\it Gaia} DR1
$G$ magnitude 
for our final light curves.  The RMS is calculated from our magnitude
light curves, which have been sigma clipped with 3-$\sigma$ clipping
iterated 3 times.  All of our final 4554 objects under consideration
except V27 are plotted here; V27's light curve did not undergo the
same processing as those of the other objects, see text for details.
The solid line shows a calculation of our expected RMS scatter and the
dashed line shows that same calculation reduced by a factor of 3; see
text for a discussion.
The collection of objects with excessive RMS values at $G{\approx}13$
are the RR Lyrae variables, though we note that our light
curve processing pipeline impacted the amplitudes of
large-amplitude variables.} 
\end{figure}

We then used the trend filtering algorithm (TFA; \citealt{kovacs-tfa})
as implemented in \texttt{VARTOOLS} (\citealt{vartools}) to clean up
systematics common across the light curves.  For each aperture,
250 light curves with at least 97\% of the maximum number of light
curves points were selected from uniform bins of source position and
magnitude to be used as the trend light curves.  For light curves with
less than 2500 points, a subset of the selected 250 trend light curves was
used in the detrending, with the number of selected trend light curves
being close to but less than
10\% the number of light curve points.  Since the KBJDs for a
given observation differed slightly depending on which TPF an object
was located (see Section~\ref{sec:photometry}) and common instrumental
effects were likely correlated based on actual observation time than
KBJD, detrending was performed based on cadence number rather than
KBJD.  Light curves from stars that were known 
to be RR Lyrae variables or saturated were not included as potential
trend light curves. All light curves were then detrended against the
trend light curves for the given aperture size with trend light
curves excluded from the detrending if they were closer than 6 pixels,
which is ${\sim}$4 FWHMs. The light curves that
resulted were the ones used in our variability search  
 and are referred to in this work as ``final light curves.''
Figure~\ref{fig:rms} shows the root-mean-square (RMS) scatter of the
sigma-clipped (3-$\sigma$ clipping, iterated three times) final light
curves for those objects included in our variability search.  Owing to
significant outlier points in our 
final light curves,  outlier removal was necessary
for our subsequent period search.
These outliers seem to be due to still-uncorrected systematics, the worst of
which occurred when the telescope changed its roll direction about
halfway through the campaign.

The photometric performance displayed in Figure~\ref{fig:rms} shows
that our sigma-clipped light curves are able to reach millimagnitude
RMS scatter down to $G{\approx}15$, 
and 0.01 mag RMS scatter down to $G{\approx}18$.  There is a large envelope
of points with significantly larger scatter than is typical for
objects of their magnitude.  Some of these are variable stars, while
the rest have excessive scatter due to the amount of blending
present in the images or also possibly due to breakdowns
of the photometric processing for individual objects.  We also note
that our saturated 
giant/bright foreground stars do not have significantly larger scatter
than, e.g., our HB stars at $G{\approx}13$.  The point at $G{\approx}9.5$ is
a star that is an intrinsic variable, hence the larger scatter.  The
clump of points with high RMS scatter at $G{\approx}13$ are the RR
Lyrae variables.

The solid line in Figure~\ref{fig:rms} shows our expected RMS
performance based on source Poisson noise and the background sky flux
as seen in our photometric reference image and the dotted line shows
the same expected RMS performance reduced by a factor of three.  We
have not entirely determined the reasons for our photometric
performance to fall as far below our expected performance as it does, but it is
perhaps attributable to some combination of an incorrect gain value,
an incorrect sky background characterization, an incorrect magnitude
zero-point determination, or outliers being
excessively clipped due to large, non-Gaussian errors.  We note that
our roll decorrelation and TFA calculations have some free parameters,
but this at most could account for only a few percent decrease of the
scatter relative to the expected.

\subsection{Skipped Images}

Now that the photometric processing pipeline has been explained,
sufficient context is available to discuss why certain cadences were
not used in our analysis.   
In what follows, the cadence numbering starts at 1 for the first
cadence in the campaign (which corresponds to the {\it Kepler} long
cadence number of 95497).
Of the 3856 cadences in Campaign 2,
39 were blank 
due to resat events, an additional 6 were blank due to other reasons (cadences 216--218 and
2856--2858), 12 were excluded due to our noticing excessive telescope
slew during the exposure (cadences 50, 191, 202, 203, 205--207, 209,
383, 863, 1535, and 1823), 68 were excluded due to being excessive
pointing outliers (1--49, 51--57, 192--201, 204, and 727), one
was excluded due to a hot pixel column we noticed (208), and six were
excluded due a majority of the light curves having large outliers (at
least 50\% off) in flux 
measurements relative to the median flux value across the whole light
curve (2150,
2151, and 2153--2156)---these all occurred around the point in
the observations when the telescope roll direction switched.  
For the pointing outliers, cadences 1--49
were all pointed in a locus
several pixels away from the main group of pointings, and this was an insufficient
number to perform our roll decorrelation just on these points;
cadences 192--201 and 204 were similarly pointed in a different
locus several pixels away from the main; cadences 51--57 were pointed in a
locus close to the main locus of pointing but not close to the pointings of its
time chunk; and similarly the pointing of cadence 727 was quite
disparate from any in its time chunk.
This is a total of 132 cadences that were
entirely removed from or not available for our consideration, 
leaving 3724 (96.6\%) of the cadences for
the final analysis.  We note that most of the these cadences were
removed from both our raw and final light curves, but that cadences
1--49 are still present in the raw light curves.

\subsection{Removal of Objects}

We removed from consideration objects with light curves
with less than 800 points (out of a maximum number of 3724 for the
final light curves).  There were 32 such objects in total,
leaving 4570 objects.
These removed objects tended to be highly blended with a much brighter
object, and this led
to many light curve points' calculations failing.  In practice, we
found that such light curves were not productive to search for
variability. The selected cutoff of 800 was rather conservative and still
permitted other relatively sparse and blended light curves that were
not useful, so the removal of these objects is not likely to remove
anything that might be detected as a variable.

\subsection{Additional Data Used for Analysis}

We used the {\it Gaia} DR2 \texttt{gaia\_dr2.dr1\_neighbourhood}
crossmatch catalog to
inform us which of the examined {\it Gaia} DR1 sources were
duplicates.  There were 16 DR2
sources matched to two entries in the DR1 
source catalog.  So that the photometric aperture used corresponded as
closely as possible to the DR2 source position, in each case we kept
whichever of the two DR1 
sources was closest in position to the corresponding DR2 source.  This
also happened to correspond in each case with the DR1 source with the
best ``RANK'' value---a calibrated measure of how close a DR1 source
is to a DR2 source in both position and magnitude---between the two
DR1 sources.  We removed the 16 extraneous DR1 sources from the
analysis and were left with a final set of 
4554 objects with usable light curves. Information on these objects
and their light curves is
presented in
Table~\ref{tab:ids}.

\begin{deluxetable*}{cccccccccc}
\tablewidth{0pc}
\tablecolumns{10}
\tablecaption{Stars Examined\label{tab:ids}}
\tablehead{
\colhead{ID\tablenotemark{a}} & \colhead{{\it Gaia} ID\tablenotemark{b}} & \colhead{R.A.\tablenotemark{c}} & \colhead{decl.\tablenotemark{c}} & \colhead{$G$\tablenotemark{d}} & \colhead{$G_\mathrm{BP}$\tablenotemark{d}} & \colhead{$G_\mathrm{RP}$\tablenotemark{d}} & \colhead{No. Pnts.\tablenotemark{e}} & \colhead{RMS\tablenotemark{f}} & \colhead{Mem. Prob.\tablenotemark{g}} \\
 \colhead{} & \colhead{} & \colhead{(hh:mm:ss)} & \colhead{(dd:mm:ss)} & \colhead{(mag)} & \colhead{(mag)} & \colhead{(mag)} & \colhead{} & \colhead{(mmag)} & \colhead{}  }
\startdata
V6 & DR2 6045478696063803648 & 16:23:25.76 & $-$26:26:16.7 & 13.25 & 13.68 & 12.63 & 3773 & 120.82 & 1.000\\
V7 & DR2 6045478391137284224 & 16:23:25.92 & $-$26:27:42.3 & 13.28 & 13.80 & 12.61 & 3762 & 285.46 & 1.000\\
V8 & DR2 6045477910100852736 & 16:23:26.12 & $-$26:29:42.0 & 13.23 & 13.64 & 12.49 & 3773 & 255.69 & 1.000\\
V9 & DR2 6045477910100361600 & 16:23:26.76 & $-$26:29:48.4 & 13.10 & 13.59 & 12.44 & 3773 & 256.94 & 1.000\\
V10 & DR2 6045478322417726848 & 16:23:29.17 & $-$26:28:54.7 & 13.19 & 13.64 & 12.53 & 3087 & 155.06 & 1.000\\
\cutinhead{...}
\enddata
\tablecomments{There is no W1873 in this table.  The identifiers beginning with ``W'' are sequential otherwise. Light curves for all of these sources are available at \cite{lightcurves}. Table~\ref{tab:ids} is published in its entirety at Princeton University's DataSpace and can be found in the \texttt{object\_information.txt} file at the URL \url{http://arks.princeton.edu/ark:/88435/dsp01h415pd368}. A portion is shown here for guidance regarding its form and content.}
\tablenotetext{a}{The identifier by which this object is known in this work. Those prepended with ``V'' are previously identified variables from the catalog of \cite{clement2001}, June 2016 edition, not marked as constant; those prepended with ``SC'' are candidate variables from \cite{stetson2014}; those prepended with ``W'' are additional {\it Gaia} DR1 sources examined in this work.}
\tablenotetext{b}{{\it Gaia} source ID, taken from DR1 or DR2 as indicated. The DR2 ID was preferentially used and only 11 objects in this table have their DR1 IDs quoted.}
\tablenotetext{c}{J2000.0; data taken from {\it Gaia} DR1 \citep{gaiaastrometry} or DR2 \citep{lindegren2018} as indicated in the ``{\it Gaia} ID'' column (see table note b).}
\tablenotetext{d}{{\it Gaia} $G$ magnitude taken from either {\it Gaia} DR1 \citep{gaiaphotometry} or DR2 \citep{riello2018} as indicated in the ``{\it Gaia} ID'' column (see table note b).  Please note that $G$ had a different definition between DR1 and DR2 \citep{evans2018}.  $G_\mathrm{BP}$ and $G_\mathrm{RP}$ are taken only from {\it Gaia} DR2 and were not included in {\it Gaia} DR1, nor are they available for all {\it Gaia} DR2 sources.}
\tablenotetext{e}{Number of points in the light curve. Raw light curves are used for objects with identifiers beginning with ``V'' and final light curves for all others. Raw light curves can include data from cadences 1--49 and so may have more points than the maximum of 3724 for the final light curves.}
\tablenotetext{f}{RMS of the light curve, with sigma clipping (3$\sigma$, iterated three times). Raw light curves are used for objects with identifiers beginning with ``V'' and final light curves for all others.}
\tablenotetext{g}{Membership probability as calculated by \cite{wallace2018}.  ``N. DR2'' means this object was not matched to a {\it Gaia} DR2 source; ``N. D.'' means this object lacked proper motion data in {\it Gaia} DR2 and its membership probability could not be calculated; ``Dup.'' means this DR1 source was matched to multiple DR2 sources.}
\end{deluxetable*}

As part of our analysis, knowledge of the cluster membership of each
of the stars was necessary.  We used the membership catalog previously
created by \cite{wallace2018} and available
at \cite{membershipcatalog2018} or on
GitHub\footnote{\url{https://github.com/joshuawallace/M4_pm_membership}}.
This catalog fitted a two-component Gaussian
mixture model to {\it Gaia} DR2 proper motions \citep{lindegren2018} to calculate a
membership probability for all {\it Gaia} DR2 sources with reported
proper motions.  A very large majority of the calculated membership
probabilities were ${<1}$\% or ${>99}$\%, essentially allowing the
catalog to function as a binary classification in all but a few cases.
Of the 4554 objects with usable light curves, 4469 of
them---98.1\%---were matched (again, using
the \texttt{gaia\_dr2.dr1\_neighbourhood} 
crossmatch catalog) to a single DR2 source with reported
proper motions and thus were able to be assigned a cluster membership
probability.  Of the remaining 85 objects, 74 were matched to DR2
sources that lacked reported proper motions, 6 were matched to more
than one DR2 source, and 5 were not matched to any DR2 sources.
Membership probabilities for these 85 objects were not calculated.  Of
the 4469 objects with reported proper motions, 3784 of them had
calculated membership probabilities of ${\geq}99$\%.

\subsection{Search for Variability}

We used three algorithms for finding periodic signals in our
data: the
Generalized Lomb-Scargle \citep[GLS;][]{lomb1976,scargle1982,zechmeister2009},
phase dispersion minimization \citep[PDM;][]{stellingwerf1978}, \
and box-fitting least squares \citep[BLS;][]{kovacs2002} algorithms.
The \texttt{astrobase} \citep{astrobase} implementations of these
algorithms were used.
With the amount of signal blending in the data, 
 we incorporated a blend search with the period
search. It is worth noting that this blend search incorporated only
data available from the section of the superstamp we examined.  Any blending or
systematics due to objects that were in the edge region of the
superstamp or beyond could not be readily 
identified. Additionally, with the amount of systematic noise remaining in
the data, it was necessary for us to employ a custom and
period-dependent SNR threshold, determined from our examination of the
data.  The code written to perform both of these
tasks, \texttt{simple\_deblend}, is available at
\cite{simpledeblend} or on 
GitHub\footnote{\url{https://github.com/simpledeblendorganization/simple_deblend}}.

The basic framework of the algorithm used by \texttt{simple\_deblend}
is as follows.  For a 
given period search method (GLS, PDM, BLS) and star, the code:
\begin{itemize}
\item Determines the best period based on the period search
\item Checks the periodogram SNR of this period against the threshold;
if below the threshold, then quits the period search
\item Phase-folds neighbor light curves at the given period and
figures out which of all the objects has the highest flux amplitude of
variability
\item Records the star as the source of that variability 
if the star has the highest flux amplitude of variability
\item Fits out the found period using a Fourier series fit to the
data, then repeats
\end{itemize}
This is repeated for the desired number of periods---three for our
analysis---or until no more robust signals are found.

As a more detailed description, for a given period search method and
star, the code runs the \texttt{astrobase} implementation of the
period search algorithm.  In each search, 
working in magnitudes (and not fluxes), 
the minimum period searched was 0.06 days
and the maximum period search was 78 days for GLS and PDM---about as
long as the maximum 
duration of the final light curves---or, for BLS, half the observation
duration of the light curve.  A frequency grid for the search was selected
automatically with the \texttt{autofreq} parameter set to true.  For
GLS and PDM, this produced a frequency grid with frequency spacing
$\Delta f = 1/(5\times L)$, with $L$ being the duration of the
observations. 
For BLS,
this produced a frequency grid with $\Delta f = 0.25\times
q_\mathrm{min}/L$, with $q_\mathrm{min}$ being the minimum transit
duration in units of fractional phase. This was set to 0.02 and the
maximum transit duration was set to 0.55.  For BLS, the 
number of phase bins also needed to be set, and was set to 200.

\begin{figure}
\begin{center}
\includegraphics{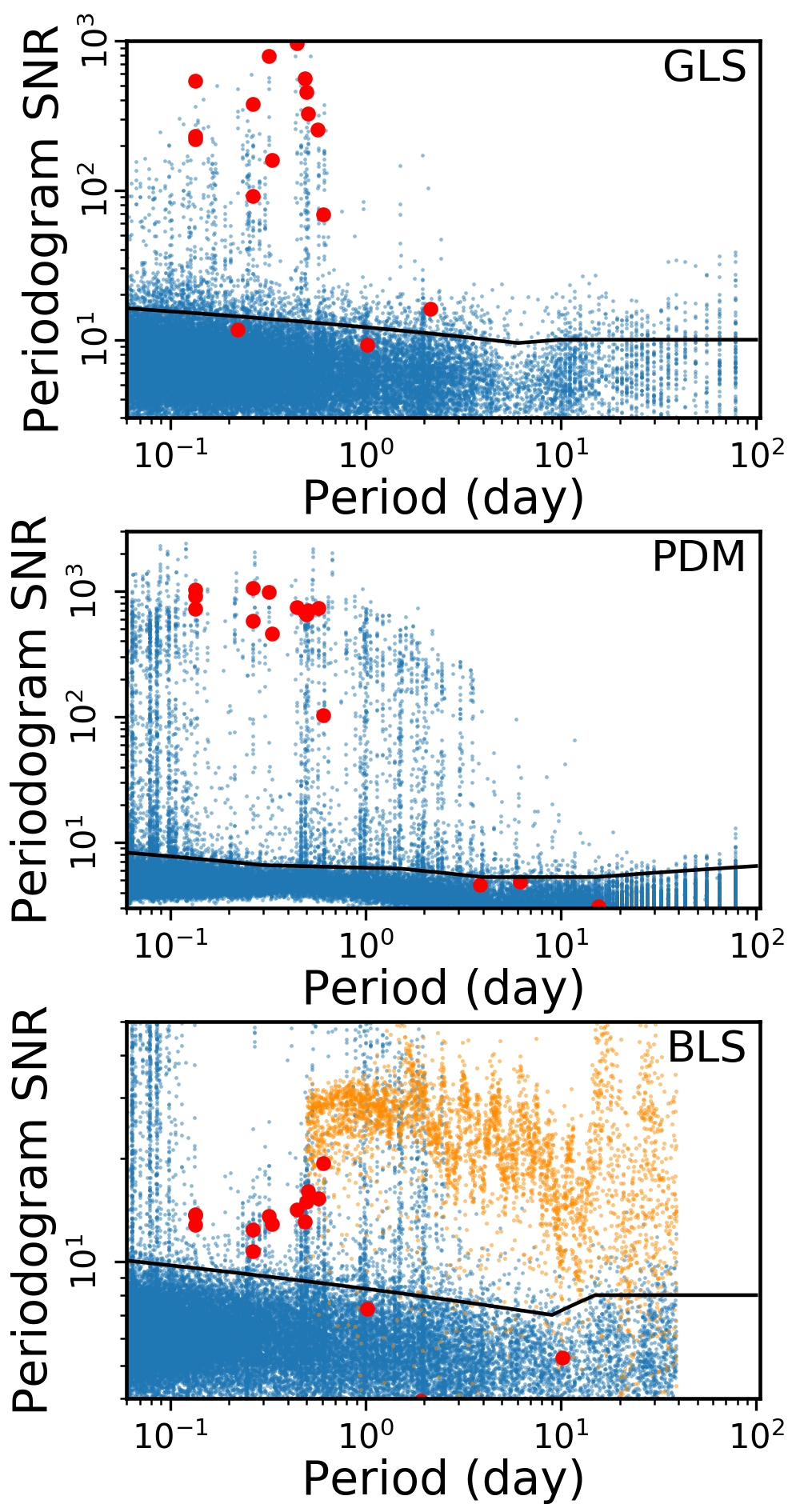}
\end{center}
\caption{\label{fig:thresholds} Thresholds for the periodogram
SNR for the three period search methods.  The
corresponding period search method is shown in the upper right of each
panel.  The blue points
show the values calculated from the best eight
periods found for each object. For BLS, the orange
dots show the values
for light curves with injected transits from transiting objects with radii between 0.3--3.5
R$_\mathrm{J}$.  For all panels, the red dots show the 
periodogram SNR values for objects and periods
we identified as being variables during some initial reconnaissance of
the data.  Not all variables are
identified by all the methods, so there are red dots missing between
the panels.  The thresholds used in our analysis are plotted with a
black line in each panel.}
\end{figure}

After running a period search, the resultant periodogram was median
filtered to correct for trends that were presumably due to non-white
noise.  For each point in the periodogram, either 40 (for GLS and PDM)
or 100 (for BLS; larger due to its smaller $\Delta f$) of the periodogram
values on each side, outside of an exclusion area that was equal to
$4/L$ on each side, were collected and were 3-$\sigma$ sigma clipped
before calculating their median, which was then subtracted to produce
the filtered
periodogram.  For PDM, which has periodogram values of one for
frequencies with no power, the filtered periodogram values had one added back on.
The peak with the highest power was then found, and the robustness
of this peak was determined using an SNR calculation on the
median-filtered periodogram values.  The noise for the ratio was
calculated using the standard deviation $\sigma_\mathrm{per}$ of
nearby periodogram values 
collected in the exact same way as described above for determining the
median filter.  The SNR value was then simply the ratio of the
periodogram value $p$ with this standard deviation,
$p/\sigma_\mathrm{per}$  or, for PDM, $(1-p)/\sigma_\mathrm{per}$.
Appropriate thresholds for this SNR were determined as a function of
period by comparing the SNR values for objects and periods with previously
determined variability and (for BLS) injected transits with the rest
of the detected periods.  This and the selected thresholds are show in
Figure~\ref{fig:thresholds}.  
If the SNR 
did not exceed the threshold, the
period is marked as not robust and the periods search for this object was done.

If the period was determined to be robust using the SNR threshold
described, the next step was to check for blends.  The light curve was
fit with a seven-harmonic Fourier series, which was then evaluated at
200 evenly spaced points.  A flux amplitude was then calculated using
the minimum and maximum of these Fourier series evaluations, converted
from magnitudes.  Subsequently, all neighbors within 12 pixels had
their flux amplitudes at the same period determined in the same
fashion.  The choice of 12 pixels was determined by choosing two RR Lyrae
variables and looking 
at all the light curves for surrounding objects to see how far their
influence extended. If the object was determined to
have the largest flux 
amplitude, then the period was considered a valid detection, and an
11-harmonic Fourier series fit at the period was
subtracted, except for the offset term, from the light curve for
subsequent period determination.  

\begin{figure*}
\begin{center}
\includegraphics[]{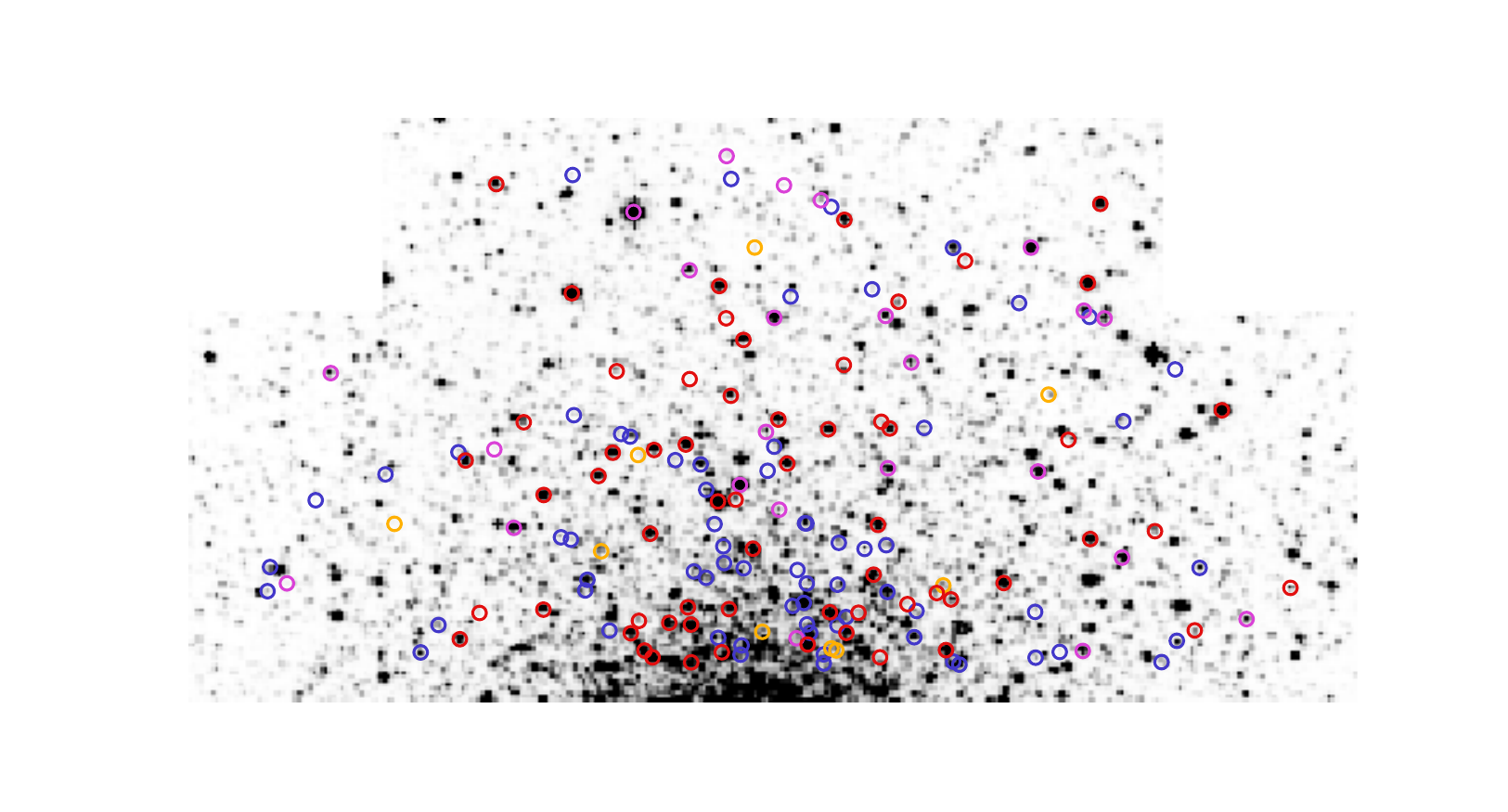}
\end{center}
\caption{\label{fig:varlocations} Locations in the superstamp images
of our detected variables and suspected variables.  This is the same
image as in Figure~\ref{fig1}.  Red circles mark the positions
of the cluster-member variables, magenta circles the positions of
variables that are not cluster members or with ambiguous cluster
membership, gold circles the positions of variables that are
indistinguishably blended (only one circle per set of blended stars), and
blue circles the positions of suspected variables irrespective of
cluster membership. Light curves were not obtained for stars in the
edges of the images and so no variables were found in those areas; see
text for details.}
\end{figure*}

\begin{figure*}
\begin{center}
\includegraphics[]{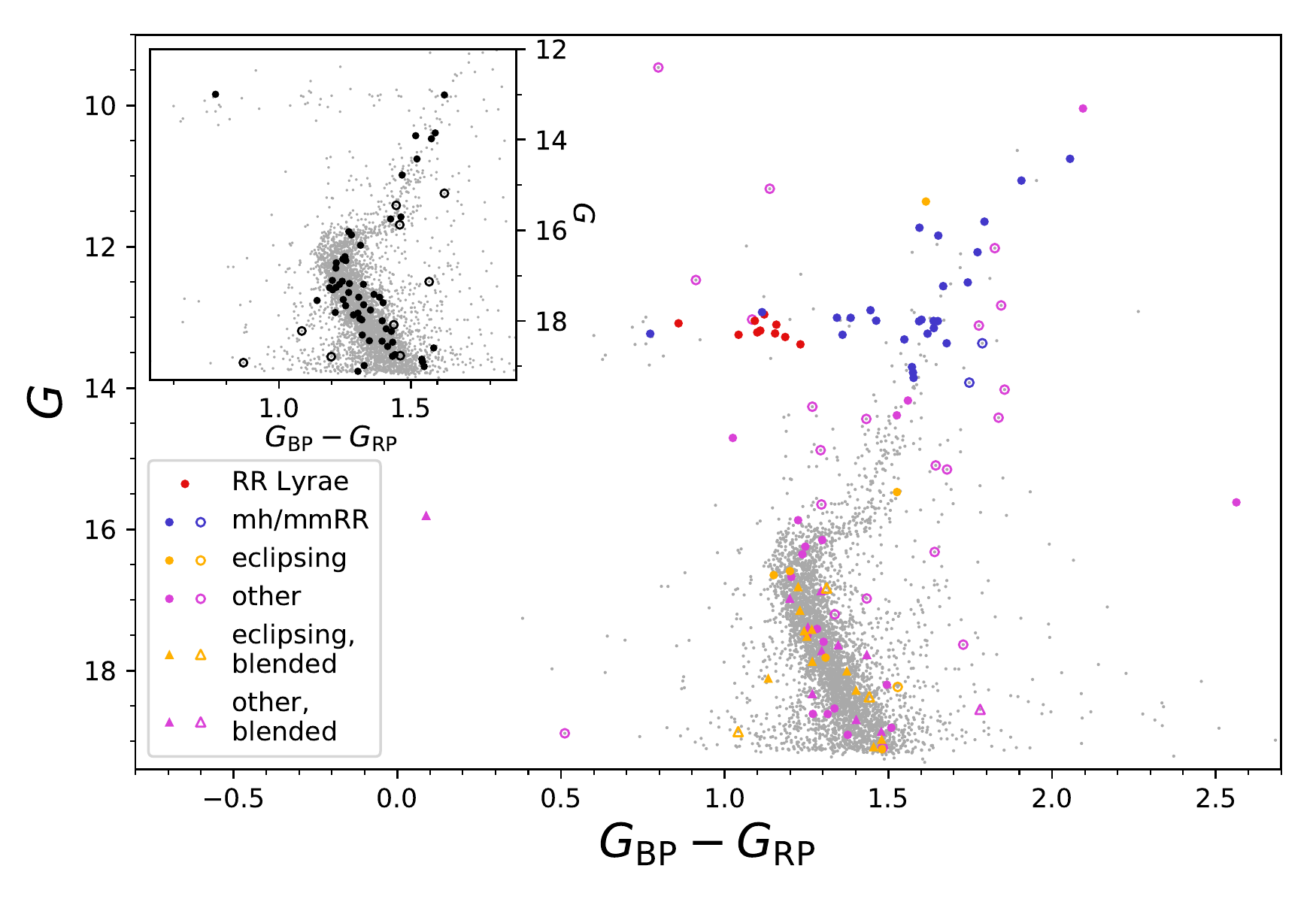}
\end{center}
\caption{\label{fig:cmd} Color-magnitude diagram for the
stars we obtained light curves for, with variables marked. The
photometric data are taken from {\it Gaia} DR2 \citep{riello2018}.
Of the 4554 objects we obtained light curves for, 11 objects did not
have any DR2 data, and 92 objects were missing $G_\mathrm{BP}$ and/or
$G_\mathrm{RP}$ data and are not included here.  None of the variables
or suspected variables were missing these data.  The gray points show the data for all the
objects.  Red points show the data for the RR Lyrae variables, blue
points the data for those objects classified as multiharmonic or
millimagnitude RR Lyrae variables, gold points the data for objects
classified as some type of eclipsing binary (EA, EB, or EW), and
magenta points the data for other types of variables.  Those variables
that are 
cluster members are marked with closed symbols and those that are
not cluster members or have ambiguous cluster membership are marked
with open symbols.  Circle symbols are those for which one object is
identified as the variable, while triangle symbols mark variables that are
indistinguishably blended.  The inset shows the same data, but with
the suspected variables marked in black, and with the same open/closed
symbol 
membership convention as the main panel. Note the differing scales
between the main panel and the inset.}
\end{figure*}

\begin{figure*}
\begin{center}
\includegraphics[]{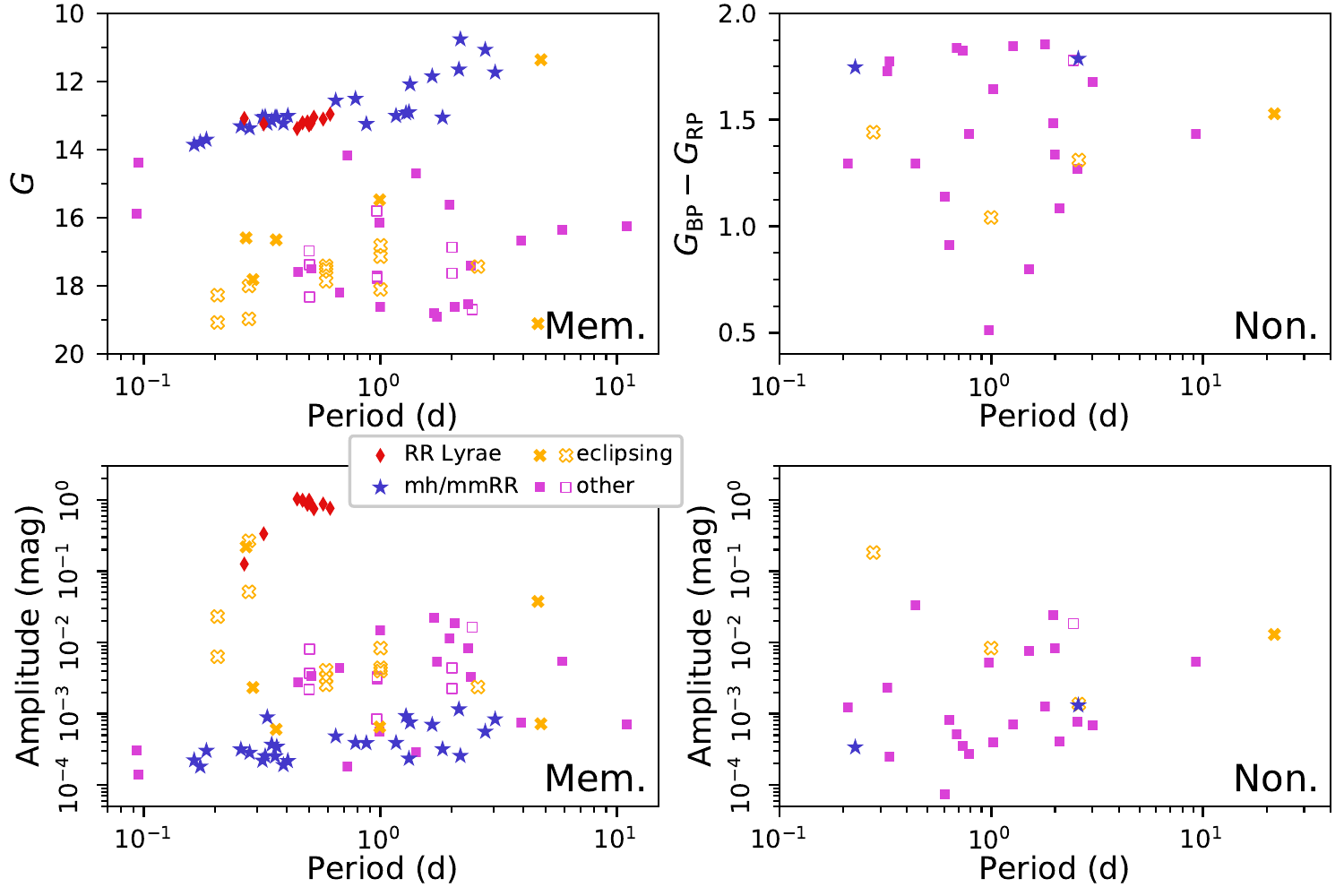}
\end{center}
\caption{\label{fig:pervsinfo} Relationships of photometric properties
and variability amplitudes with variability periods.  The left two
panels are for variables that are cluster members and the right two are
for variables that are not cluster members or have ambiguous cluster
membership (indicated by, respectively,
``Mem.'' and ``Non.'' in the lower right corners of each panel).  The
top-left panel shows $G$ 
versus period and the top-right panel shows
$G_\mathrm{BP}-G_\mathrm{RP}$ versus period.  Both the bottom-left and
bottom-right panels show variability amplitude (see
Section~\ref{sec:amplitudedetermination}) versus period, with
amplitude converted to a positive value for those few cases with
negative amplitudes as we have defined it. The x-axis scales are the same
for panels in the same column, and the y-axis scales for the
bottom-left and bottom-right panels are the same.  The legend of the
bottom-left panel applies to all panels: red diamonds are RR Lyrae
variables, blue stars are multiharmonic and mmRR variables,
gold X's are eclipsing variables, and magenta squares are all
other variables.  Solid symbols are for variables identified to
single objects and hollow symbols are for variables indistinguishably
blended with others. In the x-axis labels, ``d'' stands for ``day.''
  The period--luminosity relation of the RR Lyrae
variables is seen, and the multiharmonic variables also appear to
continue this relation to longer and shorter periods.}
\end{figure*}

We noticed two cases where 
known low-amplitude variables---specifically, the
millimagnitude RR Lyrae (mmRR) variables 
of \cite{wallace2019}---were marked as blends.
This was because their periods
were ${\sim}\twothirds$ that of some large-amplitude-variable
neighbors. Although folding these neighbors' light curves on the mmRR
variability period did not produce the ideal folding for these
neighbors' variability, the folded neighbor light curves still had a
large enough amplitude to be larger than the mmRRs' ${\sim}$mmag variability.
Because of this, if the object was determined as not
having the largest flux amplitude, then the 
neighbor with the largest flux amplitude at the given period was
checked to make sure that period corresponded to a ``real'' period of
the object.  This was determined by running the given period search method on
the neighbor's light curve and checking whether the found period
matched any of the neighbor's top 8 periods.  
If the period matched any of the neighbor's top 8 found periods, then
the period was marked as a
blend and, as for the valid period, the light curve with an
11-harmonic Fourier series fit removed (except for the offset term)
was then used for a subsequent period 
search.  This recursed until either a valid period was found, a period
was determined to not be sufficiently robust, or, in the case of
sequential finds of blending, a recursion limit was hit.  This
recursion limit was set to
be 4 for GLS and PDM and 3 for BLS. Additionally, if a particular
object and period's flux amplitude was not the greatest but was greater
than 90\% the maximum flux amplitude of its highest-amplitude
neighbor, it was marked as 
a possible source of the variability.

The 1310 objects thus determined to have robust periods 
were then searched by eye for classification and to weed out
false positives.  For this by-eye evaluation, we used
the \texttt{checkplot} submodule of \texttt{astrobase}.  After
variables and suspected variables were identified, those with similar
periods were checked against each other to look for blends by
evaluating the similar shapes and phasing of the variability.  In many
cases, nearby stars were blended with each other, but in some cases the
identified blends were quite spatially disparate and may have arisen
from some effect of our photometric processing.
Appendix~\ref{sec:blends} provides specific details on these manually
determined blends.  We had 161 variables or suspected variables
remaining after this manual step.

The periodogram SNR selection criterion as we implemented it was not
robust to detect objects with strong variability at a variety of
fairly close
periods, such as giant stars with solar-like oscillations.  This is
owing to the calculated noise being
artificially high from the variability at these other periods. In
fact, in Figure~\ref{fig:thresholds}, most of the red points that fall
below to the thresholds belong to such asteroseismically active
objects.  For simplicity and given the breadth-focused nature of this
work, we did not make a special search for such variability in those
stars for which we may have had {\it a priori} reasons for
suspecting such variability, and we know our accounting of such variables
in this work is incomplete.  Readers interested in such
variability are encouraged to download the light curves and perform
their own  searches.

\subsection{Amplitude, Epoch, and Final Period and Period Uncertainty
Determination} 
\label{sec:amplitudedetermination}

For each object determined to be a variable or a suspected variable, a
final period search was made using one of our three period search
methods with a fine frequency grid ($\Delta f=10^{-6}$) in a restricted region of
frequencies.  These frequencies corresponded to possible periods based
on the observation duration and the period originally
detected in our variability search.  The period with the strongest
power in this finer search was selected as the final period for the
object. For objects with narrow eclipses, 
a trapezoid model was instead fitted to determine the period,
amplitude (trapezoid depth; quoted as a negative number in the case of
inverse transits), epoch (center point of transit), and period
uncertainty.  For all other objects, the amplitude and epoch were
derived from a multiharmonic fit to the phase-folded light curve,
with amplitude being derived from the difference between the minimum
and maximum values of the fit and epoch being the KBJD of the minimum
of the fit.  The number of harmonics used varied from object to object, with the
most being 11 (for the RRABs) and the least being 1, and most objects
having between 1--5 harmonics for their fits.
Epochs were always adjusted to be within one period of
the KBJD of the earliest observations of our final light curves, KBJD $=$
2060.284181.  Period uncertainties were derived from bootstrap
resampling, with 100 resamplings, and with the fine-grid search described above being performed
on each resampling and the quoted uncertainty being the difference
between the 15.865 and 84.135 percentiles of the calculated periods.
Such values are more of a confidence interval than
a formal uncertainty, but we still quote them as our
period
uncertainties.  Uncertainties on epochs and amplitudes were not
determined. 


\section{Variability Search Results}
\label{sec:results}

The presentation of the results is organized based on the cluster
membership probability of the star, whether it is a horizontal branch
(HB) star, and whether a given variability
signal is certain, suspected, or indeterminably blended.  
As far as possible, we adopt the same variability classification
scheme, including 
abbreviations, as used in the General Catalog of Variable Stars
(GCVS), March
2017 edition \citep{samus2017}, with additional designations  to
describe variability not described in this classification
scheme. Other than  W1189, W3756, and the variables in
the \cite{clement2001} catalog, none of the variables
or suspected variables presented here are cross-listed in the GCVS. As
part of our breadth versus depth approach, most of our variables go
unclassified. 

\begin{deluxetable*}{ccccccccc}
\tablewidth{0pc}
\tablecolumns{9}
\tablecaption{Results for Variables from Clement et al. (2001) and Stetson et al. (2014)\label{tab:knownvars}}
\tablehead{
\colhead{ID\tablenotemark{a}} & \colhead{R.A.\tablenotemark{b}} & \colhead{decl.\tablenotemark{b}} & \colhead{$G$\tablenotemark{c}} & \colhead{Period\tablenotemark{d}} & \colhead{Per. Uncertainty\tablenotemark{e}} & \colhead{Amplitude\tablenotemark{f}} & \colhead{Epoch\tablenotemark{g}} & \colhead{Type\tablenotemark{h}} \\
 \colhead{} & \colhead{(hh:mm:ss)} & \colhead{(dd:mm:ss)} & \colhead{(mag)} & \colhead{(day)} & \colhead{(10$^{-5}$ day)} & \colhead{(mag)} & \colhead{(KBJD)} & \colhead{}  }
\startdata
V6 & 16:23:25.76 & $-$26:26:16.7 & 13.25 & 0.320500 & 0.6 & 0.33 & 2060.58 & RRC\\
V7 & 16:23:25.92 & $-$26:27:42.3 & 13.28 & 0.498787 & 0.7 & 0.99 & 2060.55 & RRAB\\
V8 & 16:23:26.12 & $-$26:29:42.0 & 13.23 & 0.50822 & 1 & 0.87 & 2060.45 & RRAB\\
V9 & 16:23:26.76 & $-$26:29:48.4 & 13.10 & 0.57192 & 2 & 0.87 & 2060.36 & RRAB\\
V10 & 16:23:29.17 & $-$26:28:54.7 & 13.19 & 0.490723 & 0.4 & 0.87 & 2060.70 & RRAB\\
V13 & 16:23:30.88 & $-$26:27:04.4 & 10.04 & ${\sim}$20--30 & \textellipsis & ${\sim}$0.1 & \textellipsis & SR\\
V15 & 16:23:31.93 & $-$26:24:18.5 & 13.38 & 0.443795 & 0.4 & 1.03 & 2060.57 & RRAB\\
V19 & 16:23:35.02 & $-$26:25:36.8 & 13.21 & 0.467809 & 0.4 & 0.99 & 2060.38 & RRAB\\
V27 & 16:23:43.14 & $-$26:27:16.7 & 12.96 & 0.612027 & 0.8 & 0.76 & 2060.74 & RRAB\\
V29 & 16:23:58.22 & $-$26:21:35.4 & 13.05 & 0.52250 & 1 & 0.75 & 2060.69 & RRAB\\
V61 & 16:23:29.72 & $-$26:29:50.7 & 13.08 & 0.265293 & 0.7 & 0.13 & 2060.49 & RRC\\
V66 & 16:23:25.53 & $-$26:29:12.1 & 16.59 & 0.269889 & 0.4 & 0.22 & 2060.29 & EW\\
SC3\tablenotemark{i} & 16:23:35.57 & $-$26:27:08.3 & 16.32 & ${\sim}$19 & \textellipsis & ${\sim}$0.1 & \textellipsis & ?\\ 
SC4\tablenotemark{i} & 16:23:44.77 & $-$26:24:29.4 & 14.88 & 0.43863 & 2 & 0.033 & 2060.62 & ?\\ 
SC5\tablenotemark{j} & 16:23:34.58 & $-$26:25:41.6 & 18.73 & \textellipsis & \textellipsis & \textellipsis & \textellipsis & \textellipsis\\ 
\enddata
\tablenotetext{a}{The identifier by which this object is known in this work, see Table~\ref{tab:ids}. Those prepended with ``V'' are previously identified variables from the catalog of \cite{clement2001}, June 2016 edition, not marked as constant, and those prepended with ``SC'' are candidate variables from \cite{stetson2014}.}
\tablenotetext{b}{J2000.0; data taken  from  {\it Gaia} DR2 \citep{lindegren2018}.}
\tablenotetext{c}{{\it Gaia} $G$ magnitude taken from {\it Gaia} DR2 \citep{riello2018}.}
\tablenotetext{d}{The period of the variability in days.}
\tablenotetext{e}{The uncertainty of the period of the variability, see Section~\ref{sec:amplitudedetermination} for details on how this is measured.}
\tablenotetext{f}{The amplitude of the variability in magnitudes, see Section~\ref{sec:amplitudedetermination} for details on how this is measured.}
\tablenotetext{g}{The epoch of the minimum of the variability, expressed in KBJD (BJD$-2454833.0$). See Section~\ref{sec:amplitudedetermination} for details on how this is measured.}
\tablenotetext{h}{Classification based on the GCVS Variability Types, fourth edition \citep{samus2017}.}
\tablenotetext{i}{Not a cluster member.}
\tablenotetext{j}{Not a cluster member; significantly blended with V19 and unable to determine its own variability.}
\end{deluxetable*}

\begin{figure*}
\begin{center}
\includegraphics[width=.9\textwidth]{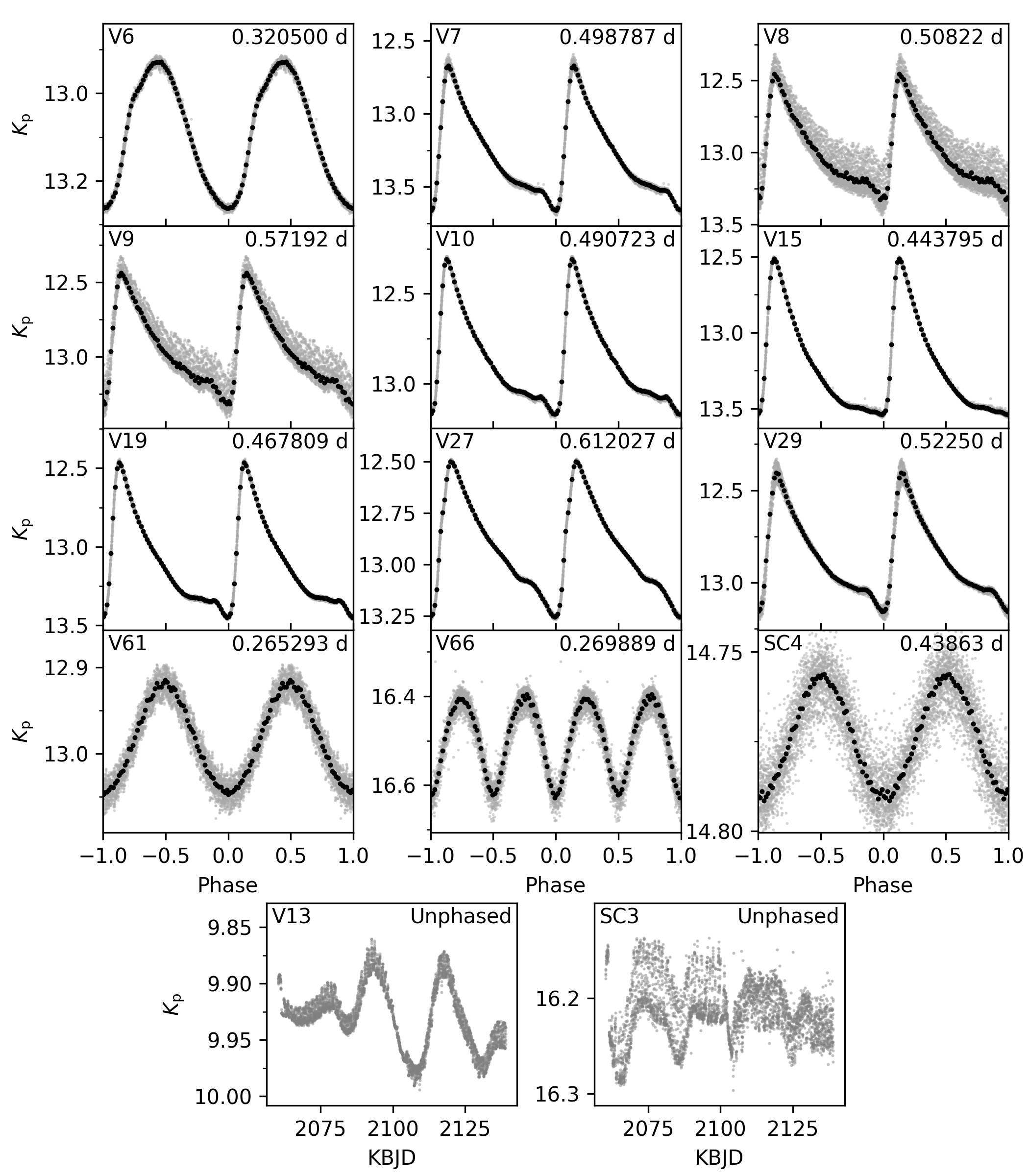}
\end{center}
\caption{\label{fig:knownvars} Light curves for 14 previously
identified variables from  \cite{clement2001}, June
2016 edition, and \cite{stetson2014} that were in the K2 superstamp
and for which we have light curves.  Phase-folded light curves are shown
in the top 12 panels, while the bottom two light curves show unphased
light curves for V13 and SC3.  The top left of each panel shows the
identifier for the associated star, and the top right shows the period
(or ``Unphased'' for V13 and SC3) at which the light curve is folded,
with ``d'' standing for ``day.''
Gray points show the individual magnitude measurements, while the black
points are binned-median values.  For all light curves except that of
SC4, the raw light curve output from our image subtraction is used.
SC4, along with almost all the other light curves presented in this work,
has the additional roll decorrelation and TFA post-processing as
described in Section~\ref{sec:postprocessing}.}
\end{figure*}

\subsection{Summary Figures}
\label{sec:sec3summary}

We first present some figures showing general results from the
variability search. Figure~\ref{fig:varlocations} shows the positions
of the variables in the superstamp images, differentiated by cluster
members, nonmembers, blended variables, and suspected
variables. Figure~\ref{fig:cmd} shows a color-magnitude diagram (CMD)
for the examined stars, with the identified variables and suspected
variables marked.  The HB is visible at $G{\approx}13$ and $0.5\lesssim
G_\mathrm{BP}-G_\mathrm{RP}\lesssim 1.6$, and the main
sequence turnoff is visible at $G{\approx}16.5$ and
$G_\mathrm{BP}-G_\mathrm{RP}{\approx}1.2$. 
We note two stars that are proper motion cluster members and are well off the
expected photometric track. The magenta triangle at
$G_\mathrm{BP}-G_\mathrm{RP}{\approx}0.0$ is W1136 and  is blended with
several other stars ({\it Gaia} DR2 source catalog has four other
stars within 5\arcsec).  However, the {\it
Gaia} DR2 data does not indicate any potential errors in the
photometric measurements: its $G_\mathrm{BP}$ flux error over mean flux is
$3.7\times10^{-3}$ and $G_\mathrm{RP}$ flux error over mean flux is
$2.2\times10^{-3}$ and \texttt{phot\_bp\_rp\_excess\_factor} of 
1.24. The magenta circle at $G_\mathrm{BP}-G_\mathrm{RP}{\approx}2.5$ is
W4490 and has no {\it Gaia} DR2 sources within 5\arcsec.  Its $G_\mathrm{BP}$ flux
error over mean flux is $7.6\times10^{-3}$ and $G_\mathrm{RP}$ flux error over mean
flux is $2.0\times10^{-3}$, while
the  \texttt{phot\_bp\_rp\_excess\_factor} is 1.46.  However, W4490 is
a unique object (likely an X-ray binary) that we discuss further in Section~\ref{sec:otherclustervar}.
Figure~\ref{fig:pervsinfo}
shows photometric data and variability amplitudes versus periods for
all of the variables.  Of particular note is the period-luminosity
relationship seen in the upper-left panel for objects with
multiharmonic variability that mirrors that seen for RR Lyrae
variables.  This will be further discussed in Section~\ref{sec:otherclustervar}.

\subsection{\cite{clement2001} and \cite{stetson2014} Variables}
\label{sec:knownvars}

This subsection focuses exclusively on the previously known variables
found in the catalog of \cite{clement2001}, June 2016 edition, with
additions from \cite{stetson2014}. 
This does not include the other
previously known variables of W1189, reported as a delta Scuti (DSCUT) variable
by \cite{yao1989}, W3756, reported as a gamma Doradus (GDOR)
variable by \cite{yao2006}, or
the asteroseismic giant stars of \cite{miglio2016};
these are discussed later. 
We also note that none of the new
variables of \cite{safonova2016}, which are not in the Clement et
al. catalog, fell on the superstamp. 
A summary of the results for sources not 
marked ``CST'' (constant) in the Clement et al.\ catalog is found in
Table~\ref{tab:knownvars}, and the associated
light curves are found in Figure~\ref{fig:knownvars}.  There are 12
variables from the \cite{clement2001} and two from \cite{stetson2014}
that fell into our observable region.  
The 12 Clement et al.\ variables were first discovered by \cite{leavitt1904}
(V6--V10, V15, V19, V27, and V29), \cite{yao1988} (V61),
and \cite{kaluzny1997} (V66; called V47 in the discovery work).
Given the variability amplitudes  for the Clement et al.\ variables, for
Figure~\ref{fig:knownvars} the raw light curves 
were used, as our implementation of the
Vanderburg-style roll decorrelation did not
perform well for objects with large-amplitude variability at
timescales shorter than our spline fit. 
As a note, we count 17 Clement et al.\ 
variables in the edge regions for which we did not obtain image subtraction
photometry.
We mention this here to show that there is still more that can be done
with the superstamp data than what is presented in this work.  For example,
simple aperture photometry could be used on those stars in the less
crowded portions of the edge region.

V6, V7, V8, V9, V10, V15, V19, V27, V29, and V61 are all RR Lyrae
variables.  V6 and V61 are RRCs, while the others are all RRABs.  
Our
period-search method did not detect any significant variability at
periods other than (sub)harmonics of the main period, but we wish to
stress that our method was focused more on deblending and primary
period finding than on a detailed analysis of small-scale variability
in these RR Lyrae variables. \cite{kuehn2017} performed such an
analysis for the RR Lyrae variables in the M4 K2 superstamp.

V8, V9, and V61 are in fairly close proximity to each
other and to a few other HB stars.  In particular, V8 and V9 are
blended and we observed a beating effect between their two periods that
created the increased scatter of their light curves seen in
Figure~\ref{fig:knownvars}. We did not correct for the blending
between these two stars, though in principle it should be possible.  We do
not know if V61's relatively larger scatter is due to blending with V8
and V9 (it is further from them than they are from each other) or
just generally higher noise in that part of the image due to the
concentration of HB stars, or perhaps something else.

\begin{figure*}
\begin{center}
\includegraphics[width=\textwidth]{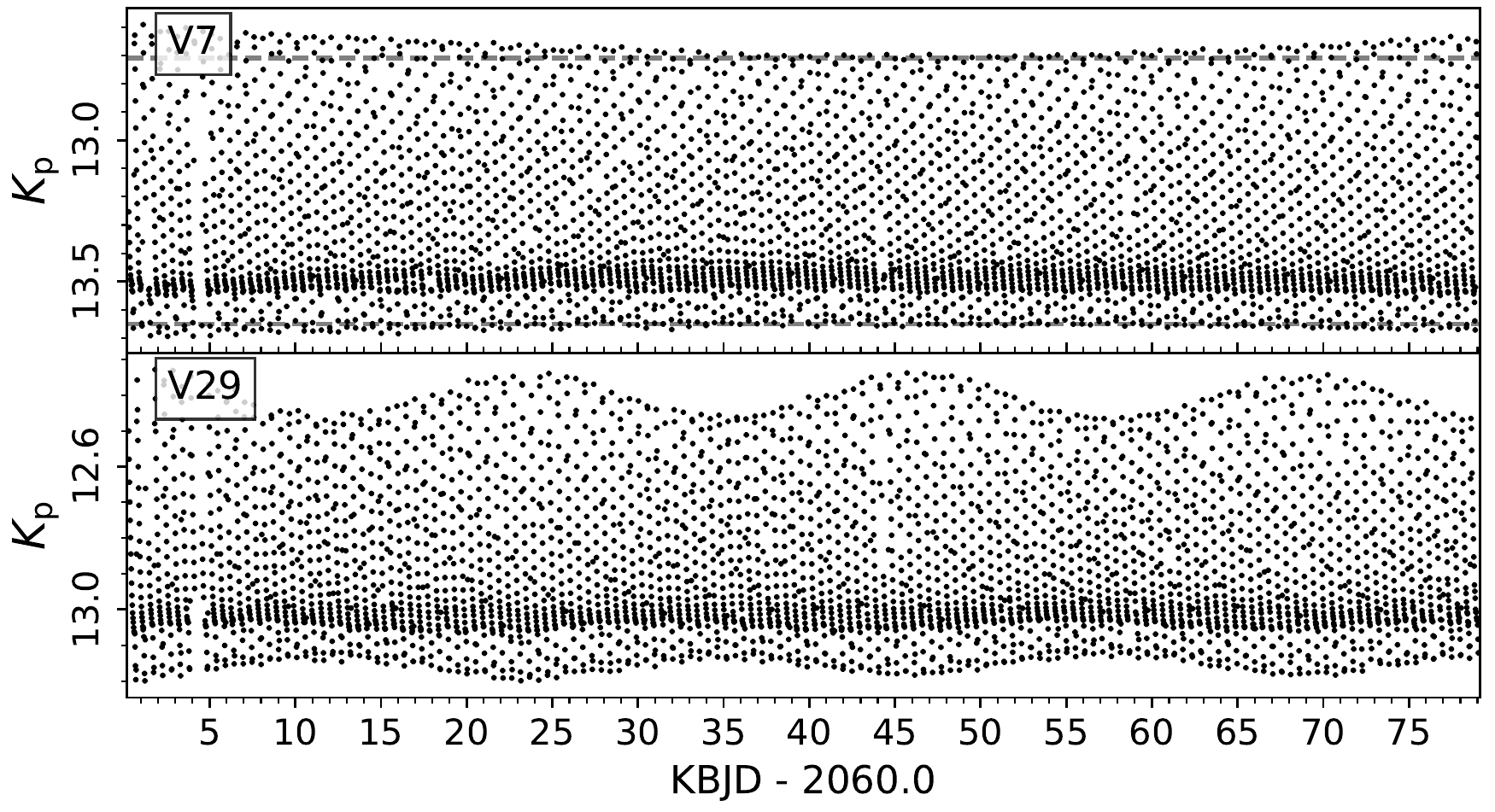}
\end{center}
\caption{\label{fig:blazhko}Light curves for V7 (candidate Blazhko variable)
and V29 (Blazhko variable). The identifier is given in the upper-left
corner of each panel.
The x-axis scale applies to both light curves. The horizontal
dashed lines in V7's panel are arbitrary lines added to help highlight the
suspected Blazhko variation.}
\end{figure*}

We checked for Blazhko variations among the RR Lyrae variables by
searching plots of the (unphased) light curves by eye.
\cite{stetson2014} reported V15 and V29
as candidate Blazhko variables. \cite{kuehn2017}, who used the same K2
superstamp data as us, reported V19 and V29 as Blazhko
variables as detected via sidepeaks in the  amplitude spectra. They
also reported the V35 of \cite{clement2001} as a Blazhko
variable, but this star appeared in our edge region and so we did not
extract a light curve for it.  Here is what we note from our analysis,
with Figure~\ref{fig:blazhko} showing the associated light curves for
V7 and V29:
\begin{itemize}
\item V7: suspected Blazhko variable, with a period longer than the
duration of the observation (most of a cycle is seen).
\item V15: Our manual vetting did not find any Blazhko variability.
As noted above, \cite{stetson2014} marked this as a candidate Blazhko
(though they did not record a period), while \cite{kuehn2017} did
not. V15 is
itself a very peculiar object, as noted by \cite{clementini1994} and
we refer interested readers to that work and its references for full
details.  In short, the star has peculiarities in its light and radial
velocity curves, which could be due either
to this star being in process of transitioning from an RRAB to an RRC
or a strong Blazhko variability.
\item V19: Our manual vetting did not find any Blazhko variability.
The sidepeak analysis of \cite{kuehn2017} found a Blazhko 
period of 16.554 days. 
\item V29: Blazhko variable, as also noted by \cite{kuehn2017} and
listed as a candidate in \cite{stetson2014}.  \cite{kuehn2017} report
a 22.419-day period, which is consistent with what we see.
\end{itemize}

V13 was first reported as a variable star in \cite{leavitt1904} and
is presently reported as being a semi-regular variable (SR).   
\cite{eggen1972}
observed a ${\sim}$40-day variability and an amplitude of $\Delta
V=0.5$ mag.  
In our raw light curve, 
we see low-amplitude variability of ${\sim}$0.1 mag, quasiperiodic with a period range
of ${\sim}$20--30 days, as can be seen in
Figure~\ref{fig:knownvars}. The star is saturated in the images, so it
is possible that systematics remain in our light curve.
We also note that our final light curve for this object did
not have any variability detected for this object, possibly due
to the spline fit fitting out the long-term variability.  We mention
this as an example of long-term variability that can go undetected by
the method employed in this work.

V66 
is a ${\sim}$0.26-day 
contact eclipsing binary of the W Ursae Majoris type (EW by the GCVS
classification). From our analysis, it was not immediately clear which of four
blended stars (V66, as well as W1347, W1380, and W1426) was the source of the variability, as all four had
approximately the same flux amplitude in our light curves.
However, the discovery observations \citep{kaluzny1997} were taken at
much higher 
resolution (median seeing FWHM ${\sim}$1\farcs0--1\farcs1 for five of
the six nights of observation) than the separations of these four
stars---which were comparable to but slightly greater than {\it Kepler}'s
${\sim}$4\arcsec pixel scale.  We thus show the light curve only for
V66 and not any of its blends.

SC3 is not a cluster member. Similar to V13, it did not have
variability detected by our 
pipeline in its final light curve, again likely owing to the long-term
and smooth nature of the 
variability being fitted out by our spline fit.  In the raw light
curve, we observe approximately the same period and amplitude of
variability as \cite{stetson2014}.

SC4, not a cluster member, was identified as a variable
by \cite{stetson2014}.  However, 
{\it Gaia} DR2  has a \texttt{phot\_variable\_flag} triggered on the
nearby W3152, which is a cluster member, and not SC4.  Our pipeline marked SC4 as the true
variable and W3152 as blended with SC4, though the flux amplitudes
are within ${\sim}$15\% of each other.  The resolution of the images
used by SC4 was sufficient to resolve these objects, which had 2\farcs7
separation, so we stick with \cite{stetson2014} in calling SC4 and not
W3152 the variable.

SC5 is reported as a 0.4197-day period object with ${\sim}$0.5 mag
amplitude and it should have easily been detected with our data and
pipeline.  However, it is separated from V19---itself having a
0.4678-day period---by 7\farcs6 and is quite
blended with it.  Our pipeline did not identify any variability for
SC5 at the reported period.  More careful 
removal of V19's signal from the data may prove fruitful for this
object, but we do not perform such an analysis here.

\begin{figure*}
\begin{center}
\includegraphics[width=.85\textwidth]{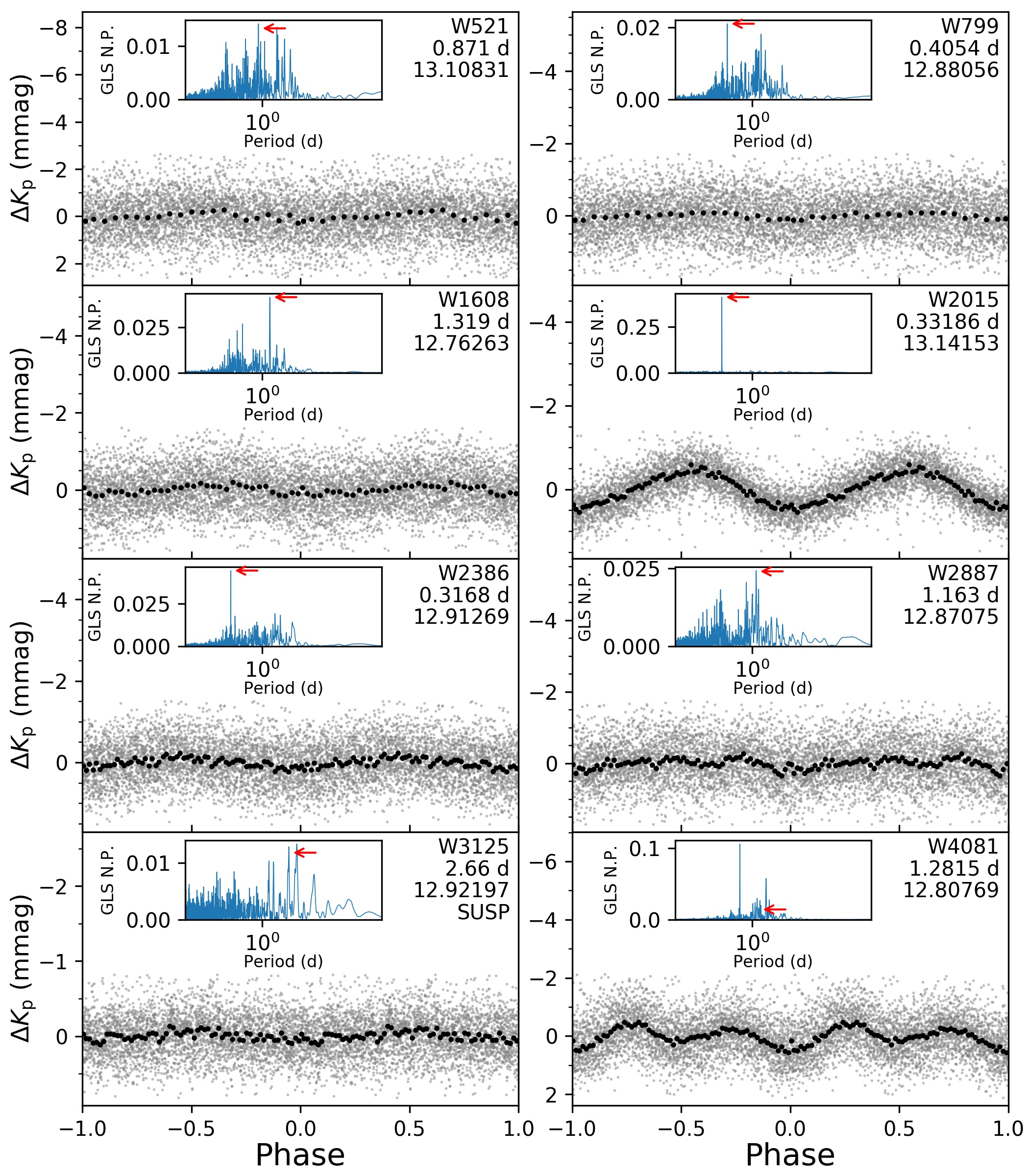}
\end{center}
\caption{\label{fig:hbvars} Phase-folded light curves and periodograms for
the eight stars identified in 
this work as variable or suspected variable HB stars that are not RR Lyrae.
Gray points show the individual magnitude measurements and the black
points are binned-median values.  The y-axis shows $K_\mathrm{p}$, in units
of millimagnitude, relative to the median $K_\mathrm{p}$ magnitude.  In the
top-right corner of each panel is shown (in order from top to bottom)
the object identifier, the period used for the phase folding, the
median magnitude subtracted off, and, for W3125, ``SUSP'' indicating
that this is a suspected variable.  For the inset periodogram in each
panel, ``GLS N.P'' stands for ``Generalized Lomb-Scargle Normalized
Power'' and the red arrow points to the location in the periodogram of
the phase-folding period. For W3125, the arrow
points slightly off the maximum value of the peak as the period used
was taken from a BLS determination of the period rather than a GLS
determination.  For W4081, the arrow is pointing at a period twice
that of the periodogram peak, since upon visual inspection of the
light curve we chose a period twice that found by GLS.
In the inset x-axis
and the listed period, ``d'' stands for ``day.'' W2015 and W2386 are
the mmRRs from \cite{wallace2019}.}
\end{figure*}

Our pipeline also produced light curves for V54 (this work: W3012),
V55 (this work: W3267), and V80 (this work: W3471), all of which
are marked ``CST'' in the \cite{clement2001} catalog, meaning that there is
uncertainty about whether they are actually variable.  Our pipeline
did not flag any significant variability for any of these objects, but
that does not mean they are not variable.  Given the caveats of our variable-search
method and the relatively low noise levels our light curves were able
to reach, we decided to take a closer look at these stars,
particularly their raw light curves.

V54 was marked
``CST'' from the time of its initial listing in the \cite{clement2001} catalog because the
first report of its variability (\citealt{yao1981}; see
also \citealt{yao1981translation} for an English translation) reported such
a small amplitude for the star and it was observed over only
a ${\sim}$2-hour time window total. V54  is a giant star
and a proper motion member of the 
cluster.  It exhibits multiharmonic variability, with the strongest
GLS power at ${\sim}$1.02-day period, with a ${\sim}$1 mmag
variability.  The reason this was not detected by our method is likely
the rich structure of the periodogram boosted the noise value used in
the periodogram SNR calculation, thus leading to an SNR value
that fell below the threshold. This variability, though, is of
${\sim}$1 mmag amplitude, much smaller than the ${\sim}$0.1--0.2 mag
seen for this star in \cite{yao1981} and is probably unrelated to what
they reported.

V55 was also first reported by \cite{yao1981,yao1981translation} and
was also marked ``CST'' from its initial entry into
the \cite{clement2001} catalog for the same reasons as V54.  
V55  is an HB star and a proper motion member of the cluster.
The variability amplitude reported
by \cite{yao1981} for V55 (${\sim}$0.1--0.2 mag) is 
larger than the ${\sim}$3 mmag RMS value we obtain for the raw light
curve or the ${\sim}$0.3 mmag RMS noise value we obtain
for the final light curve.   The strongest GLS period is ${\sim}$3.10
day, but this is somewhat weak and the periodogram overall is fairly noisy.

V80 is a subgiant member of
the cluster.  Variability was reported
by \cite{yao2007} (see \citealt{yao2007translation} for an English
translation) as variable with a period of about a day and with
amplitude of 0.05 mag in $V$. Despite our obtaining an RMS noise level of
${\sim}$0.01 mag in its raw light curve and ${\sim}$3 mmag in its
final light curve, no significant variability is seen.

Thus from our work we think V54 should be marked a low-amplitude
asteroseismic variable and V55 and V80 retain their ``CST''
designations, though it would seem the variability we observe for V54
is not the same variability, or at least significantly changed from,
what was reported by \cite{yao1981}.

\subsection{Millimagnitude RR Lyrae and the Other Horizontal Branch
Stars}
\label{sec:hbstars}

Two of the HB stars---W2015 and W2386---have been more fully examined
in \cite{wallace2019} as potential
low-amplitude RRC pulsators (millimagnitude RR Lyrae variables, or mmRRs as coined in that work).  W2015 is mmRR 1 from that work, W2386 is
mmRR2, and W4081 is G3168 briefly mentioned in that work.  We
define the HB in similar fashion as \cite{wallace2019}: 
stars with $14.3\,<\,G_\mathrm{BP}\,<\,13.0$ and
$G_\mathrm{BP}-G_\mathrm{RP}\,<\,1.5$ and a ${>}$95\% cluster
membership probability (though the membership probabilities for all
these stars are so high that a 99\% cutoff could be used with no
loss). Excluding the 10 stars previously identified as RR Lyrae variables (see
Table~\ref{tab:knownvars}), we have light curves for 24 HB stars, eight of
which we detected as significantly variable.
Information on these HB variables is found in 
Table~\ref{tab:clustervars}, and Figure~\ref{fig:hbvars} shows the
phase-folded light curves and GLS periodograms for these objects.  We
stress once again, though, that our periodogram SNR cutoff can
sometimes exclude stars with significant variability at other periods
close to the peak period, so it is entirely possible that multiharmonic
variability is to be found among many of the other 16 non-RR-Lyrae HB
stars. Indeed, a quick search that we performed revealed many of them---though
not all---to possess multiharmonic variability.  To maintain internal
consistency with our search method, we do not report them in detail
here, but do note again our light curves are available for download
and analysis at \cite{lightcurves}.  Several of these objects are blended with other bright
stars, so we advise appropriate caution in using them.  Two
particularly notable blends we noticed were W818, which is likely a
blend with W1189; and W1607, which is either blended or otherwise left
with a photometric footprint of the somewhat distant V10.  W1607 has
some power in its periodogram outside the blend period and may
possess intrinsic variability.  Likewise, W1628 and W1643 are blended with
V61 and V9 and may require a more careful analysis.

Interpreting the previously identified mmRRs in the context of these
additional HB variables is informative.  Given that the periodogram
structures seem to form 
a continuum between the strongly mono-periodic W2015 and the rich,
very multi-periodic
periodogram of W521, it is possible that what we have called mmRRs are
a transition between the asteroseismic variability of HB stars outside
of the instability strip and the RR Lyrae pulsators inside. 
We note that W2015/mmRR 1 and W3125 are blueward of the instability
strip, W4081 is inside the strip, and the remaining objects are
redward. 
There still remain many questions.  Why does W2015 (mmRR1) have such a
single dominant period 
whereas the other HBs do not have any periods with such great
prominence? What causes the range of periods seen?   What causes
W4081's striking even-odd amplitude modulation, and why is it found in
the instability strip but not pulsating like the RR Lyrae variables?
Certainly the K2 
photometric precision and the observations of concentrations of HB
stars in GCs allows for an unprecedented look at the asteroseismic
variations of HB stars outside the instability strip in addition to
the RR Lyrae variables themselves.  We also echo our previous caveat
that other HB stars with rich periodogram structures may have been
missed by our period search method, and these may not be the only HB
stars with detectable oscillations.

\subsection{Other Cluster Variables}
\label{sec:otherclustervar}

Table~\ref{tab:clustervars} shows information for the variable
cluster members, both proper and suspected variables.  The suspected
variables are more thoroughly discussed and presented in Appendix~\ref{suspvar}.
Figures~\ref{fig:clustervars1}, \ref{fig:clustervars2},
and \ref{fig:clustervars3} show the phase-folded light curves for the
variables.  We
discuss here and in Section~\ref{sec:clustereb} some of the more
notable cluster-member variables.

W4490 has a particularly interesting light curve:  a 1.959-day
period triangular-shaped increase in brightness, with an amplitude of
${\sim}$20 mmag\footnote{The value quoted here and seen in
Figure~\ref{fig:clustervars3} is different from that reported in
Table~\ref{tab:clustervars}, since the former are taken from the raw
and the latter from the final light curves.}.  Figure~\ref{fig:clustervars3} plots the phase-folded
raw light curve instead of the processed, final light curve.  We found that
the processing cut its amplitude approximately in half.  The raw light
curve has systematic noise, most
likely due to this object's period being very close to the
resaturation period (1.962 days) and nearly an integer multiple of the
drift correction and observing cadence. \cite{verbunt2001}  reports a ROSAT
X-ray source detection 2\farcs8 away from this object (object X8 in
NGC6121/M4), with a reported position statistical error on the X-ray
source of 2\farcs6 and an additional projection error of
${\sim}$5\arcsec also at play.  This spatially coincident X-ray
source with the reported variability period have informed our
classification of this object as an X-ray binary. This portion of M4
unfortunately has not been included in fields of view of previous {\it Chandra}
observations, which have been primarily focused on the
cluster's core \citep[e.g.][]{bassa2004}. Its unusual photometry was
noted in Section~\ref{sec:sec3summary} and Figure~\ref{fig:cmd}.  As
measured by {\it Gaia} DR2, this
object is much more red than we would expect for a star of its
luminosity in the cluster.

Of the other cluster-member variables in
Figures~\ref{fig:clustervars1}--\ref{fig:clustervars3}, most
 are low-amplitude sinusoids, possibly including some 
ellipsoidal or rotational variables. Many are giant
stars showing mmRR or 
multiharmonic asteroseismic variability.  For those objects the
periods shown in the figures are typically just the
dominant sinusoidal component.  In Figure~\ref{fig:pervsinfo}, it can
be seen in the top-left panel that these stars appear to extrapolate
the period--luminosity 
relationship of the RR Lyrae, with variables of longer period than the
RR Lyrae variables continuing the relation of the RRABs (the cluster
of diamonds with period greater than 0.4 days), the handful of objects
with periods less than the RRCs (the two diamonds with periods
${\sim}$0.3 days) seeming to form a parallel trend, and objects
falling into the period range of the RR Lyrae variables themselves
having similar $G$ magnitudes as them.  Since $G$ is correlated with
 evolutionary state for these stars, and
thus with stellar density,
it is not surprising that the oscillation periods, which are
determined in part by stellar densities, are correlated with $G$ even
for smaller-amplitude oscillators than the RR Lyrae variables.  The
scatter seen in the relation is probably due to the picking up
different modes for different stars as the dominant cause of the
photometric variability.  We also note an apparent correlation between
amplitude and period in the lower-right panel of
Figure~\ref{fig:pervsinfo} for the multiharmonic and mmRR stars.

There were a
number of variable signals that were indistinguishably blended between
two or more 
stars and that were not able to be disentangled either from our data
or from referencing some other previous work of which we
knew. Table~\ref{tab:blendedvars} lists these objects, both cluster
members and nonmembers, and Figure~\ref{fig:blendedvars} shows the
associated light curves.  W283/W293, and W1318/W1335/W1246, both EWs,
are discussed in 
Section~\ref{sec:clustereb}. 

\startlongtable
\begin{deluxetable*}{cccccccccc}
\tablewidth{0pc}
\tablecolumns{10}
\tablecaption{Newly Discovered Cluster Variables\label{tab:clustervars}}
\tablehead{
\colhead{ID\tablenotemark{a}} & \colhead{R.A.\tablenotemark{b}} & \colhead{decl.\tablenotemark{b}} & \colhead{$G$\tablenotemark{c}} & \colhead{Period\tablenotemark{d}} & \colhead{Per. Unc.\tablenotemark{e}} & \colhead{Amp.\tablenotemark{f}} & \colhead{Epoch\tablenotemark{g}} & \colhead{Method\tablenotemark{h}} & \colhead{Type\tablenotemark{i}} \\
 \colhead{} & \colhead{(hh:mm:ss)} & \colhead{(dd:mm:ss)} & \colhead{(mag)} & \colhead{(day)} & \colhead{(10$^{-4}$ day)} & \colhead{(mmag)} & \colhead{(KBJD)} & \colhead{} & \colhead{}  }
\startdata
\cutinhead{Variables}
W491 & 16:23:11.52 & $-$26:26:41.1 & 13.37 & 0.27941 & 0.6 & 0.3 & 2060.51 & Harm. & mmRR/mh\\
W508 & 16:23:12.04 & $-$26:29:44.0 & 13.86 & 0.16280 & 0.3 & 0.2 & 2060.40 & Harm. & mh\\
W521 & 16:23:12.36 & $-$26:21:58.8 & 13.25 & 0.871 & 10 & 0.4 & 2060.45 & Harm. & mh\\
W566\tablenotemark{j} & 16:23:13.39 & $-$26:29:15.7 & 17.82 & 0.28870 & 0.9 & 2 & 2060.50 & Harm. & EW?\\ 
W689\tablenotemark{k} & 16:23:15.73 & $-$26:25:58.0 & 15.47 & 0.992 & 10 & 0.7 & 2061.21 & Harm. & EA?\\ 
W799 & 16:23:17.63 & $-$26:27:10.6 & 13.01 & 0.4054 & 7 & 0.2 & 2060.41 & Harm. & mh\\
W837 & 16:23:18.22 & $-$26:29:07.6 & 14.39 & 0.09472 & 0.1 & 0.1 & 2060.37 & Harm. & shortperiod\\
W869 & 16:23:18.68 & $-$26:23:43.6 & 10.76 & 2.177 & 20 & 0.3 & 2061.98 & Harm. & mh\\
W1091 & 16:23:21.68 & $-$26:26:47.2 & 13.05 & 0.3583 & 1 & 0.3 & 2060.44 & Harm. & mh\\
W1154\tablenotemark{k} & 16:23:22.5 & $-$26:24:59.4 & 16.15 & 0.991 & 20 & 0.6 & 2060.66 & Harm. & ?\\ 
W1165 & 16:23:22.64 & $-$26:26:22.5 & 12.08 & 1.3338 & 5 & 0.8 & 2061.10 & Harm. & mh\\
W1349\tablenotemark{l} & 16:23:24.98 & $-$26:29:25.3 & 13.23 & 0.3884 & 2 & 0.2 & 2060.43 & Harm. & mh\\ 
W1582\tablenotemark{m} & 16:23:27.84 & $-$26:29:11.9 & 13.78 & 0.17275 & 0.4 & 0.2 & 2060.36 & Harm. & mh\\ 
W1601 & 16:23:28.07 & $-$26:25:02.2 & 19.11 & 4.6337 & 9 & 38 & 2063.48 & Trap. & EA\\
W1608 & 16:23:28.13 & $-$26:26:08.9 & 12.90 & 1.319 & 10 & 0.2 & 2061.60 & Harm. & mh\\
W1735 & 16:23:29.5 & $-$26:29:12.0 & 11.65 & 2.146 & 20 & 1 & 2062.10 & Harm. & mh\\
W1763 & 16:23:29.81 & $-$26:23:25.6 & 12.51 & 0.7836 & 8 & 0.4 & 2060.63 & Harm. & mh\\
W1848 & 16:23:30.51 & $-$26:23:57.9 & 17.59 & 0.4486 & 4 & 3 & 2060.33 & Harm. & ?\\
W1912 & 16:23:31.28 & $-$26:25:16.1 & 14.18 & 0.7247 & 8 & 0.2 & 2060.98 & Harm. & ?\\
W1978 & 16:23:31.99 & $-$26:29:38.1 & 18.61 & 2.06 & 100 & 18 & 2062.19 & Harm. & ?\\
W2005 & 16:23:32.21 & $-$26:27:01.4 & 16.35 & 5.9 & 5000 & 5 & 2060.66 & Harm. & ?\\
W2015 & 16:23:32.3 & $-$26:28:53.5 & 13.23 & 0.33186 & 0.3 & 0.9 & 2060.39 & Harm. & mmRR\\
W2162 & 16:23:33.79 & $-$26:27:50.0 & 13.15 & 0.3464 & 2 & 0.4 & 2060.62 & Harm. & mh\\
W2386 & 16:23:35.93 & $-$26:26:20.9 & 13.05 & 0.3168 & 1 & 0.2 & 2060.59 & Harm. & mmRR/mh\\
W2631 & 16:23:38.46 & $-$26:29:23.9 & 11.84 & 1.65 & 200 & 0.7 & 2061.29 & Harm. & mh\\
W2665 & 16:23:38.84 & $-$26:25:43.1 & 12.56 & 0.6464 & 2 & 0.5 & 2060.55 & Harm. & mh\\
W2678 & 16:23:38.93 & $-$26:22:09.8 & 13.05 & 0.36401 & 0.8 & 0.3 & 2060.32 & Harm. & mmRR?/mh\\
W2740 & 16:23:39.68 & $-$26:24:36.7 & 18.20 & 0.6711 & 9 & 4 & 2060.60 & Harm. & ?\\
W2772 & 16:23:39.97 & $-$26:28:49.3 & 11.06 & 2.773 & 30 & 0.6 & 2062.79 & Harm. & mh\\
W2887 & 16:23:41.33 & $-$26:29:09.1 & 13.00 & 1.163 & 30 & 0.4 & 2060.44 & Harm. & mh\\
W2951 & 16:23:42.14 & $-$26:28:47.7 & 16.67 & 3.94 & 200 & 0.8 & 2063.90 & Harm. & ?\\
W3014 & 16:23:42.83 & $-$26:25:31.6 & 17.41 & 2.414 & 80 & 3 & 2060.64 & Harm. & ?\\
W3033 & 16:23:43.08 & $-$26:28:07.8 & 13.06 & 1.83 & 300 & 0.3 & 2061.75 & Harm. & mh\\
W3070 & 16:23:43.47 & $-$26:23:28.7 & 16.24 & 11.0 & 7000 & 0.7 & 2066.27 & Harm. & ?\\
W3073 & 16:23:43.51 & $-$26:25:37.8 & 13.70 & 0.18344 & 0.2 & 0.3 & 2060.32 & Harm. & mh\\
W3114 & 16:23:44.02 & $-$26:29:31.8 & 18.81 & 1.69 & 100 & 22 & 2061.77 & Harm. & ?\\
W3259 & 16:23:45.81 & $-$26:28:35.4 & 18.91 & 1.734 & 60 & 5 & 2060.84 & Harm. & ?\\
W3407 & 16:23:47.97 & $-$26:28:21.9 & 18.54 & 2.352 & 30 & 8 & 2061.13 & Harm. & ?\\
W3430 & 16:23:48.3 & $-$26:22:42.6 & 17.50 & 0.5107 & 3 & 3 & 2060.37 & Harm. & ?\\
W3480 & 16:23:48.98 & $-$26:29:19.6 & 13.31 & 0.25657 & 0.7 & 0.3 & 2060.29 & Harm. & mmRR?/mh\\
W3485 & 16:23:49.08 & $-$26:28:27.4 & 14.71 & 1.411 & 30 & 0.3 & 2061.51 & Harm. & ?\\
W3742 & 16:23:52.99 & $-$26:28:06.9 & 13.03 & 0.32514 & 0.8 & 0.3 & 2060.37 & Harm. & mmRR/mh\\
W3957 & 16:23:57.1 & $-$26:25:36.5 & 18.61 & 0.995 & 20 & 15 & 2061.20 & Harm. & ?\\
W3996 & 16:23:57.71 & $-$26:22:56.1 & 11.73 & 3.054 & 50 & 0.8 & 2061.88 & Harm. & mh\\
W4081 & 16:23:59.3 & $-$26:27:15.8 & 12.93 & 1.2815 & 9 & 0.9 & 2061.10 & Harm. & mmRR/mh\\
W4237 & 16:24:04.17 & $-$26:27:03.1 & 15.87 & 0.09282 & 0.2 & 0.3 & 2060.35 & Harm. & shortperiod\\
W4333 & 16:24:07.73 & $-$26:28:41.4 & 16.65 & 0.3618 & 3 & 0.6 & 2060.30 & Harm. & EA?\\
W4361 & 16:24:08.57 & $-$26:24:55.5 & 11.36 & 4.768 & 30 & 0.7 & 2065.02 & Trap. & EB\\
W4490 & 16:24:14.75 & $-$26:27:51.2 & 15.62 & 1.959 & 10 & 11 & 2060.99 & Harm. & xrb\\
\cutinhead{Suspected Variables}
W58 & 16:22:57.25 & $-$26:28:44.3 & 18.78 & 0.2228 & 3 & 39 & 2060.44 & Harm. & \textellipsis\\
W267 & 16:23:05.52 & $-$26:27:01.1 & 17.82 & 2.76 & 100 & 2 & 2060.99 & Harm. & \textellipsis\\
W371 & 16:23:09.14 & $-$26:30:00.4 & 15.70 & 0.2461 & 2 & 0.3 & 2060.48 & Harm. & \textellipsis\\
W435 & 16:23:10.35 & $-$26:29:31.1 & 16.58 & 0.2468 & 2 & 0.3 & 2060.37 & Harm. & \textellipsis\\
W461 & 16:23:10.94 & $-$26:26:33.3 & 17.64 & 3.90 & 300 & 13 & 2061.24 & Harm. & \textellipsis\\
W829 & 16:23:18.1 & $-$26:21:44.1 & 18.17 & 7.9 & 8000 & 5 & 2064.10 & Harm. & \textellipsis\\
W901 & 16:23:19.17 & $-$26:27:52.4 & 17.12 & 0.2121 & 2 & 0.7 & 2060.40 & Harm. & \textellipsis\\
W920 & 16:23:19.49 & $-$26:25:47.2 & 17.17 & 0.3321 & 2 & 0.7 & 2060.52 & Harm. & \textellipsis\\
W1056 & 16:23:21.29 & $-$26:28:44.9 & 17.95 & 25.66 & 500 & 21 & 2079.96 & Trap. & \textellipsis\\
W1068 & 16:23:21.4 & $-$26:28:33.9 & 13.85 & 1.256 & 20 & 0.2 & 2060.88 & Harm. & \textellipsis\\
W1208 & 16:23:23.17 & $-$26:26:02.9 & 18.23 & 0.315311 & 0.06 & 3 & 2060.37 & Harm. & \textellipsis\\
W1222 & 16:23:23.35 & $-$26:29:24.2 & 18.60 & 11.679 & 80 & 15 & 2064.54 & Trap. & \textellipsis\\
W1263 & 16:23:23.87 & $-$26:26:04.9 & 16.33 & 5.830 & 50 & 1 & 2064.15 & Trap. & \textellipsis\\
W1539 & 16:23:27.42 & $-$26:26:25.5 & 17.60 & 0.13900 & 0.9 & 2 & 2060.31 & Harm. & \textellipsis\\
W1717 & 16:23:29.37 & $-$26:26:28.0 & 13.98 & 0.12580 & 0.3 & 0.1 & 2060.40 & Harm. & \textellipsis\\
W1725 & 16:23:29.43 & $-$26:28:17.7 & 16.67 & 0.18434 & 0.8 & 0.9 & 2060.44 & Harm. & \textellipsis\\
W1809 & 16:23:30.13 & $-$26:21:36.7 & 18.56 & 14.14 & 200 & 4 & 2071.16 & Trap. & \textellipsis\\
W1834 & 16:23:30.39 & $-$26:28:23.2 & 17.53 & 9.29 & 300 & $-$2\tablenotemark{n} & 2065.57 & Trap. & \textellipsis\\ 
W1864 & 16:23:30.74 & $-$26:27:27.7 & 17.48 & 4.3 & 2000 & 15 & 2060.93 & Harm. & \textellipsis\\
W1938 & 16:23:31.53 & $-$26:27:49.8 & 17.48 & 3.4391 & 10 & 4 & 2061.64 & Trap. & \textellipsis\\
W1947 & 16:23:31.63 & $-$26:29:23.6 & 18.99 & 1.597 & 60 & 12 & 2061.19 & Harm. & \textellipsis\\
W1953 & 16:23:31.68 & $-$26:28:06.8 & 18.00 & 2.34 & 400 & 1\tablenotemark{o} & 2061.42 & Harm. & \textellipsis\\ 
W2109 & 16:23:33.19 & $-$26:28:10.7 & 17.19 & 0.5065 & 2 & 4 & 2060.69 & Harm. & \textellipsis\\
W2126 & 16:23:33.42 & $-$26:29:39.2 & 17.55 & 2.66 & 200 & 5 & 2062.22 & Harm. & \textellipsis\\
W2127 & 16:23:33.45 & $-$26:29:29.7 & 17.98 & \textellipsis\tablenotemark{p} & \textellipsis & ${\sim}-40$ & ${\sim}2096$ & \textellipsis & \textellipsis\\ 
W2233 & 16:23:34.48 & $-$26:26:29.6 & 18.91 & 0.46817 & 0.3 & 5 & 2060.29 & Harm. & \textellipsis\\
W2272 & 16:23:34.85 & $-$26:26:04.6 & 18.74 & 2.223 & 70 & 26 & 2062.05 & Harm. & \textellipsis\\
W2324 & 16:23:35.27 & $-$26:23:31.2 & 17.76 & 1.53 & 200 & 2 & 2060.89 & Harm. & \textellipsis\\
W2499 & 16:23:37.11 & $-$26:28:45.6 & 16.72 & 3.832 & 40 & 10 & 2062.90 & Harm. & \textellipsis\\
W2515 & 16:23:37.28 & $-$26:28:08.5 & 19.11 & 1.408 & 10 & 8 & 2061.36 & Harm. & \textellipsis\\
W2543 & 16:23:37.6 & $-$26:27:20.3 & 16.10 & 31.07 & 500 & 1 & 2073.32 & Trap. & \textellipsis\\
W2556 & 16:23:37.7 & $-$26:27:20.4 & 16.03 & 33.96 & 900 & 1 & 2079.36 & Trap. & \textellipsis\\
W2577 & 16:23:37.94 & $-$26:28:41.7 & 13.01 & 3.20 & 100 & 0.4 & 2062.23 & Harm. & \textellipsis\\
W2616 & 16:23:38.3 & $-$26:29:03.4 & 18.45 & 3.81 & 200 & 40 & 2060.81 & Harm. & \textellipsis\\
W2641 & 16:23:38.58 & $-$26:29:11.7 & 14.43 & 0.8678 & 7 & 2 & 2061.11 & Harm. & \textellipsis\\
W2747 & 16:23:39.74 & $-$26:29:32.8 & 17.19 & 7.091 & 80 & 4 & 2067.02 & Harm. & \textellipsis\\
W2753 & 16:23:39.78 & $-$26:29:42.1 & 18.44 & 0.5474 & 5 & 9 & 2060.42 & Harm. & \textellipsis\\
W2790 & 16:23:40.27 & $-$26:27:37.8 & 17.42 & 6.36 & 200 & $-$2 & 2064.21 & Trap. & \textellipsis\\
W2800 & 16:23:40.4 & $-$26:28:20.6 & 17.87 & 1.84 & 200 & 5 & 2061.85 & Harm. & \textellipsis\\
W2819 & 16:23:40.61 & $-$26:29:02.1 & 18.31 & 1.66 & 200 & 120 & 2060.77 & Harm. & \textellipsis\\
W2876 & 16:23:41.22 & $-$26:28:53.0 & 16.64 & 4.203 & 90 & 5 & 2062.41 & Harm. & \textellipsis\\
W2893 & 16:23:41.41 & $-$26:23:18.0 & 17.83 & 1.28 & 200 & 1 & 2060.63 & Harm. & \textellipsis\\
W2966 & 16:23:42.25 & $-$26:27:42.2 & 15.75 & 0.372 & 20 & 0.2 & 2060.63 & Harm. & \textellipsis\\
W3105 & 16:23:43.88 & $-$26:27:36.8 & 16.83 & 0.6060 & 3 & 4 & 2060.33 & Harm. & \textellipsis\\
W3125 & 16:23:44.21 & $-$26:28:24.4 & 13.00 & 2.66 & 500 & 0.1 & 2061.01 & Harm. & \textellipsis\\
W3282 & 16:23:46.11 & $-$26:25:34.7 & 17.37 & 2.50 & 200 & 1 & 2061.92 & Harm. & \textellipsis\\
W3313 & 16:23:46.53 & $-$26:28:41.7 & 17.66 & 3.978 & 40 & 6 & 2063.97 & Trap. & \textellipsis\\
W3371 & 16:23:47.31 & $-$26:22:30.4 & 13.91 & 0.966 & 20 & 0.2 & 2060.42 & Harm. & \textellipsis\\
W3521 & 16:23:49.59 & $-$26:29:30.8 & 18.84 & 2.692 & 100 & 22 & 2060.45 & Harm. & \textellipsis\\
W3552 & 16:23:50.08 & $-$26:29:33.4 & 14.78 & 9.0 & 7000 & 1 & 2063.41 & Harm. & \textellipsis\\
W3887 & 16:23:55.53 & $-$26:28:34.0 & 17.26 & 2.443 & 30 & 1 & 2062.30 & Trap. & \textellipsis\\
W3901 & 16:23:55.82 & $-$26:29:20.8 & 17.11 & 0.522 & 30 & 0.9 & 2060.79 & Harm. & \textellipsis\\
W4014 & 16:23:58.03 & $-$26:23:30.4 & 18.47 & 1.88 & 400 & 12 & 2060.28 & Harm. & \textellipsis\\
W4143 & 16:24:01.17 & $-$26:25:13.6 & 17.27 & 3.403 & 20 & 2 & 2063.22 & Trap. & \textellipsis\\
W4250 & 16:24:04.81 & $-$26:24:17.4 & 19.01 & 1.784 & 20 & 33 & 2060.90 & Harm. & \textellipsis\\
W4268 & 16:24:05.39 & $-$26:29:16.2 & 17.31 & 1.983 & 10 & 2 & 2061.75 & Harm. & \textellipsis\\
W4301 & 16:24:06.43 & $-$26:28:53.3 & 18.70 & 4.283 & 20 & 5 & 2062.69 & Trap. & \textellipsis\\
\enddata
\tablecomments{Classifications are not attempted for the suspected variables.  Explanations regarding why these are reported as suspected instead of discovered variables can be found in Appendix~\ref{suspvar}.}
\tablenotetext{a}{The identifier by which this object is known in this work, see Table~\ref{tab:ids}.}
\tablenotetext{b}{J2000.0; data taken from {\it Gaia} DR2 \citep{lindegren2018}. All entries in this table are DR2 sources, so none of the information presented is from {\it Gaia} DR1.}
\tablenotetext{c}{{\it Gaia} $G$ magnitude taken from {\it Gaia} DR2 \citep{riello2018}.  All entries in this table are DR2 sources, so none of the information presented is from {\it Gaia} DR1.}
\tablenotetext{d}{The period of the variability in days.}
\tablenotetext{e}{The uncertainty of the period of the variability in $10^{-4}$ days, see Section~\ref{sec:amplitudedetermination} for details on how this is measured.}
\tablenotetext{f}{The amplitude of the variability in millimagnitudes, see Section~\ref{sec:amplitudedetermination} for details on how this is measured. A negative amplitude means that the light curve shows a box-like signal that is a brightening, rather than the more common eclipse-based dimmings for such signals.}
\tablenotetext{g}{The epoch of the minimum of the variability, expressed in KBJD (BJD$-2454833.0$). See Section~\ref{sec:amplitudedetermination} for details on how this is measured.}
\tablenotetext{h}{Method used for determining amplitude and epoch. ``Harm.'' means a harmonic fit was used and ``Trap.'' means a trapezoid fit was used.}
\tablenotetext{i}{Classification based on the GCVS Variability Types, fourth edition \citep{samus2017}, where possible.  Additional designations used: ``mmRR'', millimagnitude RR Lyrae; ``mh'', multiharmonic variability; ``shortperiod'', sinusoidal variability of ${<}0.1$-day period; ``xrb'', a likely X-ray binary, but not classified as ``X'' since we do not know of variability in the X-ray emission.}
\tablenotetext{j}{Six other stars observed with same variability; this chosen as variable since it was most robust detection; see paper for details.}
\tablenotetext{k}{These two stars (W689 and W1154) are 27 pixels apart but have consistent periods and, based on our analysis, may phase with each other.}
\tablenotetext{l}{Slightly blended with V8. This detected variability is not a (sub)harmonic of that variability, so we are confident this belongs to the star itself.}
\tablenotetext{m}{Slightly blended with V10. This detected variability is not a (sub)harmonic of that variability, so we are confident this belongs to the star itself.}
\tablenotetext{n}{The trapezoid model appeared to fail to fit the full amplitude of the signal.  Actual amplitude may be ${\sim}$2--3 times larger.}
\tablenotetext{o}{Epoch and possibly amplitude may be inaccurate owing to PDM being employed to fold these transits and a harmonic fit being used to determine epoch and amplitude.}
\tablenotetext{p}{Single event.}
\end{deluxetable*}

\begin{figure*}
\begin{center}
\includegraphics[]{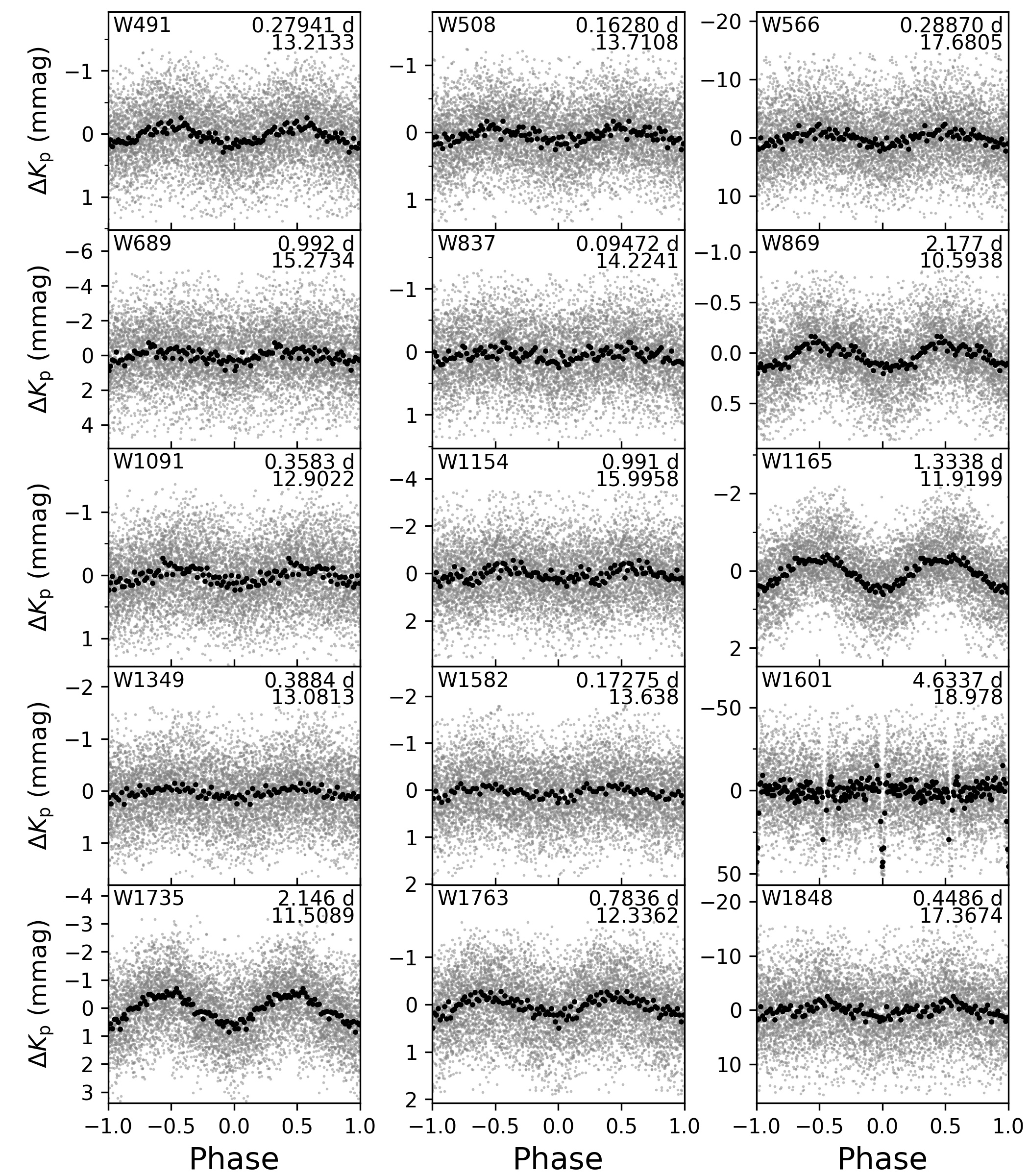}
\end{center}
\caption{\label{fig:clustervars1} Phase-folded light curves for
cluster members that, other than W2665 and W3033, are newly identified
as variable stars in 
this work. W2665 and W3033 were previously identified
by \cite{miglio2016}.  The panels are ordered by the target identifier.  Here we
show the first 15 cluster variables.  Additional variables are shown
in Figures~\ref{fig:clustervars2} and \ref{fig:clustervars3}.
Gray points show the individual magnitude measurements and the black
points are binned-median values.  The y-axis shows $K_\mathrm{p}$, in units
of millimagnitude, relative to the median $K_\mathrm{p}$ magnitude. 
In each panel, the identifier of the star is shown in the upper left
corner, and (from top to bottom) the folding period and subtracted
median magnitude are shown in the upper right corner.
For the listed period, ``d'' stands for ``day.''}
\end{figure*}

\begin{figure*}
\begin{center}
\includegraphics[]{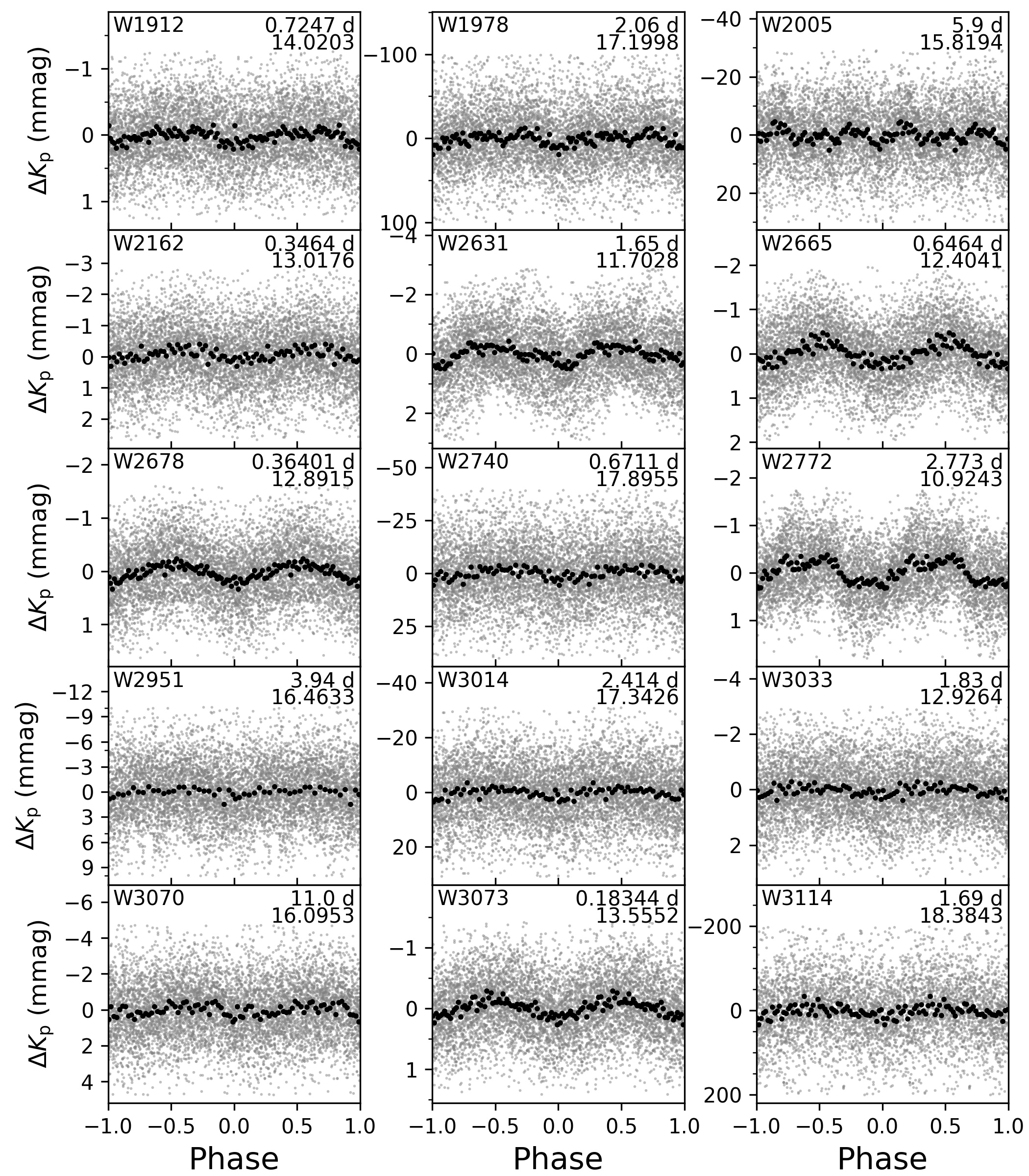}
\end{center}
\caption{\label{fig:clustervars2} Same as
Figure~\ref{fig:clustervars1}, but for additional cluster member variables.}
\end{figure*}

\begin{figure*}
\begin{center}
\includegraphics[]{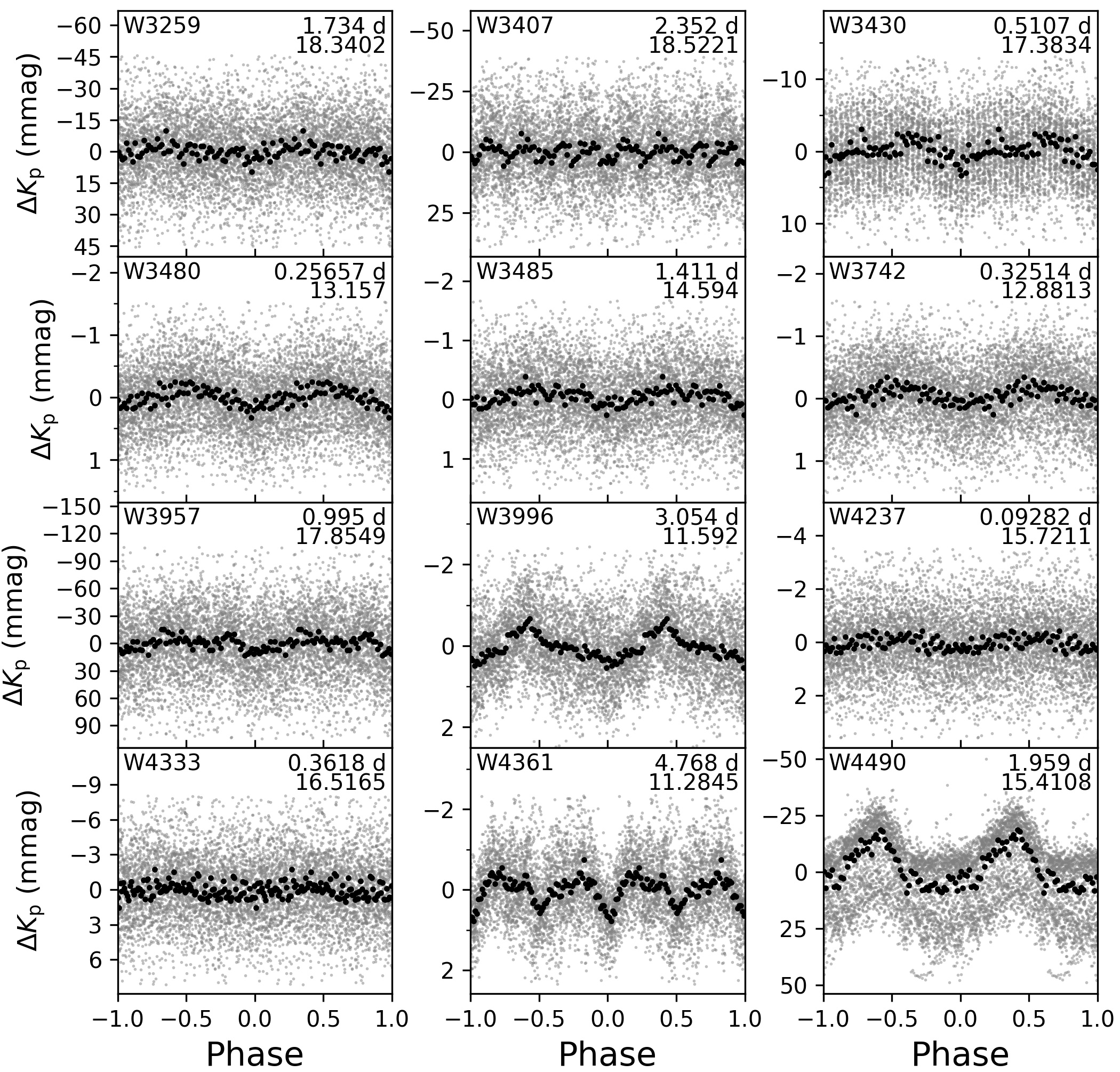}
\end{center}
\caption{\label{fig:clustervars3} Same as
Figures~\ref{fig:clustervars1} and \ref{fig:clustervars2}, but for
additional cluster member variables. The data from W4490 are taken from
its raw light curve.}
\end{figure*}

\begin{deluxetable*}{cccccccccc}
\tablewidth{0pc}
\tablecolumns{10}
\tablecaption{Newly Discovered Variables that are Indeterminable Blends\label{tab:blendedvars}}
\tablehead{
\colhead{ID\tablenotemark{a}} & \colhead{R.A.\tablenotemark{a}} & \colhead{decl.\tablenotemark{a}} & \colhead{$G$\tablenotemark{a}} & \colhead{Period\tablenotemark{a}} & \colhead{Per. Unc.\tablenotemark{a}} & \colhead{Amp.\tablenotemark{a}} & \colhead{Epoch\tablenotemark{b}} & \colhead{Type\tablenotemark{a}} & \colhead{Mem. Prob.\tablenotemark{c}} \\
 \colhead{} & \colhead{(hh:mm:ss)} & \colhead{(dd:mm:ss)} & \colhead{(mag)} & \colhead{(day)} & \colhead{(10$^{-4}$ day)} & \colhead{(mmag)} & \colhead{(KBJD)} & \colhead{} & \colhead{}  }
\startdata
\cutinhead{Blended Stars All Cluster Members}
W283 & 16:23:06.14 & $-$26:27:45.8 & 18.28 & 0.20450 & 0.3 & 6 & 2060.40 & EW & 1.00\\
W293 & 16:23:06.46 & $-$26:27:50.9 & 19.08 & 0.20450 & 0.1 & 23 & 2060.40 & EW & 1.00\\
\hline
W1129 & 16:23:22.23 & $-$26:28:00.3 & 17.71 & 0.967 & 20 & 3 & 2060.60 & ? & 1.00\\
W1136 & 16:23:22.29 & $-$26:28:03.7 & 15.80 & 0.967 & 10 & 0.8 & 2060.55 & ? & 1.00\\
W1146 & 16:23:22.38 & $-$26:27:59.3 & 17.77 & 0.969 & 20 & 3 & 2060.46 & ? & 1.00\\
\hline
W2262 & 16:23:34.78 & $-$26:29:15.1 & 17.38 & 0.5003\tablenotemark{d} & 10 & 4 & 2060.63 & ? & 1.00\\ 
W2282 & 16:23:34.95 & $-$26:29:14.2 & 18.32 & 0.50016 & 0.5 & 8 & 2060.61 & ? & 1.00\\
W2289 & 16:23:34.99 & $-$26:29:10.4 & 18.86 & \textellipsis\tablenotemark{e} & \textellipsis & \textellipsis & \textellipsis & \textellipsis & 1.00\\ 
W2300 & 16:23:35.08 & $-$26:29:12.7 & 16.97 & 0.4989 & 1 & 2 & 2060.65 & ? & 1.00\\
\hline
W2826 & 16:23:40.66 & $-$26:29:27.8 & 16.87 & 2.004 & 40 & 2 & 2060.53 & ? & 1.00\\
W2830 & 16:23:40.69 & $-$26:29:30.4 & 17.63 & 2.00 & 100 & 4 & 2060.54 & ? & 1.00\\
\hline
W3431 & 16:23:48.31 & $-$26:28:13.4 & 17.41 & 0.5887 & 4 & 3 & 2060.47 & EW? & 1.00\\
W3436 & 16:23:48.42 & $-$26:28:13.7 & 17.51 & 0.5889 & 4 & 3 & 2060.47 & EW? & 1.00\\
W3456 & 16:23:48.64 & $-$26:28:14.8 & 17.87 & 0.5887 & 4 & 4 & 2060.46 & EW? & 1.00\\
\cutinhead{Blended Stars Mixed Between Cluster Members and Non- or Ambiguous Members}
W1318 & 16:23:24.57 & $-$26:26:23.0 & 18.97 & 0.277389 & 0.07 & 270 & 2060.34 & EW & 1.00\\
W1335 & 16:23:24.86 & $-$26:26:22.8 & 18.37 & 0.27742 & 0.9 & 180 & 2060.33 & EW & \textellipsis\tablenotemark{f}\\ 
W1346 & 16:23:24.95 & $-$26:26:28.8 & 18.00 & 0.277415 & 0.09 & 51 & 2060.33 & EW & 1.00\\
\hline
W2006 & 16:23:32.23 & $-$26:22:48.3 & 18.69 & 2.440 & 40 & 16 & 2062.01 & ? & 1.00\\
W2013 & 16:23:32.29 & $-$26:22:44.2 & 18.55 & 2.440 & 20 & 18 & 2062.03 & ? & 0.00\\
\hline
W2761 & 16:23:39.86 & $-$26:29:24.5 & 16.81 & 1.0002 & 10 & 4 & 2060.68 & EW? & 1.00\\
W2779 & 16:23:40.17 & $-$26:29:26.1 & 18.11 & 1.000 & 20 & 8 & 2060.54 & EW? & 1.00\\
W2793 & 16:23:40.28 & $-$26:29:26.2 & 17.14 & 1.000 & 20 & 4 & 2060.68 & EW? & 1.00\\
W2813 & 16:23:40.55 & $-$26:29:23.5 & 18.86 & 0.998 & 40 & 8 & 2060.76 & EW? & 0.00\\
\hline
W3883\tablenotemark{g} & 16:23:55.35 & $-$26:24:51.0 & 17.43 & 2.583 & 60 & 2 & 2062.05 & EA? & 1.00\\ 
W3894 & 16:23:55.62 & $-$26:24:52.6 & 16.83 & 2.590 & 60 & 1 & 2062.10 & EA? & 0.73\tablenotemark{h}\\ 
\enddata
\tablecomments{All amplitudes and epochs calculated using a harmonic fit, compared to Tables \ref{tab:clustervars} and \ref{tab:nonclustervars} where some were determined with a trapezoid fit}
\tablenotetext{a}{See table notes for Table~\ref{tab:clustervars} for details on these columns.}
\tablenotetext{b}{The epoch of the minimum of the variability, expressed in KBJD (BJD$-2454833.0$). See Appendix~\ref{suspvar} for details on how this is measured. Significant differences in epochs between blended objects are due to differences in the fitted harmonics for each case; these objects do phase up.}
\tablenotetext{c}{Cluster membership probability, as calculated by \cite{wallace2018}.}
\tablenotetext{d}{The best period found for this object in our period search was 1.500 days, which may be a modulation of the ${\sim}$0.5 day period.}
\tablenotetext{e}{The best period found for this object was 0.38587 days.  We were unable to get a good fit on the ${\sim}$0.5 day period, but this object does have visible variability when folded on this period and has an image locations very close to the other stars in this blended group. It may be that there is more than one variable in this group.}
\tablenotetext{f}{No proper motion data available; probable photometric cluster member.}
\tablenotetext{g}{Both W3883 and W3894 have similar period and epoch as W4084, but are ${\sim}$66 pixels away.}
\tablenotetext{h}{Probable photometric cluster member.}
\end{deluxetable*}

\begin{figure*}
\begin{center}
\includegraphics[]{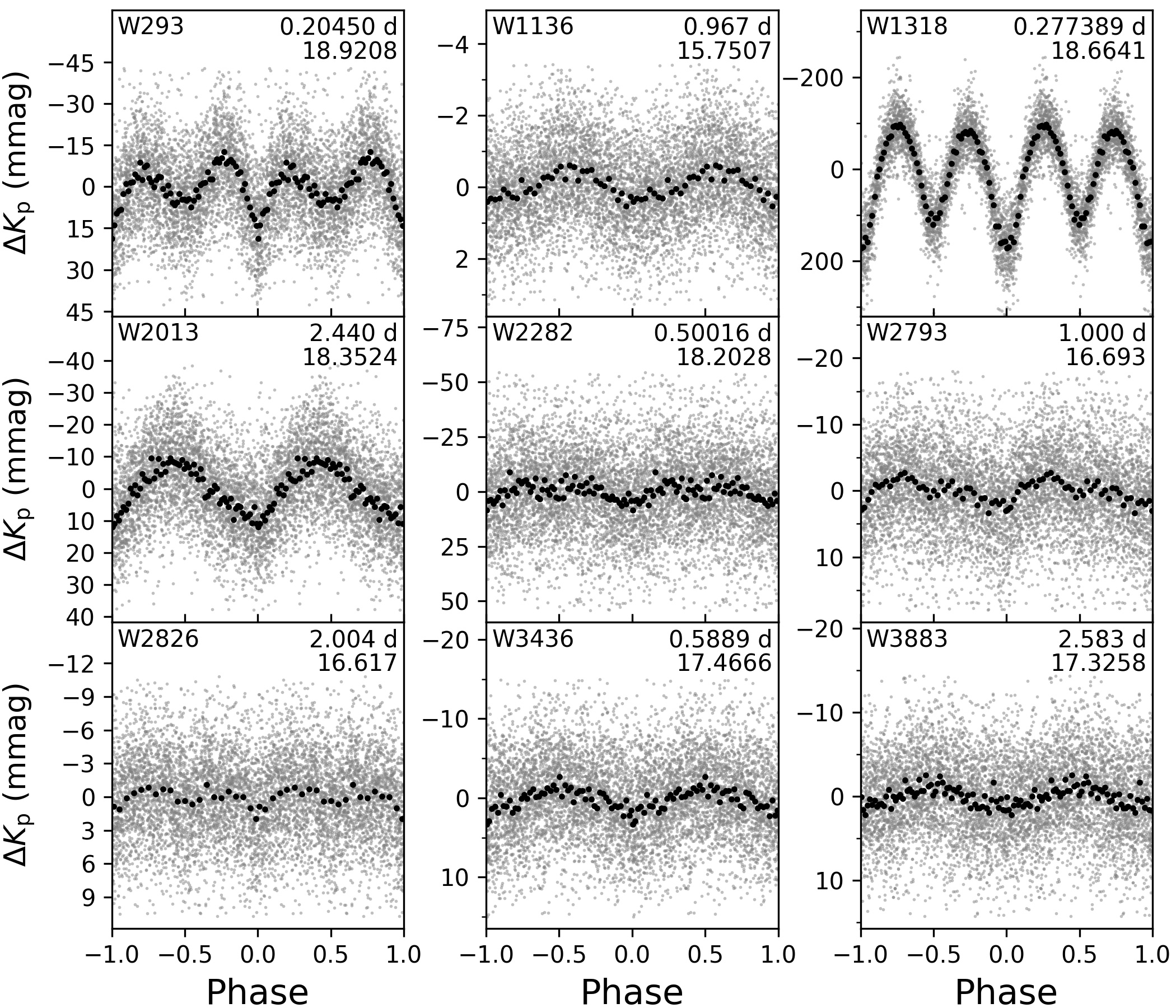}
\end{center}
\caption{\label{fig:blendedvars} Same as
Figure~\ref{fig:clustervars1}, but for stars with signals that are
indistinguishably 
blended in our data.  In each case, only one star from each set of
blended stars is chosen to represent the light curve.  See
Table~\ref{tab:blendedvars} for more information.}
\end{figure*}
\clearpage

\subsection{Cluster Eclipsing and Contact Binaries}
\label{sec:clustereb}

W1601, shown in Figure~\ref{fig:clustervars1},
is a detached eclipsing binary with a 4.6337-day period.  The
phase difference between the primary and secondary eclipses reveals
the system to be slightly eccentric. 
The system appears to be grazing, with a primary eclipse depth
of $0.038\pm0.005$ mag, a fractional duration of $0.046\pm0.003$, and
an eclipse ingress fractional duration of $0.016\pm0.004$, making the
eclipse very triangular.  There is also a sinusoidal variability on
top of the eclipses, suggesting ellipsoidal variability, not terribly
surprising considering the short period of the binary. This informs
our classification of this as an Algol-type eclipsing binary
(EA).  Based on the Clement et al.\ catalog, this is the sixth EA known in M4,
with the note that the two EAs of \cite{safonova2016} are not cluster
members based on the proper motions reported there.

W4361, shown in
Figure~\ref{fig:clustervars3}, is possibly another eclipsing binary.
In this case the system appears to be  semi-detached or maybe
even contact binary.  The eclipses are very triangular.  The depth of
the primary eclipse based on our trapezoid fit is $0.72\pm0.05$ mmag,
with a fractional eclipse duration of $0.21\pm0.01$ and fractional
ingress duration of $0.07\pm0.01$.  This is a red giant star, with a radius that
should be much larger than the ${\sim}$15 R$_\sun$ implied by the
orbital period and the ${\sim}$0.8 M$_\sun$ maximum expected masses
for each of the stars given their membership in the cluster.  Perhaps
W4361 is simply blended with a background eclipsing binary or even
another binary in the cluster.

W293, blended with W283, is a clear example of an EW, having a period of 0.20450 days and a primary
eclipse depth of ${\sim}$30 mmag and a secondary eclipse depth of
approximately half that.  Both stars are cluster members.  
Similarly, W1318, blended with W1335 and
W1346, is also a clear EW.  The orbital period is 0.277389 days and
the primary eclipse depth is ${\sim}$20 mmag and the secondary eclipse
depth ${\sim}$10 mmag.  W1318 and W1346 are proper motion members of
the cluster, but W1335 does not have reported proper motions in {\it
Gaia} DR2.  However, based on its CMD location ($G{\approx}18.4$, $G_\mathrm{BP}-G_\mathrm{RP}{\approx}1.44$), it
is a probable 
cluster member, and so we report a high degree of certainty that this
EW also belongs to the cluster.  There are also two other suspected
EWs: W3431/W3436/W3456, all three of which are cluster members, and
W2761/W2779/W2793/W2813, of which all but the last are cluster
members.

\subsection{Variables Not In M4}

\begin{figure*}
\begin{center}
\includegraphics[]{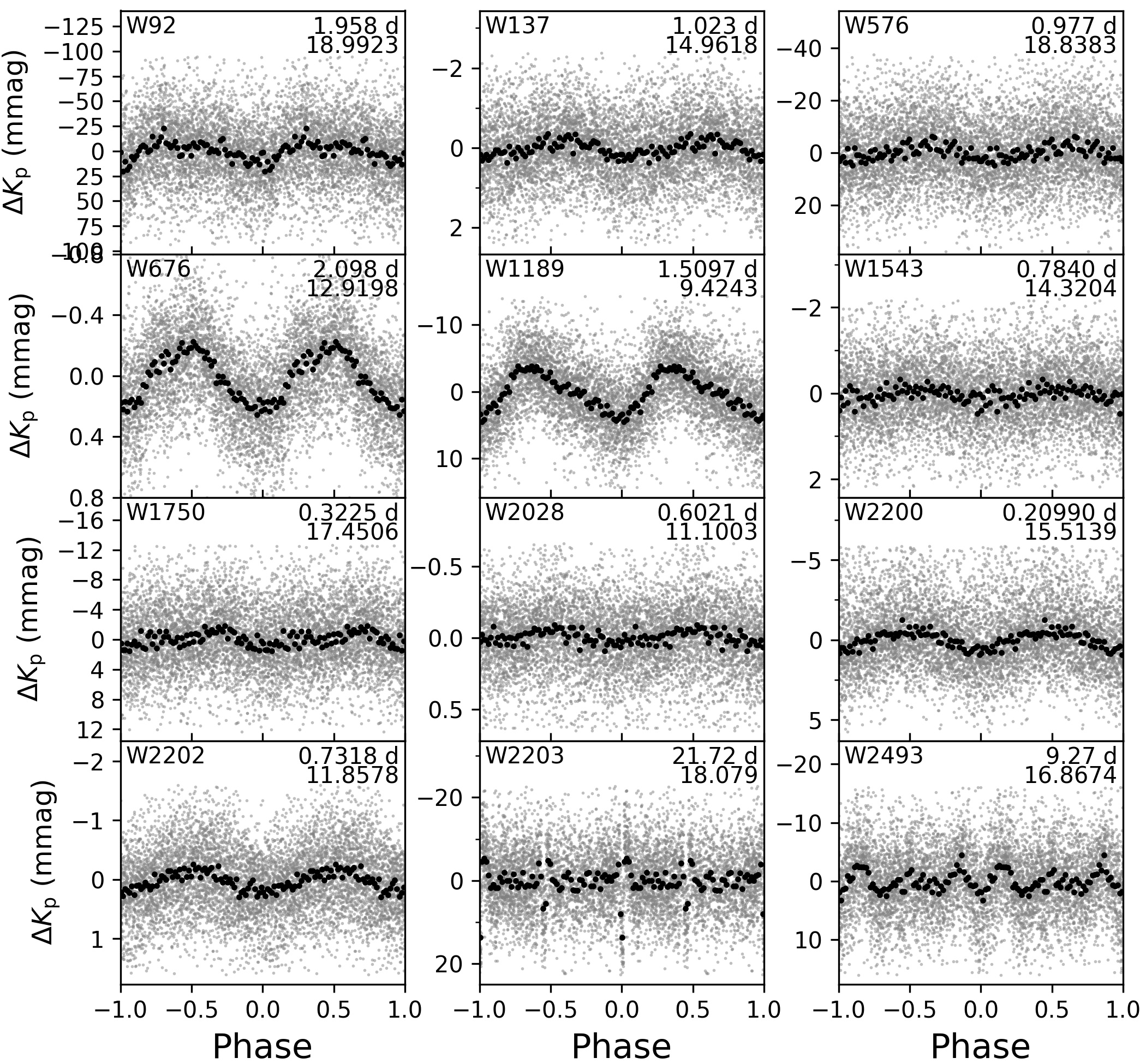}
\end{center}
\caption{\label{fig:nonclustervars1} Same as
Figure~\ref{fig:clustervars1}, but for variables that are not cluster
members. 12 variables are shown in this figure, and
Figure~\ref{fig:nonclustervars2} shows the remaining 10.}
\end{figure*}

\begin{figure*}
\begin{center}
\includegraphics[]{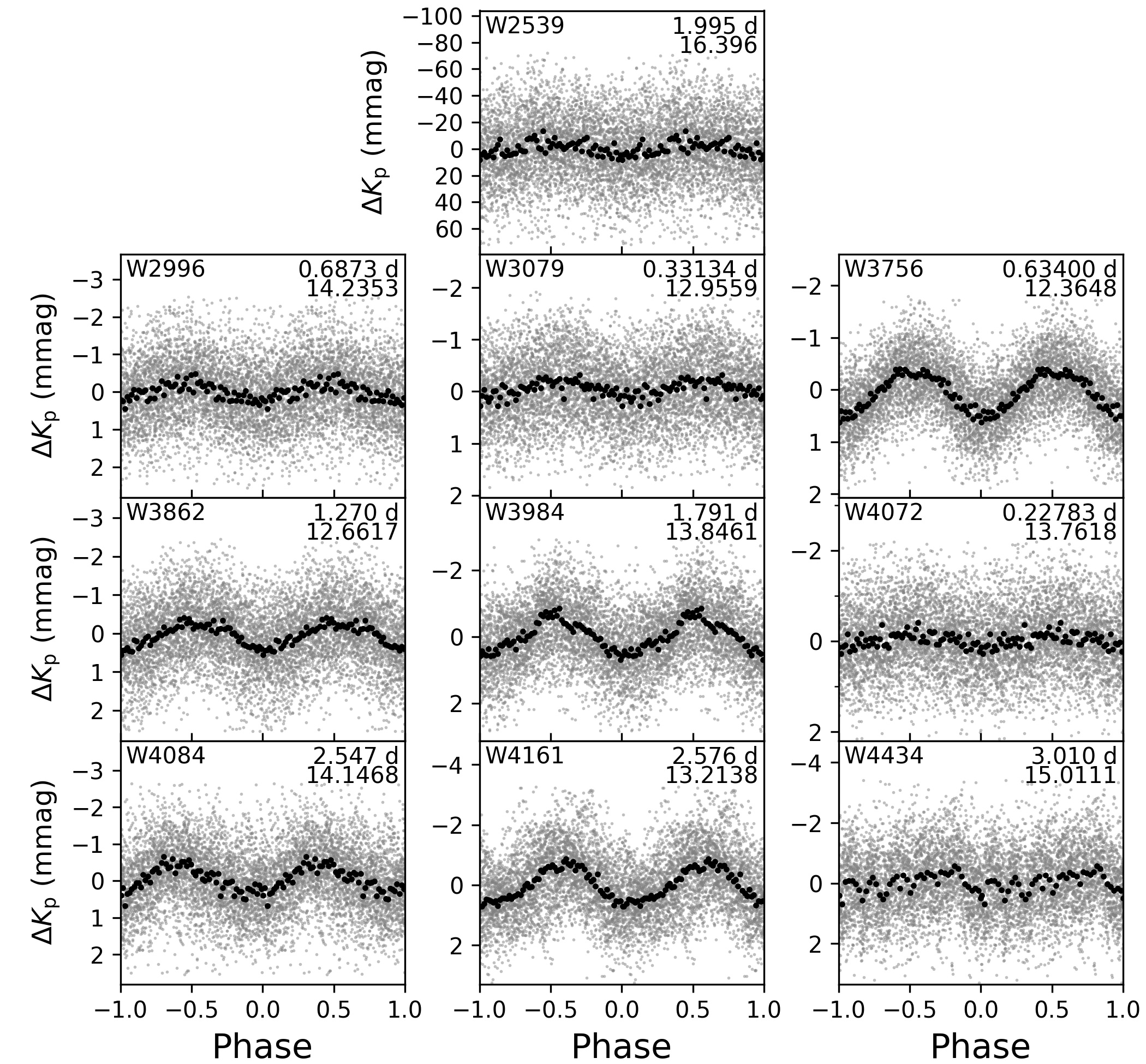}
\end{center}
\caption{\label{fig:nonclustervars2} Same as
Figure~\ref{fig:nonclustervars1}, but for additional variables that are
not cluster members.}
\end{figure*}

Included with the rich variety of cluster-member variables are many
variables that were not cluster members.
Table~\ref{tab:nonclustervars} shows information for these variable
stars, and  Figures~\ref{fig:nonclustervars1}
and \ref{fig:nonclustervars2} show the phase-folded light curves.
The suspected 
variables will be more thoroughly discussed in Appendix~\ref{suspvar}.

At ${\sim}$1.8 kpc in distance, and also being relatively close to the
Galactic center ($l{\approx}351\degr, b{\approx}16\degr$), the non-cluster-member stars
in the direction of M4 are a mixture of both foreground and background
objects.   We will touch on only two of the field variables here.

W1189 is also HD 147491 and V972 Sco of the
GCVS.  \cite{yao1989} 
reported this star as being a DSCUT variable with a ${\sim}$0.02-day
period; however, we do not see any ${\sim}$0.02-day
variability, and the 1.5097 day period we find is too
long for a DSCUT.  We think it is more likely that
this is a gamma 
Doradus variable (GDOR).  This is also the brightest star in the M4
superstamp, with {\it Gaia} DR2 $G=9.46$.

W3756 is also V1331 Sco of the GCVS.  \cite{yao2006} identified
a ${\sim}$15 mmag, 1.03-day period variability in this star based on
$V$-band observations taken in 1990 and 1991 and
classified it as a GDOR.  We see a ${\sim}$1 mmag amplitude and a
0.634-day period.  There is also power in our GLS, PDM, and BLS
periodograms for this object at a period ${\sim}$0.97 days (compare
with the
original 1.03-day period in the discovery), which
is the dominant periodogram peak when the main period and its
harmonics are removed. GDOR variability can change in amplitude and
dominant frequency over time.  This combined with the differences between the
$K_\mathrm{p}$ and $V$ bandpasses make it unsurprising for us to see
a different amplitude and dominant period relative to
the \cite{yao2006} observations, made over 23 years prior the K2
observations.

W2203 is a detached eclipsing binary with a 21.72-day period and
what appears to be reflections or other brightening events just before
and after both the primary and secondary eclipses.  The primary
eclipse depth is $0.013\pm0.001$ mag, with a fractional eclipse
duration of $0.028\pm0.003$ and a fractional ingress duration of
$0.005\pm0.002$.

Finally, we remind the reader of the blended variables in
Table~\ref{tab:blendedvars} and Figure~\ref{fig:blendedvars} that are
not cluster members: W2013 (blended with W2006) and W2813 (blended with
W2761, W2779, and W2793).


\clearpage

\startlongtable
\begin{deluxetable*}{cccccccccc}
\tablewidth{0pc}
\tablecolumns{10}
\tablecaption{Newly Discovered Variables that are Not Cluster Members\label{tab:nonclustervars}}
\tablehead{
\colhead{ID\tablenotemark{a}} & \colhead{R.A.\tablenotemark{a}} & \colhead{decl.\tablenotemark{a}} & \colhead{$G$\tablenotemark{a}} & \colhead{Period\tablenotemark{a}} & \colhead{Per. Unc.\tablenotemark{a}} & \colhead{Amp.\tablenotemark{a}} & \colhead{Epoch\tablenotemark{a}} & \colhead{Method\tablenotemark{a}} & \colhead{Type\tablenotemark{a}} \\
 \colhead{} & \colhead{(hh:mm:ss)} & \colhead{(dd:mm:ss)} & \colhead{(mag)} & \colhead{(day)} & \colhead{(10$^{-4}$ day)} & \colhead{(mmag)} & \colhead{(KBJD)} & \colhead{} & \colhead{}  }
\startdata
\cutinhead{Variables}
W92 & 16:22:58.62 & $-$26:28:59.7 & 19.09 & 1.958\tablenotemark{b} & 60 & 24 & 2061.98 & Harm. & ?\\ 
W137 & 16:23:00.83 & $-$26:25:22.2 & 15.10 & 1.023 & 20 & 0.4 & 2061.26 & Harm. & ?\\
W576 & 16:23:13.64 & $-$26:26:27.8 & 18.89 & 0.977 & 10 & 5 & 2061.13 & Harm. & ?\\
W676 & 16:23:15.54 & $-$26:27:46.3 & 13.03 & 2.098 & 10 & 0.4 & 2061.24 & Harm. & ?\\
W1189 & 16:23:22.91 & $-$26:22:16.0 & 9.46 & 1.5097 & 5 & 8 & 2060.99 & Harm. & GDOR?\tablenotemark{c}\\
W1543 & 16:23:27.47 & $-$26:23:11.9 & 14.44 & 0.7840 & 7 & 0.3 & 2060.94 & Harm. & ?\\
W1750 & 16:23:29.66 & $-$26:21:13.9 & 17.63 & 0.3225 & 2 & 2 & 2060.55 & Harm. & ?\\
W2028 & 16:23:32.46 & $-$26:26:45.0 & 11.18 & 0.6021 & 5 & 0.07 & 2060.37 & Harm. & ?\\
W2200\tablenotemark{d} & 16:23:34.14 & $-$26:25:50.2 & 15.65 & 0.20990 & 0.5 & 1 & 2060.29 & Harm. & ?\\ 
W2202 & 16:23:34.15 & $-$26:23:53.9 & 12.02 & 0.7318 & 3 & 0.4 & 2060.44 & Harm. & ?\\
W2203 & 16:23:34.17 & $-$26:21:39.3 & 18.23 & 21.72 & 100 & 13 & 2064.83 & Trap. & E/EA?\\
W2493 & 16:23:37.04 & $-$26:21:51.7 & 16.98 & 9.27 & 300 & 5 & 2069.21 & Harm. & ?\\
W2539\tablenotemark{e} & 16:23:37.58 & $-$26:29:18.4 & 17.20 & 1.995 & 40 & 8 & 2061.83 & Harm. & ?\\ 
W2996 & 16:23:42.58 & $-$26:23:43.9 & 14.42 & 0.6873 & 4 & 0.5 & 2060.76 & Harm. & ?\\
W3079 & 16:23:43.58 & $-$26:26:18.4 & 13.12 & 0.33134 & 1 & 0.2 & 2060.55 & Harm. & ?\\
W3756 & 16:23:53.21 & $-$26:22:24.5 & 12.47 & 0.63400 & 1 & 0.8 & 2060.47 & Harm. & GDOR\tablenotemark{f}\\
W3862\tablenotemark{g} & 16:23:54.98 & $-$26:26:10.8 & 12.83 & 1.270 & 30 & 0.7 & 2061.27 & Harm. & ?\\ 
W3984 & 16:23:57.56 & $-$26:23:24.5 & 14.02 & 1.791 & 10 & 1 & 2061.13 & Harm. & ?\\
W4072 & 16:23:59.18 & $-$26:23:30.8 & 13.92 & 0.22783 & 0.5 & 0.3 & 2060.47 & Harm. & mh\\
W4084 & 16:23:59.36 & $-$26:29:10.5 & 14.26 & 2.547 & 30 & 0.8 & 2062.71 & Harm. & ?\\
W4161 & 16:24:01.82 & $-$26:27:32.4 & 13.37 & 2.576 & 40 & 1 & 2061.39 & Harm. & mh\\
W4434 & 16:24:11.6 & $-$26:28:26.1 & 15.15 & 3.010 & 80 & 0.7 & 2063.07 & Harm. & ?\\
\cutinhead{Suspected Variables}
W55 & 16:22:57.19 & $-$26:29:09.1 & 18.22 & 1.85 & 100 & 7 & 2061.59 & Harm. & \textellipsis\\
W126 & 16:23:00.36 & $-$26:27:32.7 & 18.92 & 0.041004 & 0.03 & 8 & 2060.29 & Harm. & \textellipsis\\
W951 & 16:23:19.93 & $-$26:27:54.2 & 17.13 & 5.8 & 1000 & 1 & 2064.83 & Harm. & \textellipsis\\
W1779 & 16:23:29.93 & $-$26:26:53.5 & 15.88 & 4.76 & 700 & 4 & 2063.40 & Harm. & \textellipsis\tablenotemark{h}\\ 
W2571 & 16:23:37.87 & $-$26:21:57.6 & 18.77 & 2.02 & 300 & 6 & 2061.34 & Harm. & \textellipsis\\
W2588\tablenotemark{e} & 16:23:38.06 & $-$26:28:21.7 & 18.09 & 2.36 & 100 & 5 & 2060.74 & Harm. & \textellipsis\\ 
W3311\tablenotemark{i} & 16:23:46.5 & $-$26:29:08.7 & 15.18 & 15.07 & 300 & 0.8 & 2073.72 & Trap. & \textellipsis\\ 
W3717 & 16:23:52.62 & $-$26:23:21.4 & 19.22 & 0.11776 & 0.3 & 3 & 2060.35 & Harm. & \textellipsis\\
W3989 & 16:23:57.62 & $-$26:29:13.2 & 18.79 & 0.112 & \textellipsis\tablenotemark{j} & 16.69 & 2060.32 & Trap. & \textellipsis\\ 
W4337 & 16:24:07.75 & $-$26:27:37.3 & 15.45 & 0.4821 & 3 & 0.3 & 2060.72 & Harm. & \textellipsis\\
\enddata
\tablecomments{Classifications are not attempted for the suspected variables.  Explanations regarding why these are reported as suspected instead of discovered variables can be found in Appendix~\ref{suspvar}.}
\tablenotetext{a}{See table notes for Table~\ref{tab:clustervars} for details on these columns.}
\tablenotetext{b}{Period very close to a systematic period, but this object was kept as a variable owing to the strength of the signal.}
\tablenotetext{c}{Classified as DSCUT by \cite{yao1989}, but we
do not 
observe the same variability they report, and we
think a GDOR classification is more likely to be correct.}
\tablenotetext{d}{Blended with V19; this period appears in the data only after removing V19's blended signal.}
\tablenotetext{e}{Lacked proper motion data to calculate membership probability.}
\tablenotetext{f}{Classification from \cite{yao2006}.}
\tablenotetext{g}{Blended with W3825, which our code also marked as a variable; however, using a small aperture to evaluate differences in local flux amplitudes revealed this star to be the source of the variability.}
\tablenotetext{h}{Cluster membership probability is 0.067.}
\tablenotetext{i}{Eclipse was not identified by our main period-finding pipeline but was noticed in our by-eye vetting.}
\tablenotetext{j}{The trapezoid model struggled to fit well, and the calculated uncertainty on the period was unrealistically small and we decided to not report it.}
\end{deluxetable*}


\section{Discussion}
\label{sec:discussion}

To our knowledge, only two other published works (other
than \citealt{wallace2019}, which was based on the work presented in
this paper) have presented results based on the 
K2 superstamp images of M4.  \cite{miglio2016} performed
asteroseismology of K giants in M4, and \cite{kuehn2017} looked at the
RR Lyrae variables. We have already compared our results with those of
\cite{kuehn2017} in Section~\ref{sec:knownvars}, and we compare our
results with those of \cite{miglio2016} here.
\cite{miglio2016} found evidence of
solar-like oscillations in 8 stars from their chosen set of 28 (chosen
based on $B - I > 1.7$ and $V < 14$), or 29\% of the stars.  Making
comparable cuts based on {\it Gaia} DR2 magnitudes and colors, $G_\mathrm{BP}
- G_\mathrm{RP} > 1.25$ and $G < 14.0$, as well as
including only those stars that have a $>$99\% membership probability
(see Table~\ref{tab:ids} and \citealt{wallace2018}), we end up with 55 stars in
our chosen sample.  Out of those stars, we 
find asteroseismic variability in 24 of them (W491, W508, W521,
W799, W869, W1091, W1165, W1349, 
W1582, W1608, W1735, W1763, W2162, W2386, W2631, W2665,
W2678, W2772, W2887, W3033, W3073, W3480, W3742, and W3996), 
or 44\% of the stars,  plus four suspected variables
(W1068, W1717, W2577, and W3371).  Note that five of these
variables---W521, W799, W1608, W2386, W2887---are included in the
presentation of the HB variables in Section~\ref{sec:hbstars} and
Figure~\ref{fig:hbvars}. 
Restricting further to focus only on the largest giants,
selecting those stars with 
$G < 12.7$ with the same color and membership cut as
before, we end up with
18 stars in our sample, of which 8 are identified as asteroseismic
(multiharmonic) variables (W869, W1165, W1735, W1763, W2631, W2665,
W2772, and W3996), or 44\%.  It would appear
we were able to identify more asteroseismically active stars, both in
number and in percentage, than \cite{miglio2016}.  Of the eight stars
they identified,  
their S1, S6, and S7 are in our edge region so we do not have light
curves for them.  For the others, we match their S2 to our W2022, S3
to W2665, S4 
to W760, S5 to W3033, and S8 to W3929.  Our procedure detected
variability for only S3/W2665 and S5/W3033, though looking at
the periodogram results for the other three, we would have definitely
caught them
had their periodograms been presented during a manual variability
vetting.  These objects did not make it to the by-eye portion of our
variability search because they did
not have sufficiently large periodogram SNRs, probably because of the very rich
structure of the periodograms and the small differences in amplitude
between the top periodogram peak and nearby peaks.  

Other than these two papers, and our previous work in
\cite{wallace2019}, no other published work has used the M4
superstamp data.  Given that it has been publicly available for over
four years and has such rich potential, of which we believe this
work has only scratched the surface, this is
surprising. More generally, the cluster superstamps of K2 have 
received rather sparse attention, at least in terms of general
variable searches (there have been a good number of searches targeted 
at specific stars).  To our knowledge, the exhaustive list of general
variability searches among K2 cluster superstamps is: work by
\cite{lacourse2015}, \cite{libralato2016}, and \cite{soaresfurtado} 
for M35 and NGC 2185 in K2 Campaign 0;
the work of \cite{nardiello2016} for M67 in K2 Campaign 5; 
and the work of \cite{libralato2016b} for Praesepe (M44) in K2
Campaign 5. Similar, though limited, work has been done for the K2
Campaign 9 microlensing superstamp \citep[e.g.][]{zhu2017}.

The incredible results from these cluster superstamp searches speak
for themselves: \cite{libralato2016} presented a list of 2133
variables (out of 60,000 stars searched) for M35 and NGC 2158 and the
work of \cite{soaresfurtado} found 1151 variable stars from the
same data
(Soares-Furtado, private communication), 
 \cite{libralato2016b} found 1680 variable stars---of which 1071 were
 new discoveries---in M44, and \cite{nardiello2016} found 451 variable
 stars---of which 299 were new discoveries---in M67, not to mention
 the 94 variables in this work (including the two mmRRs
 of \citealt{wallace2019}), of which 76 are new, and 67 suspected 
 variables, all of which are new.  These new
 discoveries are valuable not just for better understanding the
 variable phenomena and/or the associated stars themselves, but
 with many belonging to either open or globular clusters, they can also help us
 learn more about these unique and astrophysically important
 environments.  Focusing specifically on GCs like M4, eclipsing
 binaries---sometimes referred to as the ``royal road'' to stellar
 astrophysics \citep{russell1948}---can shed important light on the
 precise masses and radii of stars belonging to a (more or less)
 monolithic, metal-poor environment.  Asteroseismic measurements can
 provide similar constraints on stellar properties for the evolved stars.  Additionally, the
 as-yet elusive detection of a transiting exoplanet in a GC (despite
 previous efforts made by
 \citealt{gilliland2000,weldrake2005,weldrake2008} and
 \citealt{nascimbeni2012}) could provide valuable clues on the
 dynamical and environmental histories of GCs. We do not attempt a focused transiting exoplanet search in this work, but we do have one
 in progress.

Even more, M4 is not the only GC that has been observed by K2.  M80
was observed concurrent with M4 during Campaign 2; M9, M19, NGC 6293,
NGC 6355, and Terzan 5 were all observed during Campaign 11; and NGC
5897 was observed during Campaign 15.  Given the increased distance of
all of these clusters relative to M4, the data will be of lower
quality and more crowded, but these are still
potentially rich datasets nonetheless, for the giant stars if not for
anything else.  This untapped potential of the K2 cluster superstamps
was recognized by \cite{barentsen2018}.  Despite the crowding and the
distance, the continuous nature and high precision of the observations
make them very valuable datasets.

And finally, K2 will not be the end of such crowded, low-resolution,
continuously observed data.  The full frame images from the {\it
Transiting Exoplanet Survey 
Satellite} \citep[{\it TESS}][]{ricker2015} are
providing similar data that, by the primary mission's end, will cover
nearly the whole sky.  At approximately five times larger pixel scale
than {\it Kepler}, observations of objects of similar crowdedness to
M4 will probably be hopelessly blended, but the outskirts of such
objects as well as the cores of less compact objects will provide rich
datasets, with important discoveries for the making, if we can learn
how to deal with such crowdedness at scale.

To this end, we wish to reiterate some of the weaknesses of our
present approach.  We do this not just to provide caveats to our
present analysis but also to provide a springboard for the community
to improve upon our and others' approaches as we look to make best use
of {\it TESS}'s crowded data.
\begin{itemize}
\item Our roll-decorrelation procedure, based on that
of \cite{vanderburg2014}, does not work with large-amplitude
variables, so our analysis of all of the \cite{clement2001} variables
(see Table~\ref{tab:knownvars}) could be improved by, e.g., a
simultaneous fit of the variability signal with the roll pattern.

\item Also, our roll-decorrelation procedure fits out a B-spline with
  breakpoints set nominally every 1.5 days, which we do not add back
  in to the light curve.  This is likely to remove any long-term
  variability that may exist, and indeed in two of the cases we
  examined (V13 and SC3), our final light curves did not exhibit
 long-period variability that was apparent in the raw light curves.

\item Since our primary variability selection criterion was based on
  periodogram SNR, those objects with significant variability at a
  variety of periods may have low periodogram SNR for otherwise robust
  variability owing to extra noise included in the calculation.
   As was discussed earlier in
  this Section, we know this is a problem for at
  least three of the asteroseismic oscillators that our code did not
  mark as robustly variable (W760, W2022, and W3929) and some HB stars
  (see Section~\ref{sec:hbstars}) and we expect there are others.

\item The blend identification and removal procedure in our code can
  be improved.  One such improvement would be a more
  nuanced selection of fit for removal of signals---whether intrinsic
  or blended---for searching for additional variability.  We employed
  an 11-harmonic Fourier series fit for removal of these signals.  The
  reason we chose such a high-harmonic fit was to fit RRAB signals
  well, but in many cases having so many harmonics led to overfitting
  of the signal and introduced spurious signals of same period but
  different shape into our light curves.  Another improvement would be to include
  a more precise determination of variability period during the period
  search instead of after, since we found some cases where the
  detected period was off slightly from the true variability period,
  leaving significant signal of similar period in the residual
  due to the not-quite-correct period being used.

\item Fainter stars that were very closely blended with considerably brighter,
  large-amplitude variable stars, as a result of the image subtraction
  photometric calculation, often had light curves with 
   exceptionally high scatter.  The stars also had many light curve points
  that were unable to be calculated (e.g., image subtraction
  determined that at a variable minimum, the fainter star would have
  to have a negative flux to match the observed flux deficit and the
  calculation would thus fail).  This itself is expected as a part of
  the image subtraction.  However, because of this high scatter
  and systematically missing data, our seven-harmonic Fourier fit to
  determine the variability amplitude would often give an egregiously
  large value for the amplitude that would far exceed the amplitude
  for the variable itself.  This meant that many of our
  highest-amplitude variables, for which detection should be most
  robust, were being marked as blends.  We fixed this by requiring
  $\Delta F / F_0 < 3$ for an amplitude measure to be considered
  realistic and ignoring the amplitude otherwise.  Better ways of
  avoiding this situation could certainly be implemented, such as
  determining from the light curves and position information prior to
  the variability search which objects are likely to have these 
  hopelessly blended, extreme light curves that would produce poor results
  in an amplitude determination.

\item The harmonic fit method of amplitude determination did not
  always work  well for the eclipsing binaries with narrow eclipses.
  For BLS searches, a more robust determination of signal amplitudes
  for comparison with neighbors and blend determination
  would be eclipse depth, determined either from the BLS fit itself or
  from another model fit, e.g. a trapezoid model.  We reran our BLS
  search with eclipse depth as the amplitude determination but did not
  find any additional variables. This modification
  to \texttt{simple\_deblend} is not yet implemented in the main
  branch, which is why we mention it here.

\item While our selection of which aperture to use for an object of a
  given magnitude was based on a superstamp-wide evaluation of light
  curve scatter versus magnitude, it may be that in the more crowded
  regions, smaller-than-globally-expected apertures produce less
  scatter. A more robust determination of this could be useful.

\item We do not treat saturated stars in any special way.

\item Our variability search produced 1310 objects (out of 4554
  searched) with purported robust variability.  Our by-eye selection
  and manual blend determination reduced this to 161. Relying so
  heavily on a manual and qualitative final vetting step is less than
  ideal and likely to lead to incorrect determinations in some of the
  marginal cases.  Reducing the amount of manual work involved in
  variable identification and classification is, of course, a
  long-standing problem in variable astronomy, and much headway is
  being made.  Specific for these data, it is likely that additional
  quantitative quality cuts could be determined to further pare down
  the number of objects that need to be searched by eye.

\item We only examined objects with a {\it Gaia} DR1 $G<19$.  While in the
  crowded regions all fainter objects were essentially included since
  the apertures for the included sources overlapped and covered the
  whole image, many stars of potential interest in the less crowded
  regions of the images were not included.  Since we discovered
  variables all the way down to the $G=19$ cut we made (see, e.g., the 
  blended pair W283 and W293 in Table~\ref{tab:blendedvars}), there
  may very well be other variables, both cluster members and
  nonmembers, to be discovered in this fainter population.

\end{itemize}

As mentioned in Section~\ref{sec:intro}, this work is intended primarily as a work of
breadth rather than depth.  The light curve processing and
results are presented, but analysis of the individual variable objects
is limited to only a very few of them, and the analysis is very
limited at that.  There is much that could be done with these data,
and since our light curves are publicly available at \cite{lightcurves}, we invite any and
all interested in these objects to perform their own analyses in
further depth.  Some potential jumping off points include: detailed
analysis of the RR Lyrae variables and further comparison with
\cite{kuehn2017}; detailed analysis of the asteroseismically active
giants and comparison with \cite{miglio2016}; further analysis of the
asteroseismically active HB stars and their connections with what we
have called mmRRs, and what connection (if any) these may have with
the RR Lyrae variables; cross-matching our identified
non-cluster-member variables with available photometric catalogs to
see if their variability could be classified; searching for
long-period variables via a different light curve processing pipeline;
observational follow up on our blended objects
(Table~\ref{tab:blendedvars}) to determine which are the actual
sources of variability; radial velocity follow up of the eclipsing
binaries; follow up, perhaps with an X-ray 
telescope, of our likely X-ray binary; and spectroscopic follow up and
characterization of all new variables presented in this work.  These
light curves represent 
the longest continuously observed GC with reduced data and, as such,
have a myriad of potential uses.

This work and the others mentioned here that worked on the K2 open
clusters demonstrates the efficiency of superstamp-style observations
of crowded regions.  For the 40,000 pixels of the K2 superstamp, we
derived light curves for 4554 objects, or ${\sim}$8.8 pixels per
object.  This is not including the objects in our edge region for
which one could still extract light curves.  To be comparably
efficient, the stamp size for observing isolated targets would have to
be ${\sim}$3 pixels by ${\sim}$3 pixels. This demonstrates how, for missions
with limited data downlink bandwidth, observations of crowded regions can be an
efficient way to maximize stars observed per pixel of data, with
the tradeoff of blending.


\section{Conclusion}
\label{sec:conclusion}
We extracted light curves for 4554 objects in the GC M4 from the K2
superstamp data of the cluster.  With ${\sim}$78 days of continuous
observations represented in the final light curves these are, by far,
the longest continuous light curves ever reduced for a GC, and
monitored at the high precision that {\it Kepler}/K2 provides.  We
employ image subtraction to extract our raw light curves, then clean
up the data using a roll-decorrelation procedure based on that of
\cite{vanderburg2014} and removing common trends in the data using
TFA.  Our final photometric precision is 0.2 mmag for $G{\approx}12$, 1
mmag for $G{\approx}15$, and  
10 mmag for $G{\approx}18$ objects, with M4's main sequence turnoff being
around $G{\approx}16$--17.  We make these light curves publicly
available \citep{lightcurves}.

We also searched for periodic variability in our light curves using
the GLS, PDM, and BLS algorithms.  We find 66 variables and 57
suspected variables that are cluster members, 24 variables and 10
suspected variables that are not cluster members, and four where cluster
membership is ambiguous.  
Of these, 52 cluster members (when including the two mmRRs
of \citealt{wallace2019}) and 20 cluster
non-members, as well as all four of the 
variables with ambiguous membership
and all the the 67 suspected variables, are new
discoveries. 
Our number of newly discovered cluster-member variables is 
three times greater than the total number of cluster-member variables
discovered in 
this area of the sky (K2 superstamp minus the edge region) 
in all previous surveys. Of
note among cluster members are seven asteroseismically variable HB stars,
a slightly eccentric 
${\sim}$4.6-day eclipsing 
binary cluster member, a ${\sim}$0.20-day EW binary, a likely X-ray
binary with quiescent periodic optical variability, and a
${\sim}$0.27-day EW binary that is highly likely to be a 
cluster member.  Among non-cluster members, we discover a slightly
eccentric 
${\sim}22$-day eclipsing binary with apparent reflection effects just
before and after transits.

This is just the starting point for the analysis of many of these
objects.  \cite{miglio2016} performed an asteroseismic analysis for
two of the asteroseismically active giants we identified, but there
remain over 20 from this work
to be analyzed, and more to be identified.  The asteroseismic
variability of the HB stars in 
particular are of interest in understanding the mmRRs first presented
in \cite{wallace2019}, and none of the seven variable non-RR-Lyrae HB
stars (see Figure~\ref{fig:hbvars} and Section~\ref{sec:hbstars}) have
received an asteroseismic analysis.  Additional analysis is needed to understand
the large number of unclassified variables we present in this work,
both in and out of the cluster.  The results of this work are the
longest continuously observed light curves ever derived for general GC
stars, and we anticipate much to come from the data.


\acknowledgements

We thank Kento Masuda and Andrew Vanderburg for very helpful
conversations, and John Hoffman and Melinda Soares-Furtado for 
coding assistance.
This paper includes data collected by the K2 mission. Funding for the
K2 mission is provided by the NASA Science Mission directorate. 
Some of the data presented in this paper were obtained from the
Mikulski Archive for Space Telescopes (MAST). STScI is operated by the
Association of Universities for Research in Astronomy, Inc., under
NASA contract NAS5-26555. Support for MAST for non-HST data is
provided by the NASA Office of Space Science via grant NNX09AF08G and
by other grants and contracts. 
This paper has made use of data from the European Space Agency (ESA)
mission {\it Gaia} (\url{https://www.cosmos.esa.int/gaia}), processed by
the {\it Gaia} Data Processing and Analysis Consortium (DPAC,
\url{https://www.cosmos.esa.int/web/gaia/dpac/consortium}). Funding
for the DPAC has been provided by national institutions, in particular
the institutions participating in the {\it Gaia} Multilateral Agreement.
This research has made use of NASA's Astrophysics Data System
Bibliographic Services.
This research has made use of the SIMBAD and VizieR databases, as well
as Aladin, operated at CDS, Strasbourg, France.

\facilities{{\it Gaia}, {\it Kepler}}

\software{\texttt{astrobase} \citep{astrobase}, 
  \texttt{astropy} \citep{astropy}, 
  \texttt{FITSH} \citep{pal}, 
  \texttt{k2mosaic} \citep{barentsen2016},
  \texttt{matplotlib} \citep{hunter2007},
  \texttt{numpy}  \citep{numpy},
  \texttt{scikit-learn} \citep{scikit-learn},
  \texttt{simple\_deblend} \citep{simpledeblend},
  \texttt{scipy} \citep{scipy}, 
  \texttt{VARTOOLS} \citep{vartools}.
} 

\bibliographystyle{aasjournal}
\bibliography{bibliography}

\begin{thebibliography}{}
\expandafter\ifx\csname natexlab\endcsname\relax\def\natexlab#1{#1}\fi
\providecommand{\url}[1]{\href{#1}{#1}}
\providecommand{\dodoi}[1]{doi:~\href{http://doi.org/#1}{\nolinkurl{#1}}}
\providecommand{\doeprint}[1]{\href{http://ascl.net/#1}{\nolinkurl{http://ascl%
.net/#1}}}
\providecommand{\doarXiv}[1]{\href{https://arxiv.org/abs/#1}{\nolinkurl{https:%
//arxiv.org/abs/#1}}}

\bibitem[{{Alard} \& {Lupton}(1998)}]{imagesubtraction}
{Alard}, C., \& {Lupton}, R.~H. 1998, \apj, 503, 325, \dodoi{10.1086/305984}

\bibitem[{{Astropy Collaboration} {et~al.}(2018){Astropy Collaboration},
  {Price-Whelan}, {Sip{\H o}cz}, {G{\"u}nther}, {Lim}, {Crawford}, {Conseil},
  {Shupe}, {Craig}, {Dencheva}, {Ginsburg}, {VanderPlas}, {Bradley},
  {P{\'e}rez-Su{\'a}rez}, {de Val-Borro}, {Aldcroft}, {Cruz}, {Robitaille},
  {Tollerud}, {Ardelean}, {Babej}, {Bach}, {Bachetti}, {Bakanov}, {Bamford},
  {Barentsen}, {Barmby}, {Baumbach}, {Berry}, {Biscani}, {Boquien}, {Bostroem},
  {Bouma}, {Brammer}, {Bray}, {Breytenbach}, {Buddelmeijer}, {Burke},
  {Calderone}, {Cano Rodr{\'{\i}}guez}, {Cara}, {Cardoso}, {Cheedella},
  {Copin}, {Corrales}, {Crichton}, {D'Avella}, {Deil}, {Depagne}, {Dietrich},
  {Donath}, {Droettboom}, {Earl}, {Erben}, {Fabbro}, {Ferreira}, {Finethy},
  {Fox}, {Garrison}, {Gibbons}, {Goldstein}, {Gommers}, {Greco}, {Greenfield},
  {Groener}, {Grollier}, {Hagen}, {Hirst}, {Homeier}, {Horton}, {Hosseinzadeh},
  {Hu}, {Hunkeler}, {Ivezi{\'c}}, {Jain}, {Jenness}, {Kanarek}, {Kendrew},
  {Kern}, {Kerzendorf}, {Khvalko}, {King}, {Kirkby}, {Kulkarni}, {Kumar},
  {Lee}, {Lenz}, {Littlefair}, {Ma}, {Macleod}, {Mastropietro}, {McCully},
  {Montagnac}, {Morris}, {Mueller}, {Mumford}, {Muna}, {Murphy}, {Nelson},
  {Nguyen}, {Ninan}, {N{\"o}the}, {Ogaz}, {Oh}, {Parejko}, {Parley}, {Pascual},
  {Patil}, {Patil}, {Plunkett}, {Prochaska}, {Rastogi}, {Reddy Janga},
  {Sabater}, {Sakurikar}, {Seifert}, {Sherbert}, {Sherwood-Taylor}, {Shih},
  {Sick}, {Silbiger}, {Singanamalla}, {Singer}, {Sladen}, {Sooley},
  {Sornarajah}, {Streicher}, {Teuben}, {Thomas}, {Tremblay}, {Turner},
  {Terr{\'o}n}, {van Kerkwijk}, {de la Vega}, {Watkins}, {Weaver}, {Whitmore},
  {Woillez}, {Zabalza}, \& {Astropy Contributors}}]{astropy}
{Astropy Collaboration}, {Price-Whelan}, A.~M., {Sip{\H o}cz}, B.~M., {et~al.}
  2018, \aj, 156, 123, \dodoi{10.3847/1538-3881/aabc4f}

\bibitem[{Barentsen(2016)}]{barentsen2016}
Barentsen, G. 2016, barentsen/k2mosaic: v2.0.0,  Zenodo,
  \dodoi{10.5281/zenodo.167343}.
\newblock \url{https://zenodo.org/record/167343#.W9nI15y1thE}

\bibitem[{{Barentsen} {et~al.}(2018){Barentsen}, {Hedges}, {Saunders}, {Cody},
  {Gully-Santiago}, {Bryson}, \& {Dotson}}]{barentsen2018}
{Barentsen}, G., {Hedges}, C., {Saunders}, N., {et~al.} 2018, arXiv e-prints,
  arXiv:1810.12554.
\newblock \doarXiv{1810.12554}

\bibitem[{{Bassa} {et~al.}(2004){Bassa}, {Pooley}, {Homer}, {Verbunt},
  {Gaensler}, {Lewin}, {Anderson}, {Margon}, {Kaspi}, \& {van der
  Klis}}]{bassa2004}
{Bassa}, C., {Pooley}, D., {Homer}, L., {et~al.} 2004, \apj, 609, 755,
  \dodoi{10.1086/421259}

\bibitem[{{Bedin} {et~al.}(2009){Bedin}, {Salaris}, {Piotto}, {Anderson},
  {King}, \& {Cassisi}}]{bedin2009}
{Bedin}, L.~R., {Salaris}, M., {Piotto}, G., {et~al.} 2009, \apj, 697, 965,
  \dodoi{10.1088/0004-637X/697/2/965}

\bibitem[{Bhatti {et~al.}(2017)Bhatti, Bouma, \& Wallace}]{astrobase}
Bhatti, W., Bouma, L.~G., \& Wallace, J. 2017, \texttt{astrobase},  Zenodo,
  \dodoi{10.5281/zenodo.1011188}.
\newblock \url{https://doi.org/10.5281/zenodo.1011188}

\bibitem[{Braga {et~al.}(2015)Braga, Dall'Ora, Bono, Stetson, Ferraro,
  Iannicola, Marengo, Neeley, Persson, Buonanno, {et~al.}}]{braga2015}
Braga, V., Dall'Ora, M., Bono, G., {et~al.} 2015, \apj, 799, 165

\bibitem[{{Clement} {et~al.}(2001){Clement}, {Muzzin}, {Dufton}, {Ponnampalam},
  {Wang}, {Burford}, {Richardson}, {Rosebery}, {Rowe}, \& {Hogg}}]{clement2001}
{Clement}, C.~M., {Muzzin}, A., {Dufton}, Q., {et~al.} 2001, \aj, 122, 2587,
  \dodoi{10.1086/323719}

\bibitem[{{Clementini} {et~al.}(1994){Clementini}, {Merighi}, {Pasquini},
  {Cacciari}, \& {Gouiffes}}]{clementini1994}
{Clementini}, G., {Merighi}, R., {Pasquini}, L., {Cacciari}, C., \& {Gouiffes},
  C. 1994, \mnras, 267, 83, \dodoi{10.1093/mnras/267.1.83}

\bibitem[{{Eggen}(1972)}]{eggen1972}
{Eggen}, O.~J. 1972, \apj, 172, 639, \dodoi{10.1086/151383}

\bibitem[{{Evans} {et~al.}(2018){Evans}, {Riello}, {De Angeli}, {Carrasco},
  {Montegriffo}, {Fabricius}, {Jordi}, {Palaversa}, {Diener}, \&
  {Busso}}]{evans2018}
{Evans}, D.~W., {Riello}, M., {De Angeli}, F., {et~al.} 2018, \aap, 616, A4,
  \dodoi{10.1051/0004-6361/201832756}

\bibitem[{{Gaia Collaboration} {et~al.}(2016{\natexlab{a}}){Gaia
  Collaboration}, {Brown}, {Vallenari}, {Prusti}, {de Bruijne}, {Mignard},
  {Drimmel}, {Babusiaux}, {Bailer-Jones}, {Bastian}, \& et~al.}]{gaiadr1}
{Gaia Collaboration}, {Brown}, A.~G.~A., {Vallenari}, A., {et~al.}
  2016{\natexlab{a}}, \aap, 595, A2, \dodoi{10.1051/0004-6361/201629512}

\bibitem[{{Gaia Collaboration} {et~al.}(2016{\natexlab{b}}){Gaia
  Collaboration}, {Prusti}, {de Bruijne}, {Brown}, {Vallenari}, {Babusiaux},
  {Bailer-Jones}, {Bastian}, {Biermann}, {Evans}, \& et~al.}]{gaiamission}
{Gaia Collaboration}, {Prusti}, T., {de Bruijne}, J.~H.~J., {et~al.}
  2016{\natexlab{b}}, \aap, 595, A1, \dodoi{10.1051/0004-6361/201629272}

\bibitem[{{Gaia Collaboration} {et~al.}(2018){Gaia Collaboration}, {Brown},
  {Vallenari}, {Prusti}, {de Bruijne}, {Babusiaux}, {Bailer-Jones}, {Biermann},
  {Evans}, \& {Eyer}}]{gaiadr2}
{Gaia Collaboration}, {Brown}, A.~G.~A., {Vallenari}, A., {et~al.} 2018, \aap,
  616, A1, \dodoi{10.1051/0004-6361/201833051}

\bibitem[{{Gilliland} {et~al.}(2000){Gilliland}, {Brown}, {Guhathakurta},
  {Sarajedini}, {Milone}, {Albrow}, {Baliber}, {Bruntt}, {Burrows},
  {Charbonneau}, {Choi}, {Cochran}, {Edmonds}, {Frandsen}, {Howell}, {Lin},
  {Marcy}, {Mayor}, {Naef}, {Sigurdsson}, {Stagg}, {Vandenberg}, {Vogt}, \&
  {Williams}}]{gilliland2000}
{Gilliland}, R.~L., {Brown}, T.~M., {Guhathakurta}, P., {et~al.} 2000, \apjl,
  545, L47, \dodoi{10.1086/317334}

\bibitem[{Harris(1996)}]{harriscatalog}
Harris, W.~E. 1996, \aj, 112, 1487

\bibitem[{{Hartman} \& {Bakos}(2016)}]{vartools}
{Hartman}, J.~D., \& {Bakos}, G.~{\'A}. 2016, Astronomy and Computing, 17, 1,
  \dodoi{10.1016/j.ascom.2016.05.006}

\bibitem[{{Howell} {et~al.}(2014){Howell}, {Sobeck}, {Haas}, {Still},
  {Barclay}, {Mullally}, {Troeltzsch}, {Aigrain}, {Bryson}, \&
  {Caldwell}}]{howell2014}
{Howell}, S.~B., {Sobeck}, C., {Haas}, M., {et~al.} 2014, \pasp, 126, 398,
  \dodoi{10.1086/676406}

\bibitem[{Hunter(2007)}]{hunter2007}
Hunter, J.~D. 2007, Computing In Science \& Engineering, 9, 90

\bibitem[{Jones {et~al.}(2001)Jones, Oliphant, Peterson, {et~al.}}]{scipy}
Jones, E., Oliphant, T., Peterson, P., {et~al.} 2001, {SciPy}: Open source
  scientific tools for {Python}.
\newblock \url{http://www.scipy.org/}

\bibitem[{{Kaluzny} {et~al.}(1997){Kaluzny}, {Thompson}, \&
  {Krzeminski}}]{kaluzny1997}
{Kaluzny}, J., {Thompson}, I.~B., \& {Krzeminski}, W. 1997, \aj, 113, 2219,
  \dodoi{10.1086/118432}

\bibitem[{{Kaluzny} {et~al.}(2013{\natexlab{a}}){Kaluzny}, {Thompson},
  {Rozyczka}, \& {Krzeminski}}]{kaluzny2013a}
{Kaluzny}, J., {Thompson}, I.~B., {Rozyczka}, M., \& {Krzeminski}, W.
  2013{\natexlab{a}}, \actaa, 63, 181.
\newblock \doarXiv{1306.2457}

\bibitem[{{Kaluzny} {et~al.}(2013{\natexlab{b}}){Kaluzny}, {Thompson},
  {Rozyczka}, {Dotter}, {Krzeminski}, {Pych}, {Rucinski}, {Burley}, \&
  {Shectman}}]{kaluzny2013b}
{Kaluzny}, J., {Thompson}, I.~B., {Rozyczka}, M., {et~al.} 2013{\natexlab{b}},
  \aj, 145, 43, \dodoi{10.1088/0004-6256/145/2/43}

\bibitem[{{Kov{\'a}cs} {et~al.}(2005){Kov{\'a}cs}, {Bakos}, \&
  {Noyes}}]{kovacs-tfa}
{Kov{\'a}cs}, G., {Bakos}, G., \& {Noyes}, R.~W. 2005, \mnras, 356, 557,
  \dodoi{10.1111/j.1365-2966.2004.08479.x}

\bibitem[{{Kov{\'a}cs} {et~al.}(2002){Kov{\'a}cs}, {Zucker}, \&
  {Mazeh}}]{kovacs2002}
{Kov{\'a}cs}, G., {Zucker}, S., \& {Mazeh}, T. 2002, \aap, 391, 369,
  \dodoi{10.1051/0004-6361:20020802}

\bibitem[{{Kuehn} {et~al.}(2017){Kuehn}, {Moskalik}, \& {Drury}}]{kuehn2017}
{Kuehn}, C.~A., {Moskalik}, P., \& {Drury}, J.~A. 2017, in Seismology of the
  Sun and the Distant Stars - Using Today's Successes to Prepare the Future -
  TASC2 \& KASC9 Workshop - SPACEINN \& HELAS8 Conference, Azores Islands,
  Portugal, Edited by Monteiro, M.J.P.F.G.; Cunha, M.S.; Ferreira, J.M.T.S.;
  EPJ Web of Conferences, Volume 160, id.04011, Vol. 160, 04011

\bibitem[{{LaCourse} {et~al.}(2015){LaCourse}, {Jek}, {Jacobs}, {Winarski},
  {Boyajian}, {Rappaport}, {Sanchis-Ojeda}, {Conroy}, {Nelson}, \&
  {Barclay}}]{lacourse2015}
{LaCourse}, D.~M., {Jek}, K.~J., {Jacobs}, T.~L., {et~al.} 2015, \mnras, 452,
  3561, \dodoi{10.1093/mnras/stv1475}

\bibitem[{{Leavitt} \& {Pickering}(1904)}]{leavitt1904}
{Leavitt}, H.~S., \& {Pickering}, E.~C. 1904, Harvard College Observatory
  Circular, 90, 1

\bibitem[{{Libralato} {et~al.}(2016{\natexlab{a}}){Libralato}, {Bedin},
  {Nardiello}, \& {Piotto}}]{libralato2016}
{Libralato}, M., {Bedin}, L.~R., {Nardiello}, D., \& {Piotto}, G.
  2016{\natexlab{a}}, \mnras, 456, 1137, \dodoi{10.1093/mnras/stv2628}

\bibitem[{{Libralato} {et~al.}(2016{\natexlab{b}}){Libralato}, {Nardiello},
  {Bedin}, {Borsato}, {Granata}, {Malavolta}, {Piotto}, {Ochner}, {Cunial}, \&
  {Nascimbeni}}]{libralato2016b}
{Libralato}, M., {Nardiello}, D., {Bedin}, L.~R., {et~al.} 2016{\natexlab{b}},
  \mnras, 463, 1780, \dodoi{10.1093/mnras/stw1932}

\bibitem[{{Lindegren} {et~al.}(2016){Lindegren}, {Lammers}, {Bastian},
  {Hern{\'a}ndez}, {Klioner}, {Hobbs}, {Bombrun}, {Michalik}, {Ramos-Lerate},
  {Butkevich}, {Comoretto}, {Joliet}, {Holl}, {Hutton}, {Parsons},
  {Steidelm{\"u}ller}, {Abbas}, {Altmann}, {Andrei}, {Anton}, {Bach},
  {Barache}, {Becciani}, {Berthier}, {Bianchi}, {Biermann}, {Bouquillon},
  {Bourda}, {Br{\"u}semeister}, {Bucciarelli}, {Busonero}, {Carlucci},
  {Casta{\~n}eda}, {Charlot}, {Clotet}, {Crosta}, {Davidson}, {de Felice},
  {Drimmel}, {Fabricius}, {Fienga}, {Figueras}, {Fraile}, {Gai}, {Garralda},
  {Geyer}, {Gonz{\'a}lez-Vidal}, {Guerra}, {Hambly}, {Hauser}, {Jordan},
  {Lattanzi}, {Lenhardt}, {Liao}, {L{\"o}ffler}, {McMillan}, {Mignard}, {Mora},
  {Morbidelli}, {Portell}, {Riva}, {Sarasso}, {Serraller}, {Siddiqui}, {Smart},
  {Spagna}, {Stampa}, {Steele}, {Taris}, {Torra}, {van Reeven}, {Vecchiato},
  {Zschocke}, {de Bruijne}, {Gracia}, {Raison}, {Lister}, {Marchant},
  {Messineo}, {Soffel}, {Osorio}, {de Torres}, \& {O'Mullane}}]{gaiaastrometry}
{Lindegren}, L., {Lammers}, U., {Bastian}, U., {et~al.} 2016, \aap, 595, A4,
  \dodoi{10.1051/0004-6361/201628714}

\bibitem[{{Lindegren} {et~al.}(2018){Lindegren}, {Hern{\'a}ndez}, {Bombrun},
  {Klioner}, {Bastian}, {Ramos-Lerate}, {de Torres}, {Steidelm{\"u}ller},
  {Stephenson}, \& {Hobbs}}]{lindegren2018}
{Lindegren}, L., {Hern{\'a}ndez}, J., {Bombrun}, A., {et~al.} 2018, \aap, 616,
  A2, \dodoi{10.1051/0004-6361/201832727}

\bibitem[{{Lomb}(1976)}]{lomb1976}
{Lomb}, N.~R. 1976, \apss, 39, 447, \dodoi{10.1007/BF00648343}

\bibitem[{{Masuda} \& {Hotokezaka}(2018)}]{masuda2018}
{Masuda}, K., \& {Hotokezaka}, K. 2018, arXiv e-prints, arXiv:1808.10856.
\newblock \doarXiv{1808.10856}

\bibitem[{{Miglio} {et~al.}(2016){Miglio}, {Chaplin}, {Brogaard}, {Lund},
  {Mosser}, {Davies}, {Handberg}, {Milone}, {Marino}, \&
  {Bossini}}]{miglio2016}
{Miglio}, A., {Chaplin}, W.~J., {Brogaard}, K., {et~al.} 2016, \mnras, 461,
  760, \dodoi{10.1093/mnras/stw1555}

\bibitem[{{Nardiello} {et~al.}(2016){Nardiello}, {Libralato}, {Bedin},
  {Piotto}, {Borsato}, {Granata}, {Malavolta}, \& {Nascimbeni}}]{nardiello2016}
{Nardiello}, D., {Libralato}, M., {Bedin}, L.~R., {et~al.} 2016, \mnras, 463,
  1831, \dodoi{10.1093/mnras/stw2169}

\bibitem[{{Nascimbeni} {et~al.}(2012){Nascimbeni}, {Bedin}, {Piotto}, {De
  Marchi}, \& {Rich}}]{nascimbeni2012}
{Nascimbeni}, V., {Bedin}, L.~R., {Piotto}, G., {De Marchi}, F., \& {Rich},
  R.~M. 2012, \aap, 541, A144, \dodoi{10.1051/0004-6361/201118655}

\bibitem[{Neeley {et~al.}(2015)Neeley, Marengo, Bono, Braga, Dall’Ora,
  Stetson, Buonanno, Ferraro, Freedman, Iannicola, {et~al.}}]{neeley2015}
Neeley, J., Marengo, M., Bono, G., {et~al.} 2015, The Astrophysical Journal,
  808, 11

\bibitem[{Oliphant(2006)}]{numpy}
Oliphant, T. 2006, Guide to {NumPy} (Trelgol Publishing).
\newblock \url{http://www.tramy.us/numpybook.pdf}

\bibitem[{P\'al(2012)}]{pal}
P\'al, A. 2012, \mnras, 421, 1825, \dodoi{10.1111/j.1365-2966.2011.19813.x}

\bibitem[{Pedregosa {et~al.}(2011)Pedregosa, Varoquaux, Gramfort, Michel,
  Thirion, Grisel, Blondel, Prettenhofer, Weiss, Dubourg, Vanderplas, Passos,
  Cournapeau, Brucher, Perrot, \& Duchesnay}]{scikit-learn}
Pedregosa, F., Varoquaux, G., Gramfort, A., {et~al.} 2011, Journal of Machine
  Learning Research, 12, 2825

\bibitem[{{Ricker} {et~al.}(2015){Ricker}, {Winn}, {Vanderspek}, {Latham},
  {Bakos}, {Bean}, {Berta-Thompson}, {Brown}, {Buchhave}, \&
  {Butler}}]{ricker2015}
{Ricker}, G.~R., {Winn}, J.~N., {Vanderspek}, R., {et~al.} 2015, Journal of
  Astronomical Telescopes, Instruments, and Systems, 1, 014003,
  \dodoi{10.1117/1.JATIS.1.1.014003}

\bibitem[{{Riello} {et~al.}(2018){Riello}, {De Angeli}, {Evans}, {Busso},
  {Hambly}, {Davidson}, {Burgess}, {Montegriffo}, {Osborne}, \&
  {Kewley}}]{riello2018}
{Riello}, M., {De Angeli}, F., {Evans}, D.~W., {et~al.} 2018, \aap, 616, A3,
  \dodoi{10.1051/0004-6361/201832712}

\bibitem[{{Russell}(1948)}]{russell1948}
{Russell}, H.~N. 1948, {The Royal Road of Eclipses}, Vol.~7 (Harvard
  Observatory Monographs), 181

\bibitem[{{Safonova} {et~al.}(2016){Safonova}, {Mkrtichian}, {Hasan},
  {Sutaria}, {Brosch}, {Gorbikov}, \& {Joseph}}]{safonova2016}
{Safonova}, M., {Mkrtichian}, D., {Hasan}, P., {et~al.} 2016, \aj, 151, 27,
  \dodoi{10.3847/0004-6256/151/2/27}

\bibitem[{{Samus} {et~al.}(2017){Samus}, {Kazarovets}, {Durlevich}, {Kireeva},
  \& {Pastukhova}}]{samus2017}
{Samus}, N.~N., {Kazarovets}, E.~V., {Durlevich}, O.~V., {Kireeva}, N.~N., \&
  {Pastukhova}, E.~N. 2017, Astronomy Reports, 61, 80,
  \dodoi{10.1134/S1063772917010085}

\bibitem[{{Scargle}(1982)}]{scargle1982}
{Scargle}, J.~D. 1982, \apj, 263, 835, \dodoi{10.1086/160554}

\bibitem[{{Soares-Furtado} {et~al.}(2017){Soares-Furtado}, {Hartman}, {Bakos},
  {Huang}, {Penev}, \& {Bhatti}}]{soaresfurtado}
{Soares-Furtado}, M., {Hartman}, J.~D., {Bakos}, G.~{\'A}., {et~al.} 2017,
  \pasp, 129, 044501, \dodoi{10.1088/1538-3873/aa5c7c}

\bibitem[{{Stellingwerf}(1978)}]{stellingwerf1978}
{Stellingwerf}, R.~F. 1978, \apj, 224, 953, \dodoi{10.1086/156444}

\bibitem[{Stetson {et~al.}(2014)Stetson, Braga, Dall'Ora, Bono, Buonanno,
  Ferraro, Iannicola, Marengo, \& Neeley}]{stetson2014}
Stetson, P.~B., Braga, V.~F., Dall'Ora, M., {et~al.} 2014, \pasp, 126, 521

\bibitem[{{van Leeuwen} {et~al.}(2017){van Leeuwen}, {Evans}, {De Angeli},
  {Jordi}, {Busso}, {Cacciari}, {Riello}, {Pancino}, {Altavilla}, {Brown},
  {Burgess}, {Carrasco}, {Cocozza}, {Cowell}, {Davidson}, {De Luise},
  {Fabricius}, {Galleti}, {Gilmore}, {Giuffrida}, {Hambly}, {Harrison},
  {Hodgkin}, {Holland}, {MacDonald}, {Marinoni}, {Montegriffo}, {Osborne},
  {Ragaini}, {Richards}, {Rowell}, {Voss}, {Walton}, {Weiler}, {Castellani},
  {Delgado}, {H{\o}g}, {van Leeuwen}, {Millar}, {Pagani}, {Piersimoni},
  {Pulone}, {Rixon}, {Suess}, {Wyrzykowski}, {Yoldas}, {Alecu}, {Allan},
  {Balaguer-N{\'u}{\~n}ez}, {Barstow}, {Bellazzini}, {Belokurov},
  {Blagorodnova}, {Bonfigli}, {Bragaglia}, {Brown}, {Bunclark}, {Buonanno},
  {Burgon}, {Campbell}, {Collins}, {Cross}, {Ducourant}, {van Elteren},
  {Evans}, {Federici}, {Fern{\'a}ndez-Hern{\'a}ndez}, {Figueras}, {Fraser},
  {Fyfe}, {Gebran}, {Heyrovsky}, {Holl}, {Holland}, {Iannicola}, {Irwin},
  {Koposov}, {Krone-Martins}, {Mann}, {Marrese}, {Masana}, {Munari}, {Ortiz},
  {Ouzounis}, {Peltzer}, {Portell}, {Read}, {Terrett}, {Torra}, {Trager},
  {Troisi}, {Valentini}, {Vallenari}, \& {Wevers}}]{gaiaphotometry}
{van Leeuwen}, F., {Evans}, D.~W., {De Angeli}, F., {et~al.} 2017, \aap, 599,
  A32, \dodoi{10.1051/0004-6361/201630064}

\bibitem[{{VandenBerg} {et~al.}(2013){VandenBerg}, {Brogaard}, {Leaman}, \&
  {Casagrande}}]{vandenberg2013}
{VandenBerg}, D.~A., {Brogaard}, K., {Leaman}, R., \& {Casagrande}, L. 2013,
  \apj, 775, 134, \dodoi{10.1088/0004-637X/775/2/134}

\bibitem[{{Vanderburg} \& {Johnson}(2014)}]{vanderburg2014}
{Vanderburg}, A., \& {Johnson}, J.~A. 2014, \pasp, 126, 948,
  \dodoi{10.1086/678764}

\bibitem[{Vanderburg {et~al.}(2016)Vanderburg, Latham, Buchhave, Bieryla,
  Berlind, Calkins, Esquerdo, Welsh, \& Johnson}]{vanderburg2016}
Vanderburg, A., Latham, D.~W., Buchhave, L.~A., {et~al.} 2016, \apjs, 222, 14

\bibitem[{{Verbunt}(2001)}]{verbunt2001}
{Verbunt}, F. 2001, \aap, 368, 137, \dodoi{10.1051/0004-6361:20000469}

\bibitem[{Wallace(2018{\natexlab{a}})}]{membershipcatalog2018}
Wallace, J. 2018{\natexlab{a}}, \texttt{M4\_pm\_membership: Version 1.0},
  Zenodo, \dodoi{10.5281/zenodo.1488302}.
\newblock \url{https://doi.org/10.5281/zenodo.1488302}

\bibitem[{Wallace \& Hoffman(2019)}]{simpledeblend}
Wallace, J., \& Hoffman, J. 2019, \texttt{simple\_deblend},  Zenodo,
  \dodoi{10.5281/zenodo.3248998}.
\newblock \url{https://doi.org/10.5281/zenodo.3248998}

\bibitem[{Wallace(2018{\natexlab{b}})}]{wallace2018}
Wallace, J.~J. 2018{\natexlab{b}}, Research Notes of the AAS, 2, 213

\bibitem[{{Wallace} {et~al.}(2019{\natexlab{a}}){Wallace}, {Hartman}, {Bakos},
  \& {Bhatti}}]{wallace2019}
{Wallace}, J.~J., {Hartman}, J.~D., {Bakos}, G.~{\'A}., \& {Bhatti}, W.
  2019{\natexlab{a}}, \apjl, 870, L7, \dodoi{10.3847/2041-8213/aaf8ac}

\bibitem[{{Wallace} {et~al.}(2019{\natexlab{b}}){Wallace}, {Hartman}, {Bakos},
  \& {Bhatti}}]{lightcurves}
---. 2019{\natexlab{b}}, Light Curves from a Search for Variable Stars in the
  Globular Cluster M4 with K2,  DataSpace at Princeton University.
\newblock \url{http://arks.princeton.edu/ark:/88435/dsp01h415pd368}

\bibitem[{{Watson} {et~al.}(2017){Watson}, {Henden}, \& {Price}}]{watson2017}
{Watson}, C., {Henden}, A.~A., \& {Price}, A. 2017, VizieR Online Data Catalog,
  1

\bibitem[{{Weldrake} {et~al.}(2008){Weldrake}, {Sackett}, \&
  {Bridges}}]{weldrake2008}
{Weldrake}, D.~T.~F., {Sackett}, P.~D., \& {Bridges}, T.~J. 2008, \apj, 674,
  1117, \dodoi{10.1086/524917}

\bibitem[{{Weldrake} {et~al.}(2005){Weldrake}, {Sackett}, {Bridges}, \&
  {Freeman}}]{weldrake2005}
{Weldrake}, D.~T.~F., {Sackett}, P.~D., {Bridges}, T.~J., \& {Freeman}, K.~C.
  2005, \apj, 620, 1043, \dodoi{10.1086/427258}

\bibitem[{{Yao} {et~al.}(2006{\natexlab{a}}){Yao}, {Sheng}, \& {Shi}}]{yao2006}
{Yao}, B.-A., {Sheng}, C.-J., \& {Shi}, H.-M. 2006{\natexlab{a}}, \apss, 302,
  241, \dodoi{10.1007/s10509-006-9040-0}

\bibitem[{{Yao} {et~al.}(2006{\natexlab{b}}){Yao}, {Sheng}, {Zhang}, {Hu}, \&
  {Lin}}]{yao2007translation}
{Yao}, B.-a., {Sheng}, C.-j., {Zhang}, C.-s., {Hu}, H.-m., \& {Lin}, Q.
  2006{\natexlab{b}}, \caa, 30, 351, \dodoi{10.1016/j.chinastron.2006.10.001}

\bibitem[{{Yao} {et~al.}(2007){Yao}, {Sheng}, {Zhang}, {Hu}, \&
  {Lin}}]{yao2007}
{Yao}, B.~A., {Sheng}, C.~J., {Zhang}, C.~S., {Hu}, H.~M., \& {Lin}, Q. 2007,
  Acta Astronomica Sinica, 48, 18

\bibitem[{{Yao} \& {Tong}(1989)}]{yao1989}
{Yao}, B.-A., \& {Tong}, J.-H. 1989, Information Bulletin on Variable Stars,
  3334, 1

\bibitem[{{Yao} {et~al.}(1988){Yao}, {Tong}, \& {Zhang}}]{yao1988}
{Yao}, B.~A., {Tong}, J.~H., \& {Zhang}, C.~S. 1988, Acta Astronomica Sinica,
  29, 243

\bibitem[{{Yao} {et~al.}(1981{\natexlab{a}}){Yao}, {Yin}, \& {Guo}}]{yao1981}
{Yao}, B.~A., {Yin}, J.~S., \& {Guo}, Z.~H. 1981{\natexlab{a}}, Acta
  Astrophysica Sinica, 1, 311

\bibitem[{{Yao} {et~al.}(1981{\natexlab{b}}){Yao}, {Yin}, \&
  {Guo}}]{yao1981translation}
{Yao}, B.-a., {Yin}, J.-s., \& {Guo}, Z.-h. 1981{\natexlab{b}}, \caa, 5, 476,
  \dodoi{10.1016/0275-1062(81)90015-1}

\bibitem[{{Zechmeister} \& {K{\"u}rster}(2009)}]{zechmeister2009}
{Zechmeister}, M., \& {K{\"u}rster}, M. 2009, \aap, 496, 577,
  \dodoi{10.1051/0004-6361:200811296}

\bibitem[{{Zhu} {et~al.}(2017){Zhu}, {Huang}, {Udalski}, {Soares-Furtado},
  {Poleski}, {Skowron}, {Mr{\'o}z}, {Szyma{\'n}ski}, {Soszy{\'n}ski}, \&
  {Pietrukowicz}}]{zhu2017}
{Zhu}, W., {Huang}, C.~X., {Udalski}, A., {et~al.} 2017, \pasp, 129, 104501,
  \dodoi{10.1088/1538-3873/aa7dd7}

\end{thebibliography}


\begin{appendix}
\setcounter{table}{0}
\renewcommand{\thetable}{A\arabic{table}}
\renewcommand{\theequation}{A\arabic{equation}}

\section{Notes on Identified Blends}
\label{sec:blends}

This Appendix provides a detailed look into blends that were manually
assessed and removed by us after the automatic processing
described in Section~\ref{sec:method}.  
This discussion is 
intended primarily as a record of the blends we manually assessed
and/or a reference for those who wish to more
completely understand the systematics in our search.

Despite the reasonably robust performance of our automated blend
detection method, there still remained many 
blends in the final set of detected periods.  Some reasons for the
residual blends include: blending with or photometric footprinting by
a variable object that was further away than our chosen search radius
of 12 pixels  or objects with
particularly small separations ending up with similar flux
amplitudes in their variability due to the amount of overlap in their
apertures.  In the latter case, there were
some objects for which we were able to disentangle which was the real
variable, while Table~\ref{tab:blendedvars} records those objects which
we were not 
able to disentangle.  Though the accounting here is fairly exhaustive, we
did not record all instances of stars that were clear blends with the
RR Lyrae variables based on proximity, period, and light curve properties.
Despite choosing the 12-pixel blend search radius based on
results in the neighborhood of two RR Lyrae variables in our images, there were
still some stars outside this radius for other RR Lyrae variables that
were blended with those variables.

Many stars had similar variability and the same period and phase as V19.
 These were all ${\sim}$12--18 pixels away
from the star and predominantly clustered together. We do not know for
 sure what caused this relatively 
distant blending. We checked all of the stars with period and phase
 that matched V19 to make sure none were obviously their own variable
 before excluding them from further consideration.  The stars thus
 excluded were W1820, W1836, W1838, W1995, W2007, W2205, W2264, W2316,
 W2381, W2413, W2420, W2439, W2467, W2540, W2583, W2600, W2626, W2695,
 W2701, W2748, W2774, W2776, and W2777.  There were also three stars
 that were 38--41 pixels away in rough relative proximity to each
 other  that were 180\degr\ out of phase with V19
 and had the same period. These  were also excluded after a visual check of
 their light curves: W1948, W1960, and W2201.

The following stars were all blended with each other and all have the
same period as V27.  They are also all ${\sim}$33-36 pixels away from
V27.  The signals look like inverted
RRAB signals, so it may be some systematic from our data reduction.
All of these were removed from
consideration: W3232, W3234, W3246, W3248, W3262, W3285, W3296, W4540.

W3623 has the same period and nearly same phase as V9 with a similar
shape, despite being 
over 80 pixels away.  We removed 
 W3623 from consideration because of this.

W285 has the same period as V35 from \cite{clement2001} and also looks
like an RRAB, which V35 is.  Thus we consider W285 as a blend with V35
even though we do not have a light curve for V35.

W2398 is blended with ${\sim}$0.47-day-period V19 and thus its
${\sim}$0.12-day variability detected by GLS is discounted by us and
we marked it as not a variable.  Closer examination may be able to
determine whether this is a correct call or not.

There were several stars in close proximity to each other
with variability of approximately the same
period as V29, but did not phase up with V29, and were also
${\gtrsim}$100 pixels away from V29.  However, V28 in the catalog
of \cite{clement2001} has nearly the same period as V29 and, while not
included in the K2 superstamp of M4, is only
${\sim}$11--15 pixels away from most of these stars (one was 27 pixels
away).  Based on this, we decided to mark the following stars as
blends with V28 given the proximity, after a visual check of their light curves: W3678, W3735,
W3796, W3811, W3848, W3854, W3880, and W3914.  Additionally, W2709 
phased up with V29 and was marked as a blend despite being ${\sim}$125
pixels away.

W1097 is hopelessly blended with the
bright variable W1165.  Looking at the respective light curve, W1097's
light curve was excessively noisy (likely due to blending with the much
brighter star) and the variability was not nearly as apparent as for
W1165.  We thus removed W1097 from consideration.

Many stars shared a similar ${\sim}$1.95-day period and phased up with
each other.  This period is approximately the same period (1.962 days)
as the resaturation events, producing a blank image at this period.
These stars were all assumed to share a common systematic based on the
resaturation events and removed from further consideration.  These were
W144, W221, W335, W338, W391, W528, W678, W2098, W2286, W2694,
W3178, W3785, and W3955.  Additionally, other stars were found with
this similar period that did not quite phase up with the others
(though some were 180\degr\ out of phase) but were still assumed to
have a similar systematic unless visual inspection of their light
curve revealed otherwise.  These objects were
W83, W610, W2040, W2309, W3306, W3779, W4062, W4083, W4096, W4177,
W4293, W4318, and W4534.  Upon visual inspection of the light curves,
W92 and W4268 were kept as a variable (W92) or suspected variable
(W4268) owing to the strength of their signal despite having periods
around this systematic.  W4490 was also kept as a variable owing to
its high-amplitude variability.

W321, W470, W548, W566, W569, W645, and W692 all had the same
variability period, phase, and shape, and were all in about the same
area of the image.  The apertures were not all quite overlapping.  Of
these, W566 had the most robust detection of the 
variability (detected by both GLS and PDM instead of just PDM, and
also had the highest periodogram SNR) and so we decided to call that the
variable but wanted to record here the other stars that were
blended with it.  All are ${\sim}$6--13 pixels away from W566.

W1938, W2805, and W4143 all have ${\sim}$3.4-day transits.  W1938 and
W2805 even phase up based on a sine curve fit to the variability.
However, these stars are all very separated.  W1938 and W4143 are
included as variables in this work, in Table~\ref{tab:clustervars}.


\section{Suspected Variables}
\label{suspvar}

\begin{figure*}
\begin{center}
\includegraphics[]{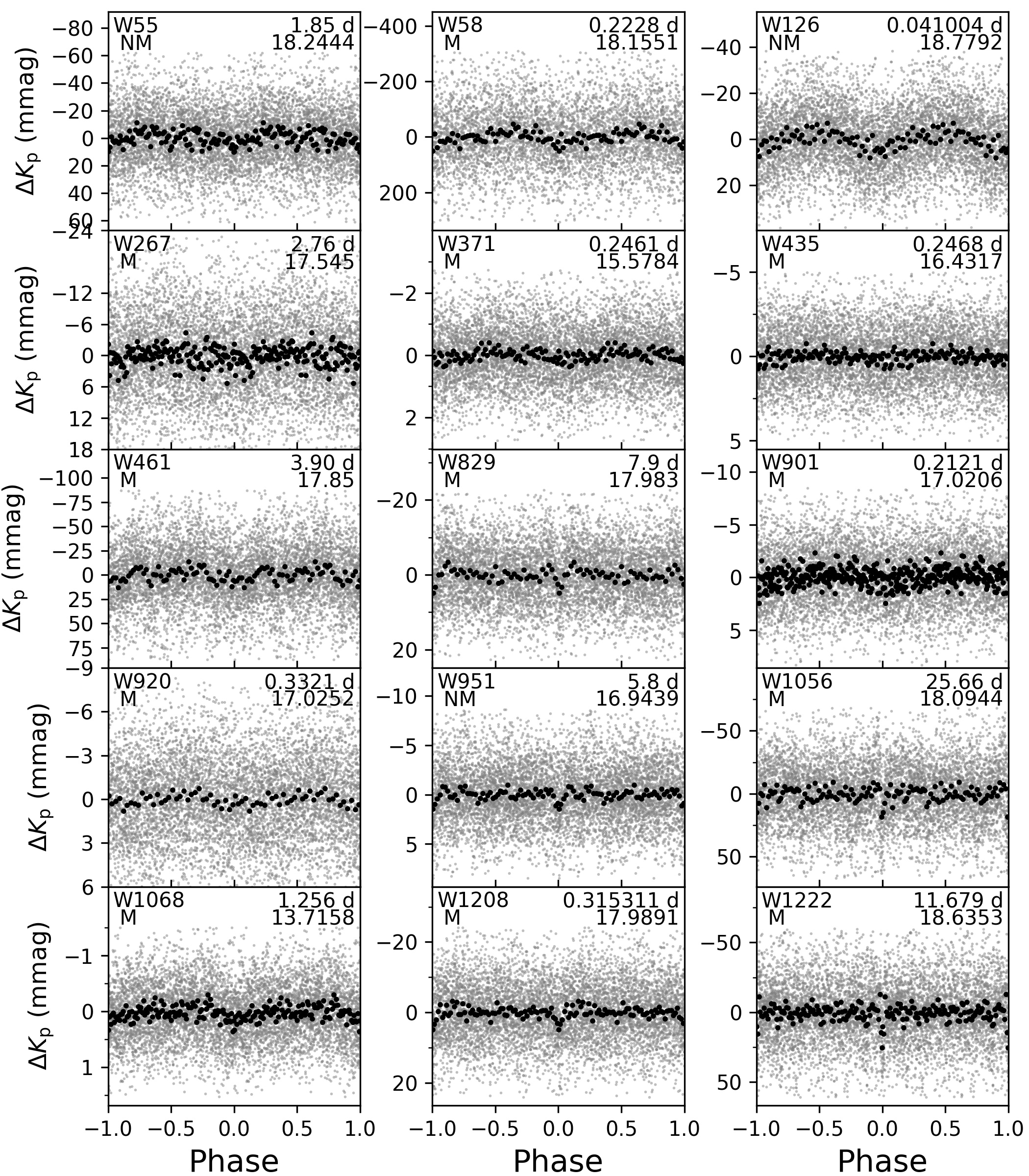}
\end{center}
\caption{\label{fig:suspected1} Same as
Figure~\ref{fig:clustervars1}, but for suspected variables and for a
mixture of cluster members and nonmembers.  The first 15 suspected
variables are shown in this figure, with the rest of the suspected
variables shown in Figures~\ref{fig:suspected2}, \ref{fig:suspected3},
\ref{fig:suspected4}, and \ref{fig:suspected5}. Cluster membership is
indicated below the object identifier in the upper right corner of
each panel: ``M'' means cluster member (specifically, that the cluster
membership probability is ${>}$99\%), while ``NM'' means not a
cluster member (specifically, that the cluster
membership probability is ${<}$1\%).}
\end{figure*}

\begin{figure*}
\begin{center}
\includegraphics[]{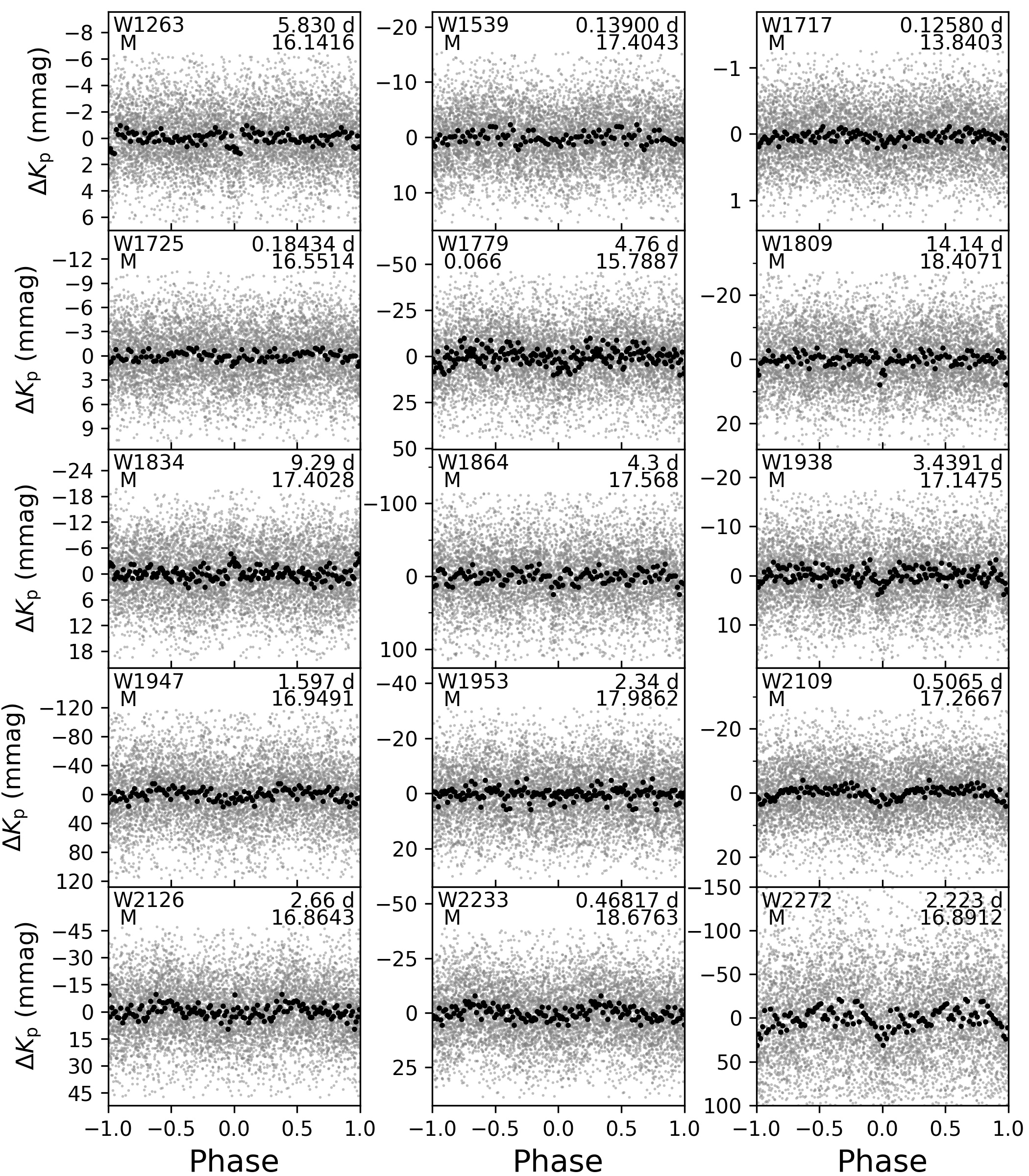}
\end{center}
\caption{\label{fig:suspected2} Same as
Figure~\ref{fig:suspected1}, but for additional suspected
variables. Instead of indicating ``M'' or ``NM'' for W1799's cluster
membership, we record the membership probability since it was not
${<}$1\% or ${>}$99\%.} 
\end{figure*}

\begin{figure*}
\begin{center}
\includegraphics[]{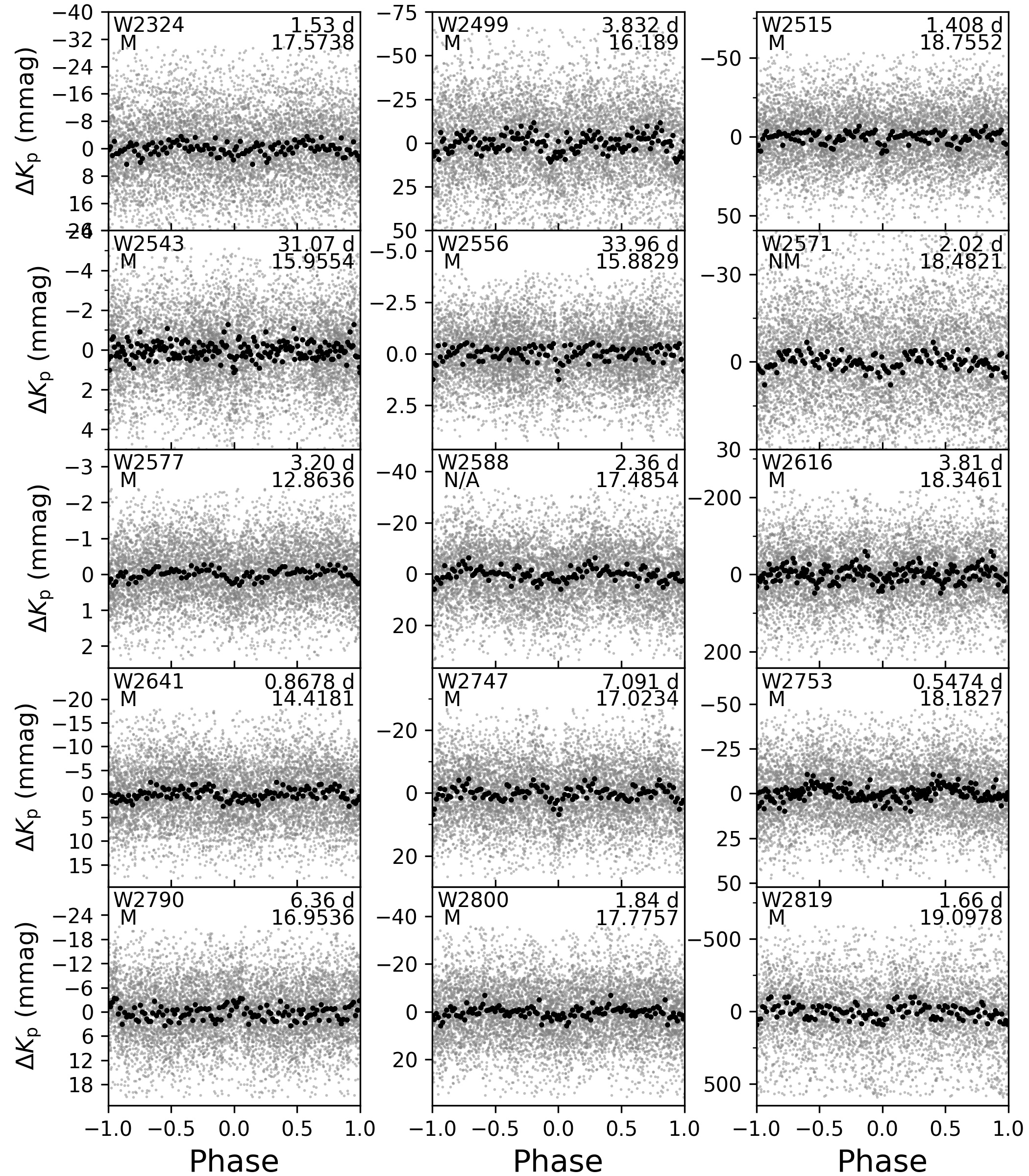}
\end{center}
\caption{\label{fig:suspected3} Same as
Figure~\ref{fig:suspected1}, but for additional suspected
variables. ``N/A'' for W2588's cluster membership status means cluster
membership information not available since there are not {\it Gaia}
DR2 proper motions reported for this object.}
\end{figure*}

\begin{figure*}
\begin{center}
\includegraphics[]{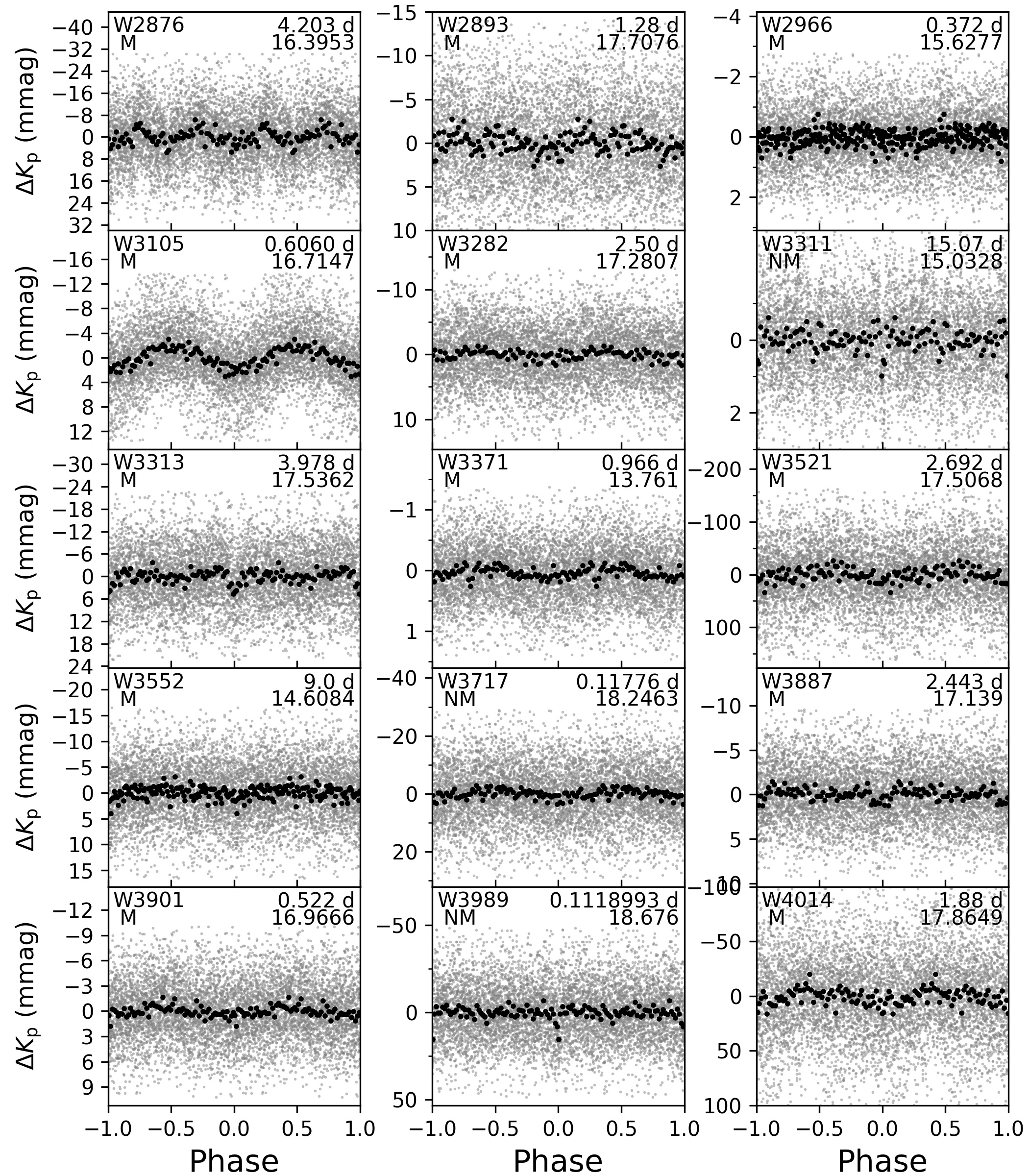}
\end{center}
\caption{\label{fig:suspected4} Same as
Figure~\ref{fig:suspected1}, but for additional suspected
variables.}
\end{figure*}

\begin{figure*}
\begin{center}
\includegraphics[]{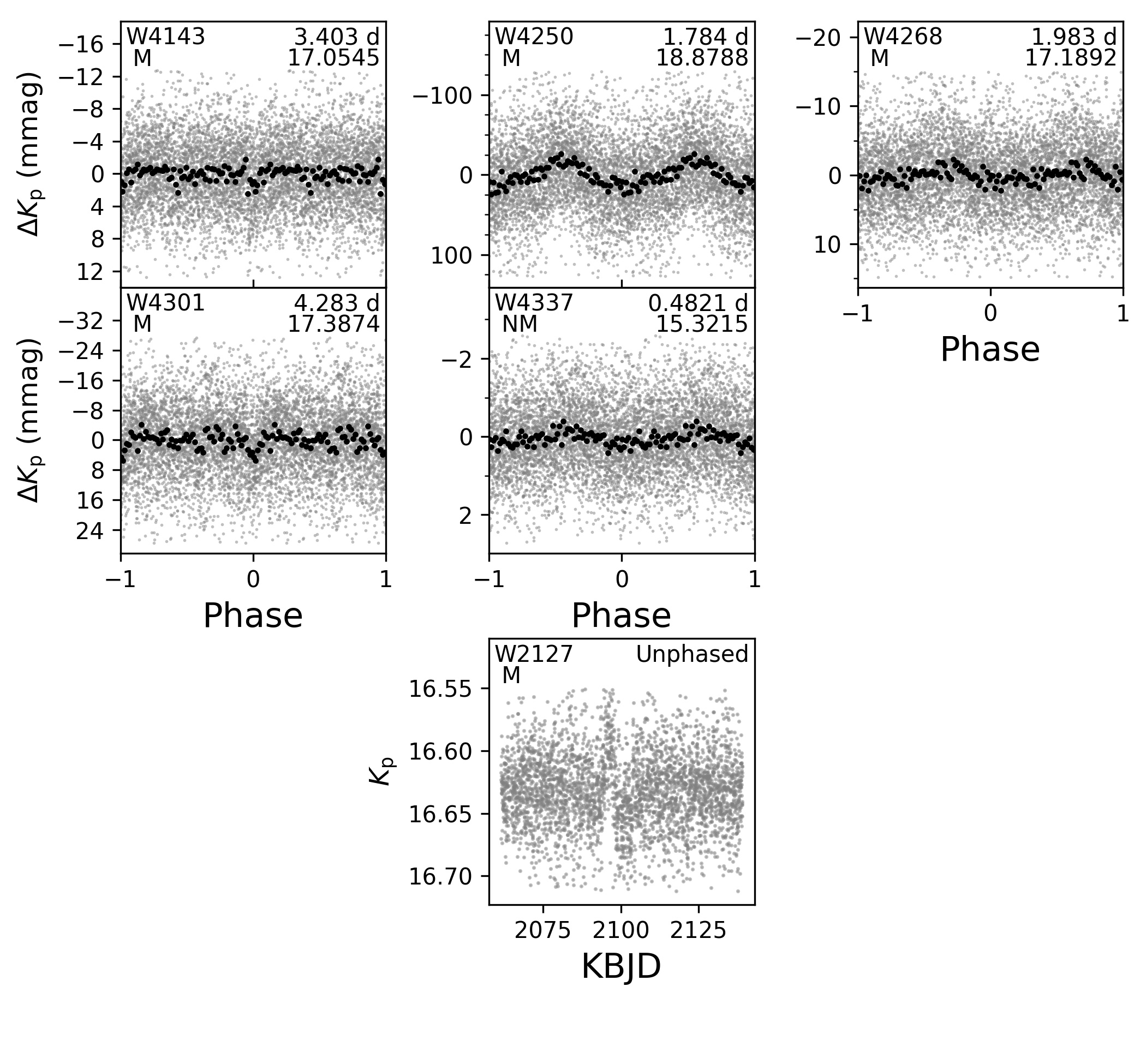}
\end{center}
\caption{\label{fig:suspected5} Same as
Figure~\ref{fig:suspected1}, but for additional suspected
variables. W2127's light curve is unphased since only a single event
was found.}
\end{figure*}

This Appendix presents results for our suspected variables.
The suspected variables can be found in the corresponding sections of
Tables~\ref{tab:clustervars} and \ref{tab:nonclustervars}.  
The
phase-folded light curves are shown in
Figures~\ref{fig:suspected1}, \ref{fig:suspected2}, \ref{fig:suspected3}, \ref{fig:suspected4},
and \ref{fig:suspected5}.  There are a few objects of particular note
in this collection.  

W1834 in Figure~\ref{fig:suspected2} is a cluster
member with a ${\sim}$5 mmag box-shaped {\it
brightening} in the light curve, occurring at a 9.29-day period.
We consider the possibility that this is a gravitational self-lens
from a neutron 
star/black hole in a binary with a main sequence star.  Figure 1
from \cite{masuda2018} shows that the amplitude and period are
consistent with self-lensing from a ${\sim}$10 M$_\sun$ black hole;
however, their equation 7 reveals that for a circular orbit such a
system would have a signal duration of ${\sim}$1 hour, much shorter than
the ${\sim}$20 hours observed.  If this is a self-lensing black hole
system, it would have to be very eccentric.  We would expect
ellipsoidal variability in such a case during a pericenter passage,
but we do not see anything larger than our ${\sim}$0.1 mag floor in
the raw light curves.

Similarly, W2127 in
Figure~\ref{fig:suspected5} is a cluster member that 
has a single observed ${\sim}$50 mmag brightening event
over a ${\sim}$5-day period. Extrapolating from their figure 1 and again
using the equation 7 from \cite{masuda2018} as before, a ${\sim}$10
M$_\sun$ black hole on a ${\sim}$250-day circular orbit would broadly match
the observed light curve. Of course, these situations would require an
orbital inclination near 90\degr, which for the wide orbit of W2127
presents something of a fine-tuning problem, as does the large
eccentricity needed for W1834.  We merely present these as possible
scenarios and do not conclude anything on the nature of the
variability on these objects.

We list here the
reasons we have for marking each of the suspected variables as
suspected rather than definite variables. 

\begin{itemize}

\item W55: Noisy periodogram; low-amplitude phase-folded light curve.

\item W58: Many light curve points from second half of campaign are missing due to blending with bright star.

\item W126: Very short period, ${\sim}$0.3\% away from twice the cadence period.

\item W267: Phase-folded light curve has low amplitude. 

\item W371: Noisy periodogram.

\item W435: Noisy periodogram and  phase-folded signal has low amplitude.

\item W461: Very nearby to W491 and might be blended, W461's period is a bit more than 14 times the period of W491.

\item W829: Small transit depth compared to light curve scatter.

\item W901: Noisy periodogram.

\item W920: Noisy periodogram and phase-folded signal has low amplitude.

\item W951: By-eye judgment call that it is unclear whether this could be a real transit or not.

\item W1056: By-eye evaluation makes it unclear whether this could be a real transit or not.

\item W1068: Noisy periodogram.

\item W1208: Noisy periodogram.

\item W1222: Binned-median points show some bright points in transit in addition to the dimmer points filling out the transit.

\item W1263: Noisy periodogram.

\item W1539: Noisy periodogram and phase-folded signal has low amplitude.

\item W1717: Noisy periodogram and phase-folded signal has low amplitude.

\item W1725: Phase-folded light curve has low amplitude, ambiguous by eye.

\item W1779: Near a saturated star; similar period to W1864, which is also near the same saturated star.

\item W1809: Small transit depth compared to light curve scatter.

\item W1834: Scatter in anti-transit portion of phase appears to be smaller than the rest of the light curve.

\item W1864: Near a saturated star; similar period to W1779, which is also near the same saturated star.

\item W1938: Noisy periodogram.

\item W1947: In a very crowded area of the image; rich, possibly noisy, periodogram.

\item W1953: Possible transit, but depth is not large and not very wide.

\item W2109: Periodogram peak similar in amplitude to other periodogram peaks, but phase-folded signal looks like it could be real.

\item W2126: Noisy periodogram; phase-folded signal has low amplitude.

\item W2127: Signal occurs close to the time the spacecraft's roll changed directions, producing systematics in other light curves around this time, but this is a stronger signal than those other systematics.

\item W2233: Noisy periodogram; looks like an RRab signal and has close to the same period as V9, but they do not quite phase up.


\item W2272: Blending with bright object, producing differing noise characteristics in second half of data relative to first half,  may be producing some kind of unique systematic.

\item W2324: Noisy periodogram; period matches W1189 and is 180\degr\ out of phase, but it is over 55 pixels away.

\item W2499: Noisy periodogram

\item W2515: Noisy periodogram.

\item W2543: Transit not very deep compared to noise.

\item W2556: Only two transits observed.

\item W2571: Noisy periodogram; strange shape to periodogram peak.

\item W2577: A bright star blended with another bright star for which we do not have light curves since they are not {\it Gaia} DR1 sources, thus unsure whether this is the source of variability (though very likely it is).

\item W2588: Noisy periodogram.

\item W2616: Noise characteristics changed halfway through campaign.

\item W2641: Noisy periodogram.

\item W2747: Low-amplitude transit signal.

\item W2753: Based on period, it might be a transformed blend of V29.

\item W2790: Phase-folded light curve has low amplitude.

\item W2800: Phase-folded light curve has low amplitude.

\item W2819: Maybe a transit present, but differing noise characteristics in second half of data relative to first half  may be producing some kind of unique systematic.

\item W2876: Noise characteristics change slightly halfway through campaign; noisy periodogram.

\item W2893: Noisy periodogram.

\item W2966: Phase-folded light curve has low amplitude. 

\item W3105: Six pixels away from and similar variability to V27, but does not phase up. However, we have seen our light curve processing transform blended RRAB signals  into sinusoidal signals with slightly different periods.

\item W3125: Noisy periodogram.

\item W3282: Phase-folded light curve of particularly small amplitude.

\item W3311: Noisy periodogram.

\item W3313: Low-amplitude transit signal.

\item W3371: Phase-folded light curve has low amplitude. 

\item W3521: Noisy periodogram.

\item W3552: Noisy periodogram.

\item W3717: Noisy periodogram; phase-folded light curve has low amplitude.

\item W3887: Small transit depth compared to light curve scatter.

\item W3901: Low-amplitude signal.  Period matches V29, but does not phase up, and is over 120 pixels away.

\item W3989: Noisy periodogram.

\item W4014: Noisy periodogram; phase-folded signal has low amplitude.

\item W4143: Small transit depth compared to light curve scatter.

\item W4250: Near to a bright star that was in the edge region. We do
  not have the light curve for the bright star to see if this signal is a blend.

\item W4268: Period falls within the 1.95-day systematic range, but we
  still decided to keep as a suspected variable based on signal strength.

\item W4301: Noisy periodogram.

\item W4337: Noisy periodogram; phase-folded light curve has low amplitude.

\end{itemize}

\end{appendix}

\end{document}